\documentclass[journal=jctcce,manuscript=article]{achemso}

\usepackage[version=3]{mhchem} 
\usepackage{amsmath,amssymb,amsthm, amsfonts, mathtools, dsfont} 
\usepackage{bbm,bm,tensor, braket}
\usepackage{eqnarray,array,enumerate}
\usepackage{csquotes}
\usepackage{siunitx}
\usepackage{booktabs}
\usepackage{graphicx,wrapfig, caption, float, subcaption, epstopdf, setspace}
\usepackage{hyperref}
\usepackage{longtable}
\usepackage[section]{placeins}
\usepackage{multicol, multirow}
\usepackage{tikz,tkz-euclide}
\usetikzlibrary{shapes.geometric,arrows,arrows.meta, calc, positioning, automata, shadows, backgrounds,decorations.markings,decorations.pathreplacing}
\usepackage{tkz-fct}
\usetikzlibrary{intersections}
\usepackage{pgfplots}
\pgfplotsset{compat=1.9}
\usepgfplotslibrary{fillbetween}
\usepackage{algorithm, algpseudocode}

\usepackage{cleveref}

\newcommand*{\citen}[1]{%
  \begingroup
    \romannumeral-`\x 
    \setcitestyle{numbers}%
    \cite{#1}%
  \endgroup   
}

\author{Arno F{\"o}rster}
\email{a.t.l.foerster@vu.nl}
\affiliation{Theoretical Chemistry, Vrije Universiteit, De Boelelaan 1083, NL-1081 HV, Amsterdam, The Netherlands}

\title{Assessment of the second-order statically screened exchange correction to the random phase approximation for correlation energies}


\DeclareUnicodeCharacter{2212}{-}
\begin{document}

\begin{abstract}
With increasing inter-electronic distance, the screening of the electron-electron interaction by the presence of other electrons becomes the dominant source of electron correlation. This effect is described by the random phase approximation (RPA) which is therefore a promising method for the calculation of weak interactions. The success of the RPA relies on the cancellation of errors, which can be traced back to the violation of the crossing symmetry of the 4-point vertex, leading to strongly overestimated total correlation energies. By addition of second-order screened exchange (SOSEX) to the correlation energy, this issue is substantially reduced. In the adiabatic connection (AC) SOSEX formalism, one of the two electron-electron interaction lines in the second-order exchange term is dynamically screened (SOSEX($W$,$v_c$)). A related SOSEX expression in which both electron-electron interaction lines are statically screened (SOSEX($W(0)$,$W(0)$)) is obtained from the $G3W2$ contribution to the electronic self-energy. In contrast to SOSEX($W$,$v_c$), the evaluation of this correlation energy expression does not require an expensive numerical frequency integration and is therefore advantageous from a computational perspective. We compare the accuracy of the statically screened variant to RPA and RPA+SOSEX($W$,$v_c$) for a wide range of chemical reactions. While both methods fail for barrier heights, SOSEX($W(0)$,$W(0)$) agrees very well with SOSEX($W$,$v_c$) for charged excitations and non-covalent interactions where they lead to major improvements over RPA. 

\end{abstract}

\maketitle

\section{Introduction}
The random phase approximation (RPA)\cite{Macke1950, Bohm1953} has found widespread use in quantum chemistry for the calculation of covalent and non-covalent interaction energies.\cite{Hesselmann2011, Eshuis2011, Eshuis2012a, Ren2012a, Chen2017, Chedid2018, Kreppel2020, Modrzejewski2020} The direct (particle-hole) RPA can be derived in the framework of the adiabatic connection (AC) fluctuation-dissipation theorem (ACFD)\cite{Langreth1975, Langreth1977, Harris1975} or as a subset of terms in the coupled cluster (CC)\cite{Coester1958, Coester1960, Cizek1966, Cizek1969, Paldus1972} singles and doubles (CCD) expansion.\cite{Scuseria2008, Scuseria2013} 

Within Many-body perturbation theory (MBPT)\cite{Abrikosov1975, Mattuck1992, Bruus2004, martin2016}, the RPA is obtained by evaluating the Klein,\cite{Klein1961} or alternatively, the Luttinger-Ward\cite{Luttinger1960a} functional with the self-energy in the $GW$ approximation (GWA) using a (non-interacting) Kohn-Sham (KS)\cite{Kohn1965} Density functional theory (DFT)\cite{Hohenberg1964} Green's function.\cite{Casida1995, Dahlen2006} In the GWA\cite{Hedin1965}, the self-energy is approximated as the first term of its expansion in terms of a screened electron-electron interaction where screening is usually calculated within a pair bubble approximation
\bibnote{The pair bubble approximation is typically also denoted as RPA. To avoid potential confusion with the expression for the correlation energy, we will use the term bubble approximation when referring to the screening.}\cite{martin2016} 
Not only for solids but also for larger molecules it becomes decisive to consider screening which is the main reason for the popularity of the $GW$ method in solid-state calculations.\cite{martin2016} 
The RPA is generally believed to describe long-range electron correlation very accurately Since charge screening is the dominant source of electron correlation in this limit.\cite{Langreth1977, martin2016}


CC and MBPT based methods describe screening by resummation of certain classes of self-energy diagrams to infinite order.\cite{Mattuck1992,Zhang2019,Keller2022} The RPA is the simplest first principle method which accounts for these effects and is implemented with $\mathcal{O}\left(N^4\right)$ scaling with system size using global density fitting (DF).\cite{Eshuis2010} Modern RPA (and $GW$) implementations typically use local density-fitting approaches to calculate the non-interacting polarizability,\cite{Wilhelm2016, 
Wilhelm2018, 
Wilhelm2021, 
Forster2020b, 
Duchemin2019, 
Duchemin2021a} leading to quadratic or cubic scaling in practice, and even effectively linearly scaling implementations (for sufficiently sparse and large systems) have been reported.\cite{Schurkus2016, Luenser2017, Vlcek2017, Graf2018} For these reasons, the RPA is considered a promising method to study weakly correlated large molecules.\cite{Lu2010, Lebegue2010, Eshuis2011, Modrzejewski2020, Nguyen2020} 

At short electron-electron distances, however, charge screening becomes less important for the description of electron correlation and taking into account higher-order contributions to the self-energy via the 4-point vertex function becomes decisive.\cite{Irmler2019a} The absence of these terms in the RPA leads to Pauli exclusion principle violating contributions to the electron correlation energy.\cite{Hummel2019} As a consequence, total correlation energies are much too high compared to exact reference values.\cite{Singwi1968, Jiang2007} 

In contrast to RPA, the approximations to the correlation energy of Møller-Plesset (MP) perturbation theory are free of Pauli principle violating terms. Especially MP2 is relatively inexpensive and can be applied routinely to systems with more than 100 atoms even close to the complete basis set limit. 
However, screening effects are entirely absent in MP perturbation theory and electron correlation is described by HF quasiparticles (QP) interacting via the bare Coulomb interaction instead, neglecting the fact that the interactions between the HF QPs are generally much weaker than the ones between the undressed electrons. This issue is also present in orbital optimized MP2 in which the HF QPs are replaced by MP2 QPs.\cite{Lochan2007, Neese2009d, Kossmann2010} Therefore, MP2 is a suitable method only for (typically small) systems in which screening effects are negligible. 
The divergence of MP perturbation theory for the uniform electron gas (see for instance chapter 10 in ref.~\citen{Mattuck1992} for a thorough discussion) is known at least since early work by Macke\cite{Macke1950} and has been demonstrated later on for metals\cite{Gruneis2010a} and recently also for large, non-covalently bound organic complexes.\cite{Nguyen2020} The divergence of the MP series for small-gap systems is directly related to this issue since the magnitude of the screening is proportional to the width of the fundamental gap.\cite{VanSchilfgaarde2006, Tal2021}  

There have been various approaches to regularize MP2 by an approximate treatment of higher-order screening effects, either using empirical regularizers\cite{Jung2004, 
Lochan2005, 
Grimme2003,
Szabados2006,
Pitonak2009, 
Sedlak2013a,
Goldey2012, 
Goldey2013, 
Goldey2015,
Lee2018,
Monino2022} or diagrammatically motivated modifications\cite{
Pittner2003, 
Keller2022,
Engel2006, 
Jiang2006} or attacking the problem from a DFT perspective.\cite{Daas2021,Daas2022} Starting from the opposite direction, there have been many attempts to correct the RPA correlation energy expression by adding additional terms to improve the description of short-range correlation. This includes range-separation based approaches,\cite{Kurth1999, Yan2000, Angyan2005,Janesko2009a, Janesko2009b, Toulouse2009, Zhu2010, Toulouse2010, Toulouse2011b, Beuerle2017} or augmentations by singles contributions.\cite{Ren2011, Paier2012, Ren2013} Via MBPT, the RPA can generally be improved upon inclusion of the 4-point vertex in the electronic self-energy, either directly, or indirectly through the kernel of the Bethe-Salpeter equation (BSE) for the generalized susceptibility. Following the latter approach, approximations often start from the ACFD and go beyond the Coulomb kernel in the BSE by adding additional terms, for instance exact exchange (exx) (often denoted as exx-RPA)\cite{Hellgren2007, Hellgren2008, Hesselmann2010, Hesselmann2011a, Bleiziffer2015, Mussard2016} and higher order contributions,\cite{Bates2013, Erhard2016, Olsen2019, Gorling2019} or the statically screened $GW$ kernel,\cite{Maggio2016, Holzer2018a, Loos2020} but also empirically tuned functions of the eigenvalues of the KS density-density response.\cite{Trushin2021,Fauser2021} Notice, that the BSE for the generalized susceptibility reduces to a Dyson equation for the density-density response function which makes local kernels very attractive from a computational perspective.

Instead of relying on the ACFD theorem, beyond-RPA energy expressions can also be introduced directly from approximations to the self-energy beyond the GWA. For instance, in RPAx\cite{Colonna2014, Colonna2016, Hellgren2018, Hellgren2021a} a local 4-point vertex obtained from the functional derivative of the \emph{local} exact exchange potential calculated within the optimized effective potential method\cite{Sharp1953, Talman1976, EngelEberhardandDreizler2013} is used in the self-energy. In Freeman's second-order screened exchange (SOSEX) correction,\cite{DavidL.Freeman1977} the HF vertex (i.e. the functional derivative of the \emph{non-local} HF self-energy with respect to the single-particle Green's function) is included in the self-energy directly but not in the screened interaction.\cite{Jansen2010, Paier2010, Paier2012, Ren2012a, Ren2013, Gruneis2009, Hummel2019} Another expression for SOSEX can be obtained by including the static $GW$ kernel in the self-energy but not in the density-density response. This possibility has not been explored until recently\cite{Forster2022} and is the main topic of this work. 

In our recent work, we have assessed the accuracy of the statically screened $G3W2$ correction to the $GW$ self-energy for charged excitations.\cite{Forster2022} This correction has first been applied by Grüneis \emph{at al.}\cite{Gruneis2014} to calculate the electronic structure of solids and is obtained by calculating the self-energy to second-order in the screened Coulomb interaction (equivalent to including the full $GW$ vertex) and then taking the static limit for both terms. The resulting energy expression fulfills the crossing symmetry of the vertex to first order in the electron-electron interaction. Preliminary results for the correlation energies of atoms have been promising.\cite{Forster2022} This realization of SOSEX is computationally more efficient than AC-SOSEX since no expensive numerical frequency integration is required. Here, we assess the accuracy of this method for bond dissociation, atomization energies, barrier heights, charged excitations and non-covalent interactions. Our results show that the statically screened SOSEX variant is comparable in accuracy to AC-SOSEX but we observe important differences in the dissociation of diatomic molecules and charged dimers. 

The remainder of this work is organized as follows. In section~\ref{sec::theory} we give a detailed derivation of the different SOSEX energy expressions. After an outline of our computational approach and implementation in section~\ref{sec::compDetails}, we present and analyze our numerical results in section~\ref{sec::results}. Finally, section~\ref{sec::conclusions} summarizes and concludes this work.

\section{\label{sec::theory}Theory} 
The central object of MBPT is the one-particle irreducible (1PI) electronic self-energy $\Sigma$. It is the sum of all 1PI skeleton diagrams (diagrams which do not contain any self-energy insertions) of $n$th order in the electron-electron interaction $v_c$. It maps the interacting single-particle Green's function $G$ to its non-interacting counterpart $G^{(0)}$ by means of Dyson's equation,\cite{Dyson1949}
\begin{equation}
\label{dyson}
    G(1,2) = G^{(0)}(1,2) + G^{(0)}(1,3)\Sigma(3,4)G(4,2) \;.
\end{equation}
Space, spin, and imaginary time indices are collected as $1 = (\bm{r}_1,\sigma_1,i\tau_1)$. One can always switch between imaginary time and imaginary frequency using the Laplace transforms\cite{Rieger1999}
\begin{equation}
\label{TtoW}
    f(i\tau) = \frac{i}{2\pi} \int d\omega F(i\omega) e^{i\omega \tau}
\end{equation}
and
\begin{equation}
\label{WtoT}
    f(i\omega) = -i \int d\tau F(i\tau) e^{-i\omega \tau} \;.
\end{equation}
In \eqref{dyson}, $G = G_1$ is defined by 
\begin{equation}
\label{greens_definition}
    G_n(1, \dots 2n) = 
    \left\langle
    \Psi^{(N)}_0
    \Big|
    \mathcal{T} 
    \left[
    \hat{\psi}^{\dagger}(1) 
    \hat{\psi}(2) 
    \dots 
    \hat{\psi}^{\dagger}(2n-1) 
    \hat{\psi}(2n) 
    \right]
    \Big| 
    \Psi^{(N)}_0
    \right\rangle \;.
\end{equation}
Here, $\Psi^{(N)}_0$ is the ground state of an $N$-electron system, $\mathcal{T}$ is the time-ordering operator and $\hat{\psi}$ is the field operator. $\Sigma$ is given by
\begin{equation}
\Sigma(1,2) = v_H(1)\delta(1,2) + \Sigma_{xc}(1,2) \;,
\end{equation}
where the second term on the \emph{r.h.s.} can be written as
\begin{equation}
\label{sigma} 
\Sigma_{xc}(1,2) = i G(1,2)W(1,2)
+ i G(1,3)W^{(0)}(1,4)\chi(6,4,5,4^+)\Gamma_{xc}^{(0)}(6,5,2,3) \;.
\end{equation}
For a detailed deviation we refer to the supporting information. We note, that Maggio and Kresse\cite{Maggio2017} and Martin et al.\cite{martin2016} used a similar expression. \Cref{sigma} combines several quantities. These are the particle-hole irreducible 4-point vertex (i.e. the sum of all diagrams which cannot be cut into parts by removing a particle and a hole line),\cite{Baym1961}
\begin{equation}
\label{vertex}
    \Gamma_{Hxc}^{(0)}(1,2,3,4) = \Gamma_H^{(0)}(1,2,3,4) + \Gamma_{xc}^{(0)}(1,2,3,4) = i \frac{\delta\Sigma_{H} (1,3)}{\delta G(4,2)} + i \frac{\delta\Sigma_{xc} (1,3)}{\delta G(4,2)} \;,
\end{equation}
the non-interacting generalized susceptibility,
\begin{equation}
\label{chi0}
    \chi^{(0)}(1,2,3,4) = -i G(1,4)G(2,3) \;,
\end{equation}
and the screened (bare) Coulomb interaction $W$ ($W^{(0)}$). These quantities are related by the Dyson equation 
\begin{equation}
\label{screened-coulomb}
   W(1,2) = W^{(0)}(1,2) + W^{(0)}(1,3)
   P(3,4) W^{(0)}(4,2) \;,
\end{equation}
with
\begin{equation}
    W^{(0)}(1,2) = v_c(\bm{r}_1,\bm{r}_2)\delta_{\sigma,\sigma'}\delta(t_1-t_2) \;,
\end{equation}
given in terms of the bare coulomb interaction $v_c$ and the reducible polarizability  \begin{equation}
\label{polarizability-1}
    P(1,2) = \chi(1,2,3,4)\delta (1,4)\delta (2,3) \;,
\end{equation}
with
\begin{equation}
\label{susceptibility}
    \chi(1,2,3,4) = -iG_2(1,2,3,4) -i G(1,2)G(3,4) \;.
\end{equation}
$\chi$ is related to its non-interacting counterpart $\chi^{(0)}$ by a Bethe-Salpeter equation (BSE),\cite{Salpeter1951, Baym1961}
\begin{equation}
\label{bse}
 \chi(1,2,3,4) = \chi^{(0)}(1,2,3,4)
    + \chi^{(0)}(1,8,3,7) \Gamma_{Hxc}^{(0)}(7,5,8,6) \chi(6,2,5,4) \;,
\end{equation}
which reduces to a Dyson equation for the polarizability $P$ when the xc-contribution to the 4-point vertex is set to zero. One can then also introduce the irreducible polarizability $P^{(0)}$ as
\begin{equation}
\label{polarizability}
    P^{(0)}(1,2) = \chi^{(0)}(1,2,3,4)\delta (1,4)\delta (2,3) \;.
\end{equation}
Using this quantity, \eqref{screened-coulomb} can also be written as 
\begin{equation}
\label{screened-coulomb-2}
   W(1,2) = W^{(0)}(1,2) + W^{(0)}(1,3)
   P^{(0)}(3,4)W(4,2) \;.
\end{equation}
Note, that the equations above are completely equivalent to Hedin's equations.\cite{Hedin1965} Their form given here has the advantages that the BSE appears explicitly and that only 2-point or 4-point quantities occur. Therefore, the resulting equations are invariant under unitary transformations of the basis, as has for instance been pointed out by Starke and Kresse.\cite{Starke2012} or in ref.~\citen{Held2011}

The xc-contribution to the self-energy defined in \eqref{sigma} can also be obtained as the functional derivative 
\begin{equation}
\label{sigma_derivative}
    \Sigma_{xc} = \frac{\delta \Phi [G]}{\delta G} \;.
\end{equation}
$\Phi$ is a universal functional of the interacting $G$ and is defined by\cite{Luttinger1960a, Caruso2013b, martin2016} 
\begin{equation}
\label{LW}
    \Phi[G] = \frac{1}{2}\sum_n \frac{1}{n} \int d1 d2 G(1,2)\Sigma_{xc}^{(n)}[G](2,1) \;.
\end{equation}
As for instance discussed in refs.~\cite{Dahlen2006,Caruso2013b},
If this expression is evaluated with a non-interacting Green's function one directly obtains the exchange-correlation energy from it. A suitable non-interacting Green's function $G^s$ can be obtained from $G^{(0)}$ by
\begin{equation}
       G^{s}(1,2) = G^{(0)}(1,2) + 
    G^{(0)}(1,3) v_{s}(3,4) G^{s}(4,1) \;,
\end{equation}
where
\begin{equation}
    v_s(1,2) = v_H(1,2)\delta(1,2) + v_{xc}(\bm{r}_1, \bm{r}_2) \delta(\tau_{12}) 
\end{equation}
and with $v_{xc}$ being a KS xc-potential mixed with a fraction of HF exchange and $\tau_{12} = \tau_1 - \tau_2$. The correlation energy
\begin{equation}
 E_{c}[G^s] = E_{Hxc}[G^s] - E_{Hx}[G^s]  
\end{equation}
is then given by\cite{Dahlen2006}
\begin{equation}
\label{LW2}
    E_{c} = \frac{1}{2}\sum_{n=2}\frac{1}{n} \int d 1 d2 G^s(1,2) \Sigma^{(n)}(2,1)[G^s] \;.
\end{equation}
The $Hx$ contribution to the electron-electron interaction energy is obtained as 
\begin{equation}
\label{hartree-exchange}
    E_{Hx}[G^s]  = \frac{1}{2} \int d 1 d 2 G^s(1,2) \Sigma^{(1)}(2,1)[G^s] \;.
\end{equation}
In case $G^s$ is the Hartree-Fock (HF) Green's function, \eqref{hartree-exchange} is the HF expression for the Hartree and exchange energy.

In the GWA, the self-energy \eqref{sigma} is approximated as $\Sigma \approx \Sigma_H + i GW$. $W$ is typically calculated within the RPA which consists in approximating 
$\Gamma^{(0)}_{Hxc} \approx \Gamma^{(0)}_{H} $ in the BSE \eqref{bse}. Making both approximations and using \cref{polarizability,screened-coulomb}, the RPA exchange-correlation energy
\begin{equation}
\begin{aligned}
     E^{RPA}_{xc} = & i\frac{1}{2} \int d 1 d2 G^s(1,2) G^s(2,1)W(2,1) \\
     = & - \frac{1}{2} \int d 1 d2 
     P^{(0)}(1,2) \;
     \left\{W^{(0)}(2,1) + \frac{1}{2}W^{(0)}(2,3)P^{(0)}(3,4)W^{(0)}(4,1) + \dots\right\} 
\end{aligned}
\end{equation}
is obtained.\cite{Caruso2013b} Isolating the exchange contribution to the Hartree-exchange energy,
\begin{equation}
    E_x = \int d1d2 \; \delta(\tau_{12})G^s(1,2)W^{(0)}(2,1) G^s(2,1) \;,
\end{equation}
we obtain the RPA correlation energy
\begin{equation}
\begin{aligned}
    E^{RPA}_c = & -\frac{1}{2}\sum_n \frac{1}{n} \int d1d2  \left\{\left[\int d3 P^{(0)}(1,3)W^{(0)}(3,2)\right]^n + \int d3 P^{(0)}(1,3)W^{(0)}(3,2) \right\}  \\ 
    = & \frac{1}{2} \int d1d2\left\{ \ln\left[\delta(1,2) - \int d3 P^{(0)}(1,3)W^{(0)}(3,2)\right] + \int d3 P^{(0)}(1,3)W^{(0)}(3,2) \right\} \;,
\end{aligned} 
\end{equation}
and using \eqref{TtoW} as well as the symmetry of the polarizability on the imaginary frequency axis, its well-known representation due to Langreth and Perdew\cite{Langreth1977} is obtained,
\begin{equation}
    \label{e_rpa}
    \begin{aligned}
    E^{RPA}_c = & \frac{1}{2\pi}\int d \bm{r}_1  \int^{\infty}_0  d \omega \left\{ 
    \ln\left[\delta(1,2) - \int d \bm{r}' P^{(0)}(\bm{r}_1,\bm{r}',i\omega)v_c(\bm{r}',\bm{r}_1)\right] \right. \\ 
    & + \left.\int d \bm{r}' P^{(0)}(\bm{r}_1,\bm{r}',i\omega)v_c(\bm{r}',\bm{r}_1) \right\} \;.
    \end{aligned}
\end{equation}
In this work, we are interested in approximations to the self-energy beyond the GWA. It follows from the antisymmetry of Fermionic Fock space that $G_2$ needs to change sign when the two creation or annihilation operators in \eqref{greens_definition} are interchanged. This property is known as crossing symmetry.\cite{Rohringer2012} In the RPA, the crossing symmetry is violated which leads to the well-known overestimation of absolute correlation energies. However, when the 4-point vertex is approximated by the functional derivative of the Hartree-exchange self-energy the crossing symmetry is fulfilled. We show this explicitly in the supporting information.

Approximations to the self-energy in Hedin's equations always violate the crossing symmetry.\cite{Rohringer2018, Krien2021}
However, with each iteration of Hedin's pentagon, the crossing symmetry is fulfilled up to an increasingly higher order in $v_c$. 
We can then expect to obtain improvements over the RPA energies expressions by choosing a self-energy which fulfills the crossing symmetry to first order in $v_c$. The easiest approximation to the self-energy of this type is obtained from the HF vertex,
\begin{equation}
    \Gamma_{xc}^{(0),HF} (1,2,3,4) = i\frac{\delta \Sigma^{HF}_{xc} (1,3)}{\delta G^s (4,2)} = - W^{(0)}(1,2) \delta(1,4)\delta(3,2) \;.
\end{equation}
Using this expression in \eqref{sigma} with \eqref{chi0} yields the AC-SOSEX contribution to the self-energy.\cite{Ren2015,Maggio2017} We first notice that within the pair bubble approximation for $W$, \eqref{sigma} becomes 
\begin{equation}
\label{ac-sosex}
    \Sigma^{\textrm{SOSEX}(W,v_c)}(1,2) = - \int d3 d4 G^s(1,3)W(1,4) G^s(3,4)G^s(4,2) W^{(0)}(3,2) \;,
\end{equation}
where we have indicated the screening of the electron-electron interaction in the SOSEX expression in the superscript on the \emph{l.h.s.} of \eqref{ac-sosex}. Here we have used the identity $W\chi^{(0)} = W^{(0)}\chi$ in \eqref{sigma} (see supporting information) which is only valid if $W$ is calculated within the RPA. Using the $GW$ self-energy in \eqref{vertex}, to first order in $W^{(0)}$ (ignoring the vriation of $W$ with respect to $G$) the screened exchange kernel is obtained, 
\begin{equation}
    \Gamma_{xc}^{(0),GW} (1,2,3,4) = i\frac{\delta \Sigma^{GW}_{xc} (1,3)}{\delta G^s (4,2)} = - W(1,2) \delta(1,4)\delta(3,2) \;.
\end{equation}
The resulting self-energy is the complete second-order term in the expansion of the self-energy in terms of the screened electron-electron interaction,\cite{Hedin1965}
\begin{equation}
\label{g3w2}
    \Sigma^{\textrm{G3W2}}(1,2) = - \int d3 d4 G^s(1,3)W(1,4) G^s(3,4)G^s(4,2) W(3,2) 
\end{equation}
and contains the AC-SOSEX self-energy. 
\begin{figure}[hbt!]
    \centering
    \includegraphics[width=1.0\textwidth]{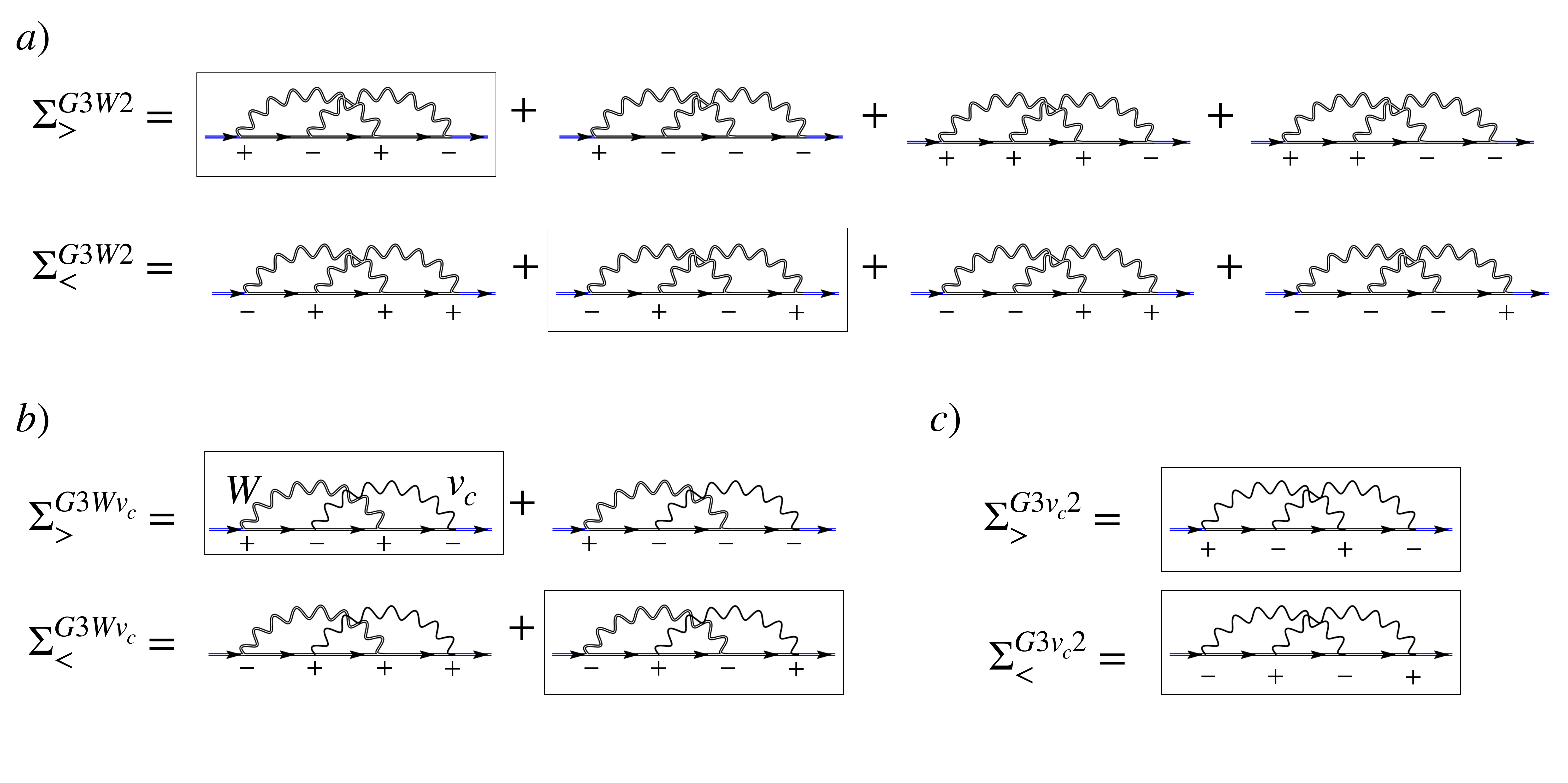}
    \caption{Diagrammatic representation of the different contributions to the second order exchange (SOX) term. pluses and minuses denote the different branches on the Keldysh contour. The double and single wiggly lines are screened and bare electron-electron interactions, respectively a) Greater and lesser contributions to the full $G3W2$ the self-energy term. b) Greater and lesser components of the SOSEX self-energy c) Greater and lesser components of the MP2 self-energy. The static approximation to the $G3W2$ self-energy is the same, with the bare electron-electron interaction lines replaced by the statically screened ones. The black part of the diagrams are the contributions to the self-energy only which, combined with the blue lines, yield the corresponding single-particle propagator.}
    \label{fig::diagrams}
\end{figure}
The $G3W2$ self-energy can be decomposed into eight skeleton diagrams on the Keldysh contour,\cite{VanLeeuwen2015} but the AC-SOSEX self-energy only into four.\cite{Stefanucci2014} Diagrammatically, this is shown in figure~\ref{fig::diagrams}a) and \ref{fig::diagrams}b), respectively. In practice, the evaluation of the resulting energy expression requires to perform a double frequency integration while the evaluation of the AC-SOSEX energy only requires a single frequency integration. Since the computation of the AC-SOSEX term is already quite cumbersome, the complete $G3W2$ energy expression is therefore not a good candidate for an efficient beyond-RPA correction. Instead, we take the static limit in both $W$ in \eqref{g3w2} to arrive at a self-energy expression similar to AC-SOSEX,
\begin{equation}
\label{sosex_w0w0}
    \Sigma^{\textrm{SOSEX}(W(0),W(0))}(1,2) = - \int d3 d4 G^s(1,3)W(1,4) G^s(3,4)G^s(4,2) W(3,2)\delta(\tau_{32}) \delta(\tau_{14}) \;,
\end{equation}
whose diagrammatic form is shown in figure~\ref{fig::diagrams}c). Due to the presence of the two $\delta$-functions, only two out of the eight diagrams of the $G3W2$ term remain. This expression is similar to the MP2 self-energy, with the only difference that the bare electron-electron interaction is replaced by the statically screened one. However, the resulting expression for the correlation energy will be different due to the factors $\frac{1}{n}$ in \eqref{LW}. Using \eqref{screened-coulomb}, eq. \eqref{sosex_w0w0} can be written as 
\begin{equation}
    \Sigma^{\textrm{SOSEX}(W(0),W(0))}(1,2) = 
    \Sigma^{\textrm{MP2-SOX}}(1,2) + 
    \Sigma^{\delta\textrm{MP2-SOX}}(1,2) \;,
\end{equation}
with the first term being the second-order exchange (SOX) term in MP2 and with the remainder accounting for the screening of the electron-electron interaction. Defining
\begin{equation}
\delta W(1,2) =  \int d3 d4 W^{(0)}(1,3)P(3,4) W^{(0)}(4,2) \;,
\end{equation}
it can be written as
\begin{equation}
\begin{aligned}
    \Sigma^{\delta\textrm{MP2-SOX}}(1,2) = &  - \int d3 d4 G^s(1,3) G^s(3,4)G^s(4,2) 
    \left[
    W^{(0)}(1,4) \delta W(3,2) \delta(\tau_{32}) \right. \\ 
    & \left. 
    \qquad +\delta W(1,4)\delta(\tau_{14}) W^{(0)}(3,2)
    + \delta W(1,4)\delta(\tau_{14})
     \delta W(3,2) \delta(\tau_{32}) \right] \;.
\end{aligned}
\end{equation}
In the same way one can see, that the statically screened $GW$ vertex contains the HF vertex. The same is obviously true for all other flavors of SOSEX, and therefore all of them fulfill the crossing symmetry of the full 4-point vertex to first order in the electron-electron interaction. Therefore, all of these approximations compensate the overestimation of the electron correlation energy in the RPA.

In contrast to the RPA which is efficiently evaluated in a localized basis, beyond-RPA energies are most easily formulated in the molecular spin-orbital basis $\left\{\phi_i(\bm{r,\sigma})\right\}$ in which the time-ordered KS Green's function is diagonal,
\begin{equation}
\begin{aligned}
\label{g0}
    G^s_{kk'}(i\tau_{12}) = & 
    \delta_{kk'}\Theta(\tau_{12}) G^{>}_{kk'}(i\tau_{12}) - 
    \delta_{kk'}\Theta(\tau_{21}) G^{<}_{kk'}(i\tau_{21}) \\
    G^{>}_{kk}(i\tau_{12}) = &
    i \left(1 - f(\epsilon_k)\right)
    e^{-\epsilon_k\tau_{12}} \\
    G^{<}_{kk}(i\tau_{12}) = &
    i f(\epsilon_k)
    e^{-\epsilon_k\tau_{12}} \;.
    \end{aligned}
\end{equation}
The $\epsilon_k$ denote KS eigenvalues which are understood to be measured relative to the chemical potential $\mu$ and $f(\epsilon_k)$ denotes the occupation number of the $k$th orbital. One can now obtain energy expressions analogous to \eqref{e_rpa}. For example, inserting the AC-SOSEX self-energy \eqref{ac-sosex} into \eqref{LW2}, we obtain
\begin{equation}
\label{sosex_lambda}
    \begin{aligned}
        E^{\textrm{SOSEX}}_c = &\frac{1}{2} \int d1 d2 d3 d4 \;
        G^s(1,2)G^s(2,3)G^s(3,4)G^s(4,1) \\
        & \times
        \left\{
        \frac{1}{2} W^{(0)}(3,1)W^{(0)}(2,4) + \frac{1}{3} \int d5 d6 
        W^{(0)}(3,1)W^{(0)}(2,5)P^{(0)}(5,6)W^{(0)}(6,4) + \dots 
        \right\} \;.
    \end{aligned}
\end{equation}
In contrast to the RPA energy expression, the terms in this equation cannot be summed exactly due to the presence of the $1/n$-terms. However, defining 
\begin{equation}
\label{lambda_eq}
    \Sigma_{Hxc}^{\lambda} = \sum_{n=1}^{\infty} \lambda^n \Sigma_{Hxc}^{(n)} \left[G^s, v_c\right] 
\end{equation}
we can rewrite \eqref{LW2} as an integral over a coupling constant $\lambda$,
\begin{equation}
    E_{c} = \frac{1}{2}\sum_{n=2}\frac{1}{n} \int d 1 d2 G^s(1,2) \Sigma_{Hxc}^{(n)}(2,1)[G^s] = \frac{1}{2} \int^1_0 \frac{d \lambda}{\lambda} 
    \int d 1 d2 G^s(1,2) \Sigma_{Hxc}^{(\lambda)}(2,1)[G^s] \;,
\end{equation}
Therefore, \eqref{lambda_eq} becomes 
\begin{equation}
    \Sigma_{Hxc}^{\lambda} = 
    \sum_{n=1}^{\infty} \Sigma_{Hxc}^{(n)} \left[G^s, \lambda v_c\right]
    = 
    \sum_{n=1}^{\infty} \Sigma_{Hxc}^{(n)} \left[G^s, W^{(\lambda)}\right] \;,
\end{equation}
where $W^{(\lambda)}$ is defined as in \eqref{screened-coulomb-2}, with $W^{(0)}$ replaced by $\lambda W^{(0)}$. Defining 
\begin{equation}
\label{overlineW}
    \overline{W} = \int^1_0 d \lambda W^{(\lambda)} \;,
\end{equation}
and 
\begin{equation}
    \overline{\Sigma} = \Sigma\left[\overline{W}\right]
\end{equation}
the correlation energy becomes
\begin{equation}
 \label{general_ec}
        E_c = \frac{1}{2} \int d 1 d2 \; G^s(1,2) \overline{\Sigma}_c(2,1) \;.
\end{equation}
The integral in \eqref{overlineW} needs to be computed numerically, but converges typically very fast when Gauss-Legendre grids are employed.\cite{Ren2013} In ref.~\citen{Rodriguez-Mayorga2021} a trapezoidal rule for the solution of this integral has been used and also ref.~\citen{Hesselmann2011} suggests that this choice is often suitable for the calculation of correlation energies within the RPA and beyond. This choice is very well justified for weakly correlated systems for which the adiabatic connection is approximately a straight line.\cite{Becke2014, Vuckovic2020} Below, we will assess the effect of such approximate coupling constant integration on absolute and relative correlation energies for non-covalent interactions. Notice, that using a trapezoidal rule, \eqref{general_ec} reduces to
\begin{equation}
\label{mp2-generate}
        E_c = \frac{1}{4} \int d 1 d2 \; G^s(1,2) \Sigma_c(2,1) \;,
\end{equation}
and when the statically screened $G3W2$ self-energy \eqref{sosex_w0w0} is used in this expression, the energy expression of ref.~\citen{Forster2022} is obtained. When additionally both $W(0)$ are replaced by $v_c$, \eqref{mp2-generate} gives the MP2 correlation energy (evaluated with $G^s$).\cite{Dahlen2006} 

Using \eqref{general_ec}, simple expressions for the AC-SOSEX energy in the basis of KS orbitals is obtained. With \cref{general_ec,ac-sosex,g0} we have 
\begin{equation}
\begin{aligned}
    E^{\textrm{SOSEX}(W,v_c)} = & \frac{i}{2}\sum_{pqrs} \int d\tau_{12} d \tau_3 
    G^s_p(\tau_{13})
    G^s_q(\tau_{31})
    G^s_r(\tau_{12})
    G^s_s(\tau_{21}) W^{(0)}_{spqr}\overline{W}_{rspq}(\tau_{23}) \\ 
    = & 
- \frac{1}{4 \pi} \sum_{pqrs} \int d \omega'
    W^{(0)}_{spqr}\overline{W}_{rspq}(i\omega')     
    \int d\tau_{12}
    G^s_r(\tau_{12}) 
    G^s_s(\tau_{21}) \\
    & \times \underbrace{ \int d \tau_3 
    e^{-i\omega' \tau_{23}}
    G^s_p(\tau_{13})
    G^s_q(\tau_{31})}_{I(i\tau_{12})} \;.
\end{aligned} 
\end{equation}
In going from the second equations, we have used \eqref{TtoW} to transform $W$ to the imaginary frequency axis. The integral over $\tau_3$ can be evaluated by splitting it at $\tau_1$ and using the definition of the KS Green's function \eqref{g0},
\begin{equation}
    I(i\tau_{12}) =  \frac{    \left[
    \left(1-f(\epsilon_p)\right)f(\epsilon_q)
    - \left(1-f(\epsilon_q)\right)f(\epsilon_p)
    \right] e^{i\omega'\tau_{12}}}{\epsilon_p - \epsilon_q + i\omega'}
    =  - e^{i\omega'\tau_{12}} 
    \frac{f(\epsilon_p) - f(\epsilon_q)}{\epsilon_p - \epsilon_q + i\omega'}
\end{equation}
The remaining integral over $\tau_{12}$ is
\begin{equation}
    I_{\tau_{12}} = -\int 
    G^{(0)}_r(\tau_{12}) 
    G^{(0)}_s(\tau_{21}) e^{i\omega'\tau_{12}} d\tau_{12} = 
   \frac{f(\epsilon_r) - f(\epsilon_s)}
   {\epsilon_r - \epsilon_s - i\omega'} \;,
\end{equation}
so that the correlation energy becomes 
\begin{equation}
\label{sosex-before}
   E^{\textrm{SOSEX}(W,v_c)} = - \frac{1}{4 \pi} \sum_{pqrs} \int d \omega'
    W^{(0)}_{spqr}\overline{W}_{rspq}(i\omega') 
   \frac{f(\epsilon_r) - f(\epsilon_s)}
   {\epsilon_r - \epsilon_s - i\omega'}
    \frac{f(\epsilon_p) - f(\epsilon_q)}{\epsilon_p - \epsilon_q + i\omega'}
\end{equation}
Each of the nominators can only give a non-vanishing contribution if one of the two occupation numbers are zero. If the difference of the occupation numbers is $-1$, we simply flip sign in the denominator. Without loss of generality we can then decide that the indices $r$ and $p$ belong to occupied and the indices $s$ and $q$ to virtual single-particle states. \Cref{sosex-before} then becomes 
\begin{equation}
\label{e_sosex}
       E^{\textrm{SOSEX}(W,v_c)} = - \frac{1}{4\pi} \sum^{occ}_{ij}\sum^{virt}_{ab} \int^{\infty}_0 d \omega
       \overline{W}_{iajb}(i\omega) W^{(0)}_{jaib}
    \frac{4(\epsilon_i - \epsilon_a)(\epsilon_j - \epsilon_b)}
    {\left[(\epsilon_i - \epsilon_a)^2 + \omega^2\right]
    \left[(\epsilon_j - \epsilon_b)^2 + \omega^2\right]} \;.
\end{equation}
For a closed-shell system we can also sum over spins which gives us an additional factor of 2. The resulting expression is then equivalent to the one of ref.~\cite{Ren2013}. In the spin-orbital basis, the SOSEX$(W(0),W(0))$ Correlation energy is obtained from \eqref{g3w2} and \eqref{g0} as
\begin{equation}
\label{e_sosex2}
\begin{aligned}
    E^{\textrm{SOSEX}(W(0),W(0))} = & - \frac{1}{4}\sum_{pqrs} \int d\tau_{12}
    G^s_p(\tau_{12})
    G^s_q(\tau_{21})
    G^s_r(\tau_{12})
    G^s_s(\tau_{21}) \\ 
    & \times \overline{W}_{spqr}(i\omega=0)\overline{W}_{rspq}(i\omega=0) \\ 
    = & 
    - \frac{1}{2}\sum^{occ}_{ij}\sum^{virt}_{ab}
    \frac{\overline{W}_{spqr}(i\omega=0)\overline{W}_{rspq}(i\omega=0)}
    {\epsilon_i + \epsilon_j - \epsilon_a - \epsilon_b} \;.
\end{aligned}
\end{equation}
This is the expression we have introduced in ref.~\citen{Forster2022}. It is completely equivalent to the exchange term in MP2 with the bare electron-electron interaction replaced by the statically screened, coupling constant averaged one. Both RPA+SOSEX variants can be understood as renormalized MP2 expressions and allow for a clear diagrammatic interpretation. In the next section, we briefly outline our implementation of these expressions, before we proceed by assessing their accuracy for correlation energies in sec.~\ref{sec::results}.

\section{\label{sec::compDetails}Technical and Computational Details} 
All expressions presented herein have been implemented in a locally modified development version of the Amsterdam density functional (ADF) engine of the Amsterdam modelling suite 2022 (AMS2022).\cite{adf2022} The non-interacting polarizability needed to evaluate \eqref{e_rpa} and \eqref{screened-coulomb-2} is calculated in imaginary time with quadratic scaling with system size in the atomic orbital basis. The implementation is described in detail in ref.~\citen{Forster2020b}. In all calculations, we expand the KS Green's functions in correlation consistent bases of Slater-type orbitals of triple- and quadruple-$\zeta$ quality (TZ3P and QZ6P, respectively).\cite{Forster2021} All 4-point correlation functions (screened and unscreened Coulomb interactions as well as polarizabilities) are expressed in auxiliary basis sets of Slater type functions which are usually 5 to 10 times larger than the primary bases. In all calculations, we use auxiliary basis sets of \emph{VeryGood} quality. The transformation between primary and auxiliary basis (for the polarizability) is implemented with quadratic scaling with system size using the pair-atomic density fitting (PADF) method for products of atomic orbitals.\cite{Krykunov2009, Wirz2017} For an outline of the implementation of this method in ADF, we refer to ref.~\citen{Forster2020}. Eq.~\eqref{e_rpa} is then evaluated in the basis of auxiliary fit functions with cubic scaling with system size. \Cref{e_sosex,e_sosex2} are evaluated with quintic scaling with system size in the canonical basis of KS states. This implementation is completely equivalent to the canonical MP2 implementation outlined in ref.~\citen{Forster2020}. 

Eq.~\eqref{overlineW} is evaluated using Gauss-Legendre grids with 8 points, except for the potential energy curves where 20 points have been used. At stretched bonds, the integrands become increasingly non-linear and a large number of integration points are necessary. As discussed in detail in the supporting information, for non-covalent interactions a single integration point does generally suffice and therefore we have used a single integration point only for all calculations for the S66 and S66x8 database. In case of a single $\lambda-$point, a trapezoidal rule is used for integration. 

Imaginary time and imaginary frequency variables are discretized using non-uniform bases $\mathcal{T} = \left\{\tau_{\alpha}\right\}_{\alpha = 1, \dots N_{\tau}}$ and $\mathcal{W} = \left\{\omega_{\alpha}\right\}_{\alpha = 1, \dots N_{\omega}}$ of sizes $N_{\tau}$ and $N_{\omega}$, respectively, tailored to each system. More precisely, \eqref{TtoW} and \eqref{WtoT} are then implemented by splitting them into sine- and cosine transformation parts as
\begin{equation}
\label{sine-cosine}
\begin{aligned}
    \overline{F}(i\omega_{\alpha}) = &  \sum_{\beta} \Omega^{(c)}_{\alpha\beta} \overline{F}(i\tau_\beta) \\
    \underline{F}(i\omega_{\alpha}) = &  \sum_{\beta} \Omega^{(s)}_{\alpha\beta} \underline{F}(i\tau_\beta) \;, 
\end{aligned}
\end{equation}
where $\overline{F}$ and $\underline{F}$ denote even and odd parts of $F$, respectively. The transformation from imaginary frequency to imaginary time only requires the (pseudo)inversion of $\Omega^{(c)}$ and $\Omega^{(s)}$, respectively. Our procedure to calculate $\Omega^{(c)}$ and $\Omega^{(s)}$  as well as $\mathcal{T}$ and $\mathcal{W}$ follows Kresse and coworkers.\cite{Kaltak2014,Kaltak2014a,Liu2016} The technical specifications of our implementation have been described in the appendix of ref.~\citen{Forster2021}. 

We use in all calculations grids of 24 points in imaginary time and imaginary frequency which is more than sufficient for convergence.\cite{Forster2020} The final correlation energies are then extrapolated to the complete basis set limit using the relation ,\cite{Helgaker1997}
\begin{equation}
    \label{helgaker}
    E_{CBS} = E_{QZ} + \frac{E_{QZ} * 4^3 - E_{TZ} * 3^3}{4^3-3^3} \;, 
\end{equation}
where $E_{QZ}$ ($E_{TZ}$) denotes the total energies at the QZ6P (TZ3P) level. The extrapolation scheme has been shown to be suitable for correlation consistent basis sets but cannot be used for KS or HF contributions.\cite{Helgaker1997, Jensen2013} Therefore, we do not extrapolate the DFT energies, but assume them to be converged on the QZ level. Since the basis set error is not completely eliminated with this approach, we also counterpoise correct all energies, taking into account 100 \% of the counterpoise correction. With these settings, we assume all our calculated values to be converged well enough to be able to draw quantitative conclusions about the performance of the methods we benchmark herein. We use the \emph{VeryGood} numerical quality for integrals over real space and distance cut-offs. Dependency thresholds\cite{Forster2020b} have been set to $5e^{-4}$. 
All Full configuration interaction calculations have been performed with the code by Knowles and Handy.\cite{Knowles1984,Knowles1989} The 1- and 2-electron integral which are required as input have been generated with ADF.

\section{\label{sec::results}Results} 

\subsection{Dissociation Curves}

The potential energy curves of small diatomic molecules serve as an important test for electronic structure methods. We first consider molecules with different bonding types for which we were able to calculate FCI reference values: \ce{H2} is covalently bound, \ce{LiH} is an ionic molecule, and \ce{He2} has a very weak, non-covalent bond. 

\begin{figure}
    \centering
    \includegraphics[width=1.0\textwidth]{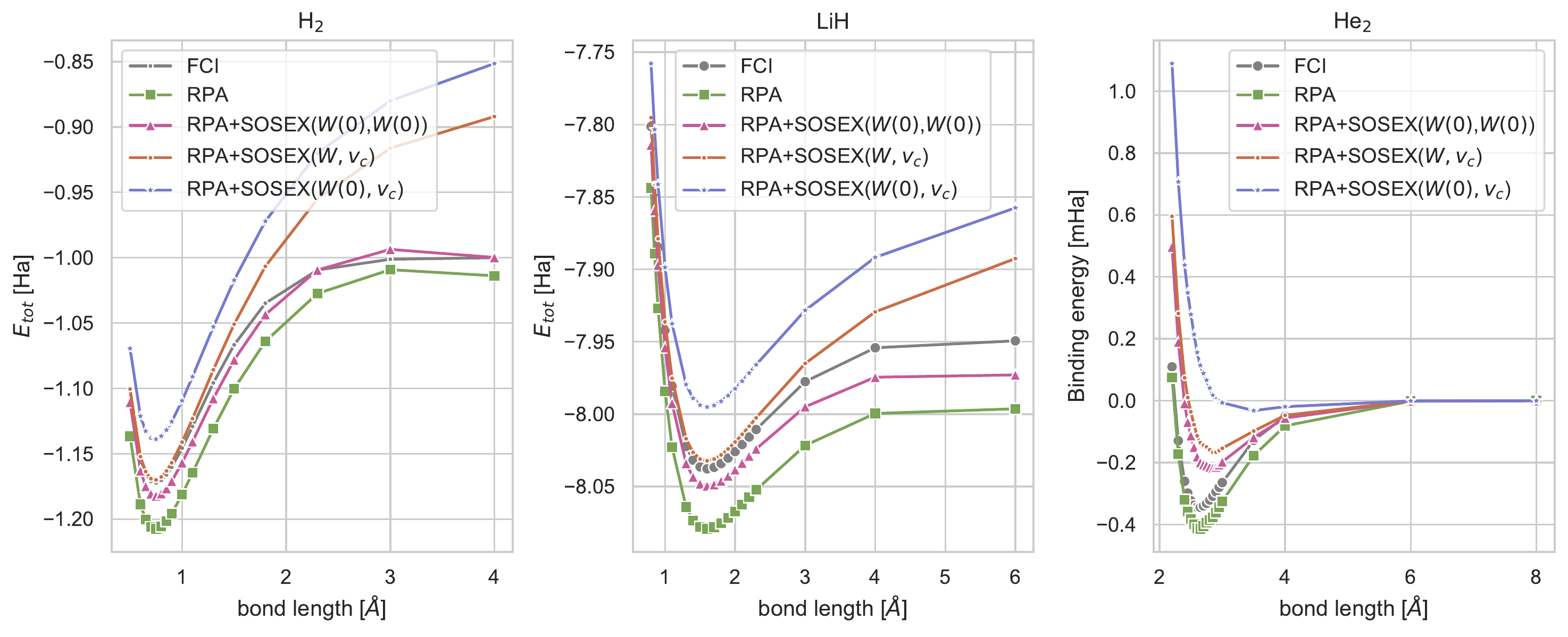}
    \caption{Potential energy curves (in Hartree) of \ce{H2} (left) and \ce{LiH} (middle), as well as binding energy (in mHartree) as a function of system size for \ce{He2} on the right using FCI, RPA@PBE and different variants of RPA+SOSEX@PBE. For \ce{H2} and \ce{He2}, all calculations have been performed with the TZ3P basis set. For \ce{Lih}, all calculations have been performed using the TZP basis set.}
    \label{fig::dissociation_curves}
\end{figure}

The dissociation curve of \ce{H2} calculated with RPA+SOSEX($W(0)$,$W(0)$)@PBE is the red line in figure~\ref{fig::dissociation_curves}. Our calculations are not converged with respect to the basis set size but comparison of our dissociation curves calculated with RPA@PBE and RPA+SOSEX($W$,$v_c$)@PBE to the ones presented in ref.~\citen{Paier2010} and~\citen{Henderson2010} clearly shows that their qualitative behavior is reproduced correctly. It is known that RPA describes the dissociation of covalently bonded molecules qualitatively correctly while RPA+SOSEX($W$,$v_c$) and other exchange-corrected RPA approaches fail.\cite{Paier2010,Henderson2010,Hesselmann2011a} Here we find that also RPA+SOSEX($W(0)$,$W(0)$) dissociates the hydrogen molecule correctly and that the potential energy curve has a similar shape than the RPA one. Henderson and Scuseria have argued that the self-correlation in the RPA mimics static correlation effects\cite{Henderson2010} whose good description is necessary to dissociate \ce{H2} correctly. The fact that in RPA+SOSEX($W(0)$,$W(0)$ the self-correlation error is eliminated to some large extent (also see table~1 in the SI) but not completely therefore explains the similarity to the RPA dissociation curve.

To rationalize this result further, we also calculated the dissociation curve within the static limit of RPA+SOSEX($W$,$v_c$), RPA+SOSEX($W(0)$,$v_c$) (blue curve). This shows that the screening of the second electron-electron interaction line is responsible for the qualitative differences between SOSEX($W$,$v_c$) and SOSEX($W(0)$,$W(0)$). It should also be noted that the RPA+SOSEX($W(0)$,$W(0)$) dissociation curve of \ce{H2} very closely resembles the one calculated by Bates and Furche using the approximate exchange kernel (AXK) correction to the RPA.\cite{Bates2013} SOSEX($W(0)$,$W(0)$) and the AXK kernel have in common that both electron-electron interaction lines are screened. For \ce{LiH}, we find a similar behavior than for \ce{H2}. For \ce{He2} (notice that we plotted here the binding energy and not the total energy) we see that all approaches give the correct dissociation limit. 

\begin{table}[hbt!]
    \centering
    \begin{tabular}{lccccc}
    \toprule 
    method & \ce{H2} & \ce{LiH} & \ce{He2} & \ce{F2} & \ce{Be2} \\ 
    \midrule
exp.                         &       &       &       & 1.413\cite{Peterson1993} & 2.320\cite{Honisch2009} \\
accurate                     & 0.741 & 1.601 & 2.626 & 1.413\cite{Peterson1993} & 2.320\cite{Roeggen2005} \\ 
RPA                      & 0.742 & 1.597 & 2.632 & 1.437  & 2.403 \\
RPA + SOSEX($W(0),W(0)$) & 0.744 & 1.605 & 2.852 & 1.444  & 2.424 \\
RPA + SOSEX($W,v_c$)    & 0.738 & 1.594 & 2.871 & 1.364  & --     \\ 
RPA + SOSEX($W(0),v_c$) & 0.735 & 1.599 & 3.542 & 1.348  & --     \\
    \bottomrule 
    \end{tabular}
    \caption{Equilibrium bond length of selected molecules. All values are in $\AA$. The bond lengths for \ce{H2}, \ce{He2}, and \ce{LiH} have been calculated using the TZ3P and TZP basis sets to make them comparable to the FCI result. The bond lengths for \ce{F2} and \ce{Be2} have been obtained using the QZ6P basis set. All RPA(+SOSEX) calculations have been performed with a PBE Green's function.}
    \label{tab::bond_length}
\end{table}

From these potential energy curves, we also extracted the equilibrium bond lengths via cubic spline interpolation. These are shown in table~\ref{tab::bond_length} Around the equilibrium distances, RPA+SOSEX($W$,$v_c$) generally gives the best energies but this does not necessarily translate into the best bond lengths. For the covalently bound molecules \ce{LiH} and \ce{F2} as well as \ce{LiH} RPA+SOSEX($W$,$v_c$) underestimates and RPA+SOSEX($W(0)$,$W(0)$) overestimates the bond lengths. Again,  RPA+SOSEX($W(0)$,$W(0)$) behaves qualitatively similar to RPA. For \ce{He2}, both approaches give similar results, while RPA+SOSEX($W(0)$,$v_c$) fails completely. On the other hand, unlike RPA+SOSEX($W(0)$,$W(0)$),  RPA+SOSEX($W$,$v_c$) does predict an unbound \ce{Be2} dimer.

\subsection{Dissociation of charged Dimers}

\begin{table}[hbt!]
    \centering
    \begin{tabular}{lcccc}
    \toprule 
    & RPA & SOSEX($W$,$v_c$) & SOSEX($W(0)$,$v_c$) & SOSEX($W(0)$,$W(0)$) \\
    \midrule 
    \ce{H2+}   \\ \cline{1-1}
       1.0   &   5.19 &  0.76 &  -2.58 &   3.09 \\
       1.25  &   7.59 & -0.26 &  -5.33 &   5.19 \\
       1.50  &  11.21 & -1.31 &  -8.23 &   8.89 \\
       1.75  &  16.15 & -2.30 & -11.14 &  14.27 \\
    \ce{He2+}   \\ \cline{1-1}  
       1.0   &  13.23 &  0.23 &  -5.30  &  14.34 \\
       1.25  &  25.40 & -2.84 & -12.91  &  27.56 \\
       1.50  &  40.60 & -5.64 & -20.32  &  44.79 \\
       1.75  &  56.76 & -7.65 & -25.76  &  63.38 \\
    \ce{(NH3)2+} \\ \cline{1-1}
       1.0   &   5.89 & 15.17 &  24.91 &  16.23 \\
       1.25  &  13.00 & 20.08 &  36.23 &  33.50 \\
       1.50  &  20.61 & 21.89 &  42.78 &  50.41 \\
       1.75  &  30.88 & 15.14 &  28.73 &  61.48 \\
    \ce{(H2O)2+} \\ \cline{1-1}
       1.0   &  10.19 & 29.79 &  51.70 &  33.79 \\
       1.25  &  20.62 & 12.16 &  21.61 &  38.68 \\
       1.50  &  31.88 &  2.35 &   4.58 &  50.58 \\
       1.75  &  42.08 &  0.50 &   5.47 &  65.61 \\
      \midrule 
    MAD & 21.95 & 8.63 & 19.22 & 33.24 \\
    \bottomrule
    \end{tabular}
    \caption{Errors in kcal/mol for the charger dimers in the SIE4x4 benchmark set calculated with RPA and different variants of RPA+SOSEX. PBE orbitals have been used in all calculations}
    \label{tab::sie4x4}
\end{table}

In table~\ref{tab::sie4x4} we investigate the dissociation of four charged dimers by means of the SIE4x4 dataset.\cite{Goerigk2017} Here, the self-correlation error of RPA leads to considerable underbinding,\cite{Paier2010, Henderson2010, Ren2012a} whereas RPA+SOSEX($W$,$v_c$) is exact,\cite{Gruneis2009} the remaining error for \ce{H2} being due to basis set errors as well as the fact that PBE orbitals have been used. Furche and coworkers have observed a catastrophic failure of RPA+SOX for \ce{(NH3)2+} and \ce{(H2O)2+}\cite{Chen2018} and also SOSEX($W(0)$,$W(0)$) considerably deteriorates the RPA results for those systems. Only for \ce{H2+}, one finds that the partial cancellation of the RPA self-correlation leads to small improvements over RPA.

\subsection{Thermochemistry and Kinetics}

We move on to assess the performance of RPA+SOSEX($W(0)$,$W(0)$) for reaction types which are relevant for thermochemistry and kinetics.  Total atomization energies, ionization potentials and electron affinities as well as barrier heights of different reactions serve hereby as important testing grounds. For this work, we calculated the atomization energies (defined as the total energy of the molecule minus the sum of the energies of the atomic fragments) of the 144 small and medium molecules in the W4-11 dataset.\cite{Karton2011}  
The reference values have been calculated using the highly accurate W4 protocol.\cite{Karton2006} For barrier heights, we use the BH76 database which is a compilation of the HTBH38\cite{Zhao2005a} and NHTBH38\cite{Zhao2005b} databases for barrier heights by Truhlar and coworkers, which are typically used in benchmarks of (beyond-)RPA methods.\cite{Paier2012,Eshuis2012a,Ren2012a, Ren2013} The reference values have been calculated with the W2-F12 protocol.\cite{Karton2012b,Goerigk2017} To benchmark the performance for ionization potentials and electron affinities we employ the G21IP and G21EA databases by Pople and coworkers and use the original experimental reference values.\cite{Curtiss1991}

\begin{figure}[hbt!]
    \centering
    \includegraphics[width=\textwidth]{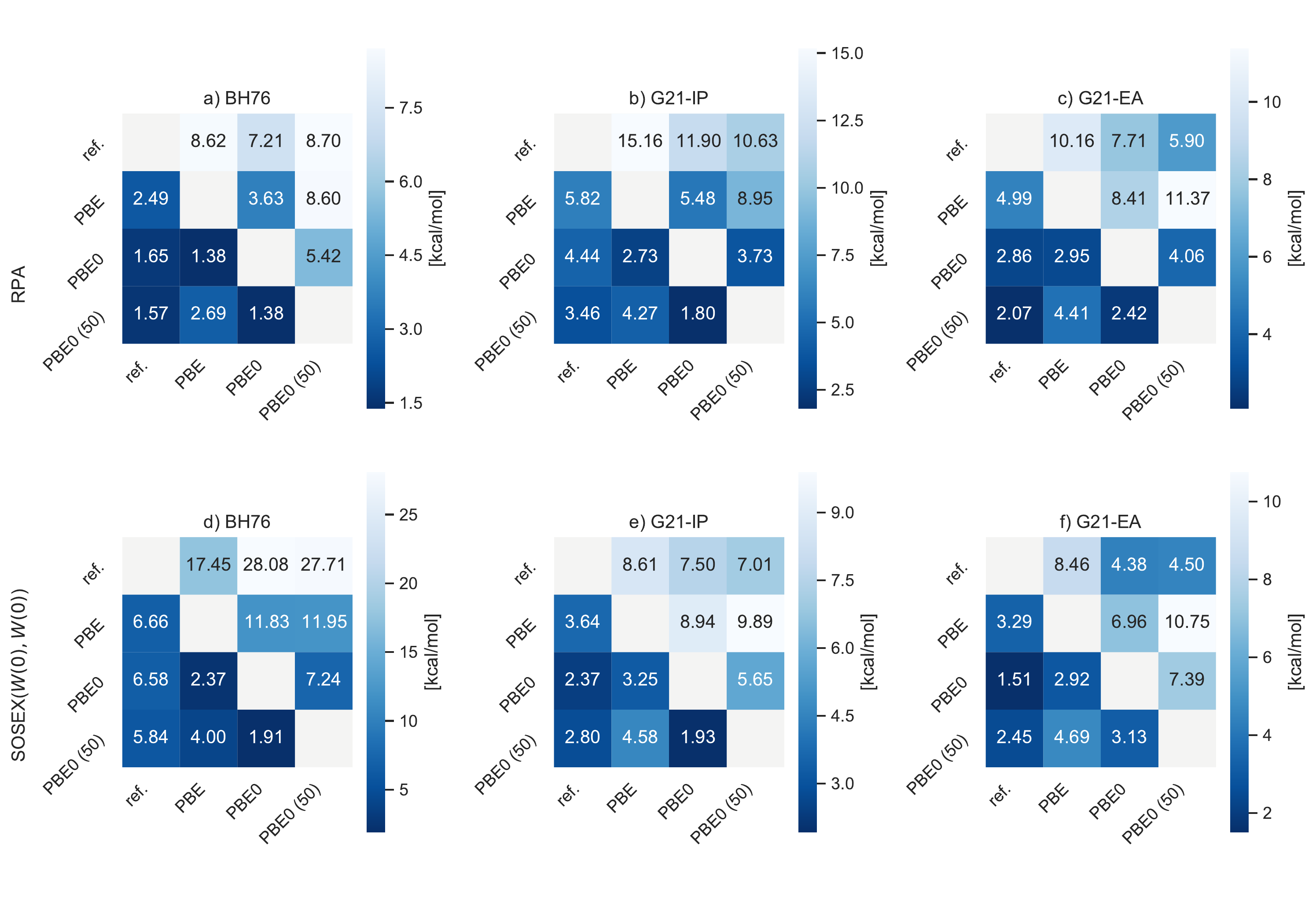}
    \caption{Mean absolute deviations (MAD) (lower triangle in each plot) and Maximum deviations (MAX) (upper triangle) with respect to the reference values as well as using different KS Green's functions as input for BH76 (left), G21-IP (middle) and G21-EA (right) for RPA (top) and RPA+SOSEX($W(0)$,$W(0)$) (bottom). All values are in kcal/mol.}
    \label{fig::startingPoint1}
\end{figure}

To start with, we assess the effect of the Green's function $G^s$ used to calculate the correlation energies. RPA calculations can in principle be performed self-consistently using a variety of approaches.\cite{
Hellgren2007, 
Hellgren2012, 
Verma2012, 
Bleiziffer2013, 
Klimes2014a,
Hellgren2015, 
Voora2019,
Graf2020,
Riemelmoser2021} (see ref.~\citen{Yu2021} for a review) This is rarely done in practice since self-consistent RPA calculations are computationally demanding and since the resulting energies are often worse than the ones evaluated using a Green's function from a generalized gradient approximation (GGA) or hybrid calculation.\cite{Bleiziffer2013} GGAs like PBE or TPSS are often used to construct $G^s$.\cite{Ren2013, Nguyen2020, Kreppel2020} Using hybrid orbitals can be seen as a pragmatic way to compensate for the lack of self-consistency in the RPA calculation and therefore we assess here whether they lead to improvements over GGA orbitals.

For W4-11, the differences between different starting points are minor, but PBE tends to give the best results. For the BH76, G21IP, and G21EA datasets, we show mean absolute deviations (MAD) and maximum deviations (MAX) with respect to the reference values and with respect to the different starting points in figure~\ref{fig::startingPoint1}. The RPA results generally improve with increasing amount of Fock exchange, while 25 \% (PBE0) generally seems to work best for RPA+SOSEX($W(0)$,$W(0)$). The differences are often substantial, for instance in case of the RPA barrier heights (fig~\ref{fig::startingPoint1}a) or the RPA+SOSEX($W(0)$,$W(0)$) electron affinities (fig~\ref{fig::startingPoint1}f). 

For charged excitations, this observation aligns very well with the experience from $G_0W_0$ calculations where hybrid functional with 25 - 50 \% are typically a much better starting point than GGAs.\cite{Caruso2016,Zhang2022} However, when $G3W2$ corrections are added to the $G_0W_0$ QP energies, using a hybrid functional with a smaller fraction of exact exchange might often be beneficial.\cite{Wang2021, Forster2022} For barrier heights, hybrid functionals with a larger fraction of exact exchange are usually required to obtain qualitatively correct barrier heights\cite{Goerigk2017,Hait2018a} and it therefore not surprising that hybrid orbitals serve as a suitable starting point for RPA calculations.

\begin{figure}[hbt!]
    \centering
    \includegraphics[width=0.5\textwidth]{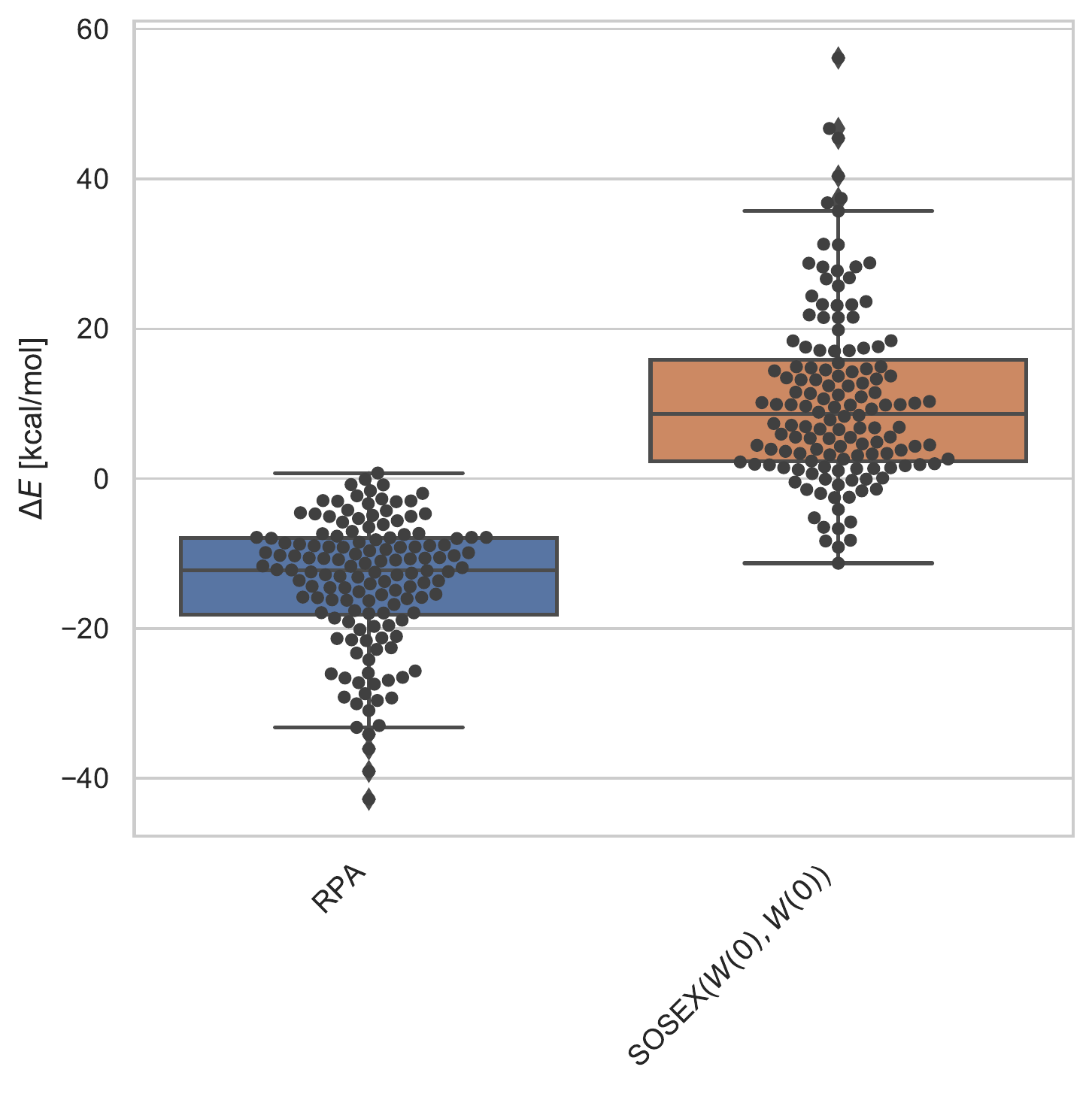}
    \caption{Errors of RPA@PBE and RPA+SOSEX($W$,$v_c$)@PBE for the atomization energies in the W4-11 dataset. Black dots denote the individual data points and the horizonal line in each box denotes the median deviation. the box contains all data points between the first quartile ($Q1$) and third quartile ($Q2$) and the whiskers are at $Q1 \pm |Q1 - Q3|$. (in case of a normal distribution, the whiskers include 99.3\% of all data points). All values are in kcal/mol.}
    \label{fig::boxes-w411}
\end{figure}

Our atomization energies for the W4-11 dataset are shown in figure~\ref{fig::boxes-w411}. It has first been observed by Furche\cite{Furche2001} that RPA underestimates atomization energies (indicated here by negative errors). This has been confirmed later by Ren at al.\cite{Ren2012a} and Paier et al.\cite{Paier2012} for the 55 covalently bound molecules in the G2-I set.\cite{Curtiss1991} The same holds for RPA+SOSEX($W$,$v_c$), but compared to RPA the magnitude of the error is reduced on average.\cite{Paier2012, Ren2012a} We observe here that unlike SOSEX($W,v_c$), the addition of SOSEX($W(0),W(0)$ substantially overcorrects the RPA atomization energies which are now much too high in magnitude.\bibnote{Notice, that our non-counterpoise corrected calculations based on (T,Q) extrapolation will still include a sizable basis set incompleteness error for atomization energies. However, our qualitative conclusions will be valid. } 
Adding bare SOX to RPA leads to underestimated correlation energies.\cite{Jiang2007} This effect is expected to be more pronounced for the molecule than for the individual atoms since more electrons are correlated in the former. Therefore, RPA+SOX will substantially overestimate atomization energies and due to underestimated screening of the SOX term in SOSEX($W(0),W(0)$, RPA+SOSEX($W(0),W(0)$ inherits this problem.

\begin{figure}[hbt!]
    \centering
    \includegraphics[width=\textwidth]{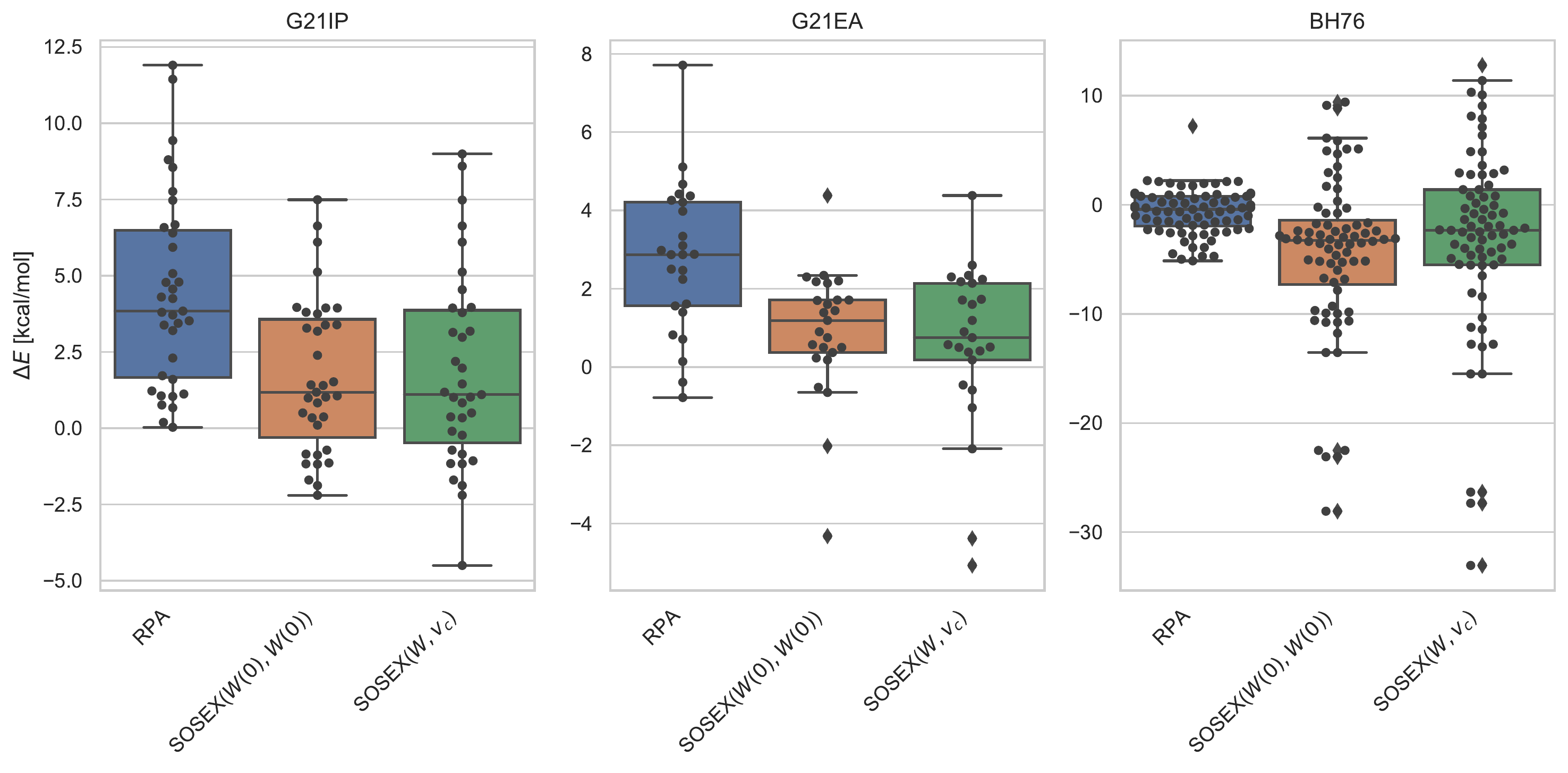}
    \caption{Errors of RPA@PBE and different RPA+SOSEX variants for barrier heights (BH76, left), ionization potentials (G21-IP, middle) and electron affinities (G21-EA, right). For an explanation of the boxplots, see the caption of figure~\ref{fig::boxes-w411}. All values are in kcal/mol.}
    \label{fig::boxes}
\end{figure}

As also shown in more detail in figure~\ref{fig::boxes}, the performance of RPA+SOSEX($W(0)$,$W(0)$) is in all cases comparable to RPA+SOSEX($W$,$v_c$), for which the trends presented here are well known:\cite{Paier2010,Paier2010a,Eshuis2012a, Ren2012a, Ren2013} RPA+SOSEX($W$,$v_c$), fails for barrier heights, where the inclusion of renormalized singles excitations is necessary to obtain good results\cite{Paier2012,Ren2012a,Ren2013} and works very well for charged excitations.\cite{Ren2012a, Eshuis2012a} We note, that RPA+SOSEX($W(0)$,$W(0)$)@PBE0 performs very well for charged excitations, with an accuracy challenging modern double-hybrid functionals.\cite{Goerigk2017}

\subsection{Non-covalent Interactions}

\subsubsection{S66 Interaction Energies}

\begin{figure}[hbt!]
    \centering
    \includegraphics[width=\textwidth]{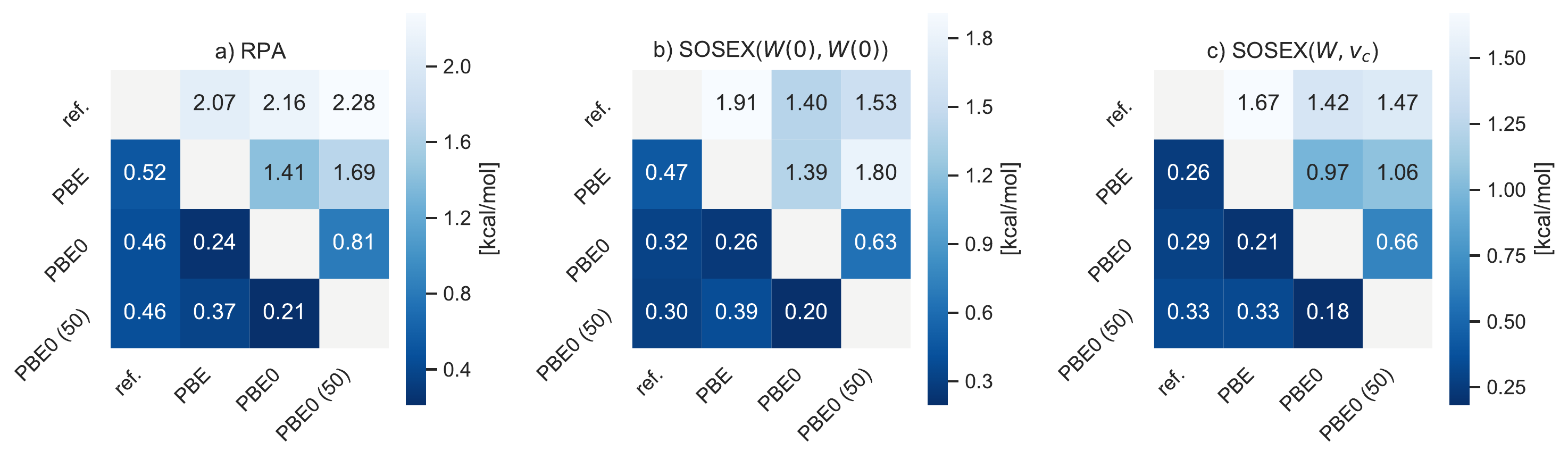}
    \caption{Mean absolute deviations (MAD) (lower triangle in each plot) and Maximum deviations (MAX) (upper triangle) for a) RPA,  b) SOSEX($W(0), W(0)$) and c) SOSEX($W(0), v_c$) interaction energies for the S66 database using different KS Green's functions as well as to the CCSD(T) reference values (ref.). All values are in kcal/mol.}
    \label{fig::startingPoint}
\end{figure}

We now turn to our benchmark results for non-covalent interactions. As for the previous datasets, we also assess the dependence of RPA and RPA+SOSEX correlation energies on the choice of the KS Green's function $G^s$. In figure~\ref{fig::startingPoint} the interaction energies in the S66 database\cite{Rezac2011} obtained using different $G^s$ are compared to each other as well as to the CCSD(T) reference values by Hobza and coworkers.\cite{Rezac2011} All values have been obtained using a single integration point for the $\lambda$-integral. As shown in the supporting information, a few outliers aside the errors arising from this approximation are small for non-covalent interactions. RPA and RPA+SOSEX($W(0), W(0)$) are equivalently independent of the choice of the KS Green's function, with MADs between 0.20 and 0.39 kcal/mol between the different functionals. However, individual values can differ by almost 2 kcal/mol which is a sizable difference, given that the largest interaction energies in the S66 database are of the order of 20 kcal/mol only. The performance of RPA compared to the CCSD(T) reference is rather insensitive to the KS Green's function, even though the hybrid starting points lead to slightly better results.\bibnote{With 0.52 kcal/mol, the MAD for RPA@PBE is in excellent agreement with the 0.61 kcal/mol MAD obtained by Nguyen \emph{et. al.} in ref.~\citen{Nguyen2020}, which has been obtained with GTO-type basis sets and 50 \% counterpoise correction instead of 100 \%. This shows, that our interaction energies are well converged with respect to the basis set size.}
The RPA+SOSEX($W(0), W(0)$) results are much better using the hybrid functionals than with PBE. RPA+SOSEX($W, v_c$)@PBE, is slightly more accurate than RPA+SOSEX($W, v_c$)@PBE0, but unlike for the datasets discussed before, the differences between the different starting points are negligibly small. Also, the dependence of SOSEX($W,v_c$) on the starting point is smaller than for SOSEX($W(0), W(0)$). 

\begin{figure}[hbt!]
    \centering
    \includegraphics[width=\textwidth]{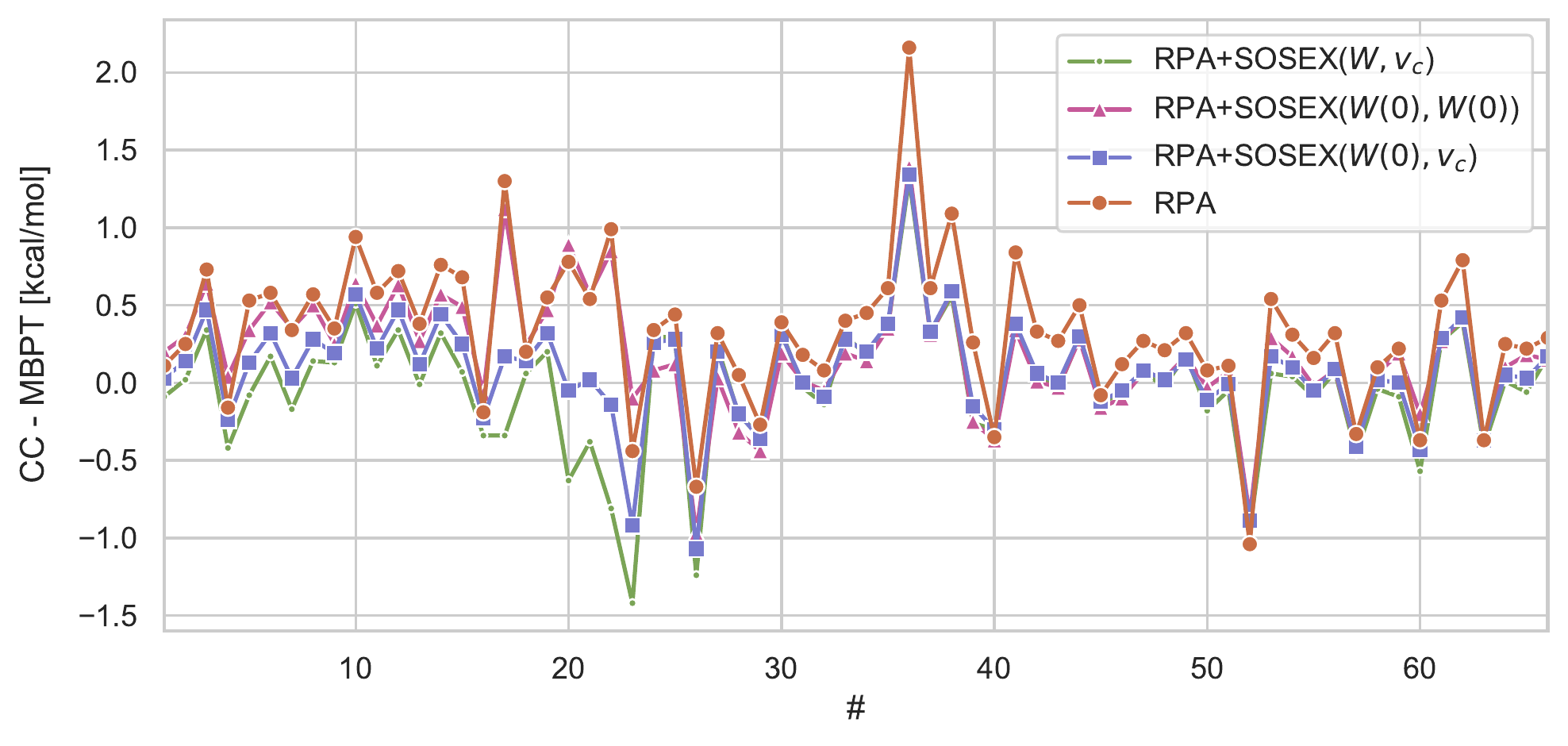}
    \caption{Deviations of RPA@PBE0 and both RPA + SOSEX@PBE0 variants for the S66 database with respect to the CCSD(T) reference. All values are in kcal/mol.}
    \label{fig::s66}
\end{figure}

\begin{table}[hbt!]
    \centering
    \begin{tabular}{lcccccccc}
    \toprule 
    & \multicolumn{8}{c}{MAD} \\
    & \multicolumn{2}{c}{S66} & \multicolumn{2}{c}{hydr. bond} & \multicolumn{2}{c}{dispersion} & \multicolumn{2}{c}{mixed} \\ 
    Method & [$\frac{kcal}{mol}$] & [\%] & [$\frac{kcal}{mol}$] & [\%] & [$\frac{kcal}{mol}$] & [\%] & [$\frac{kcal}{mol}$] & [\%] \\
    \midrule 
    SOSEX($W(0), W(0)$)@PBE0 &  0.32 &  7.28 & 0.45 & 5.76 & 0.29 & 10.33 & 0.21 & 5.50 \\
    SOSEX($W(0),v_c$)@PBE0   &  0.28 &  6.88 & 0.30 & 3.42 & 0.34 & 11.77 & 0.20 & 5.25 \\
    SOSEX($W, v_c$)@PBE0     &  0.29 &  6.85 & 0.31 & 3.39 & 0.33 & 11.63 & 0.21 & 5.33 \\
    SOSEX($W, v_c$)@PBE      &  0.26 &  6.25 & 0.23 & 3.51 & 0.33 & 10.16 & 0.17 & 4.26 \\
    RPA                      &  0.46 & 11.54 & 0.55 & 7.19 & 0.47 & 17.74 & 0.34 & 9.41 \\
    PBE0-D3(BJ)              &  0.28 &  5.09 & 0.47 & 4.80 & 0.18 & 5.09  & 0.18 & 5.42 \\
    DSD-PBE-P86-D3(BJ)       &  0.23 &  5.07 & 0.31 & 3.71 & 0.21 & 6.99  & 0.16 & 4.43 \\
    \bottomrule 
    \end{tabular}
    \caption{MADs (absolute and in \%) of different electronic structure methods with respect to the CCSD(T) reference values for the whole S66 database and for its subcategories.}
    \label{tab::s66_values}
\end{table}

Figure~\ref{fig::s66} shows the deviations of RPA and both RPA+SOSEX variants with respect to CCSD(T) for all datapoints in the S66 database. MADs and mean absolute percentage deviations (MAPD) for the whole database as well as for the individual categories are presented in table~\ref{tab::s66_values}. The interactions of the first 22 complexes in the database are dominated by Hydrogen bonds which are predominantly of electrostatic origin.\cite{Becke2014} The next 22 complexes are mostly bound by dispersion interactions and the remaining interactions are of mixed nature.\cite{Rezac2011} It is useful to distinguish between these different interaction patterns in the following comparison. 

For the whole database, RPA gives a MAPD of 11.5 \% and the SOSEX corrections sizably reduce the MAPDs with respect to the CCSD(T) reference values to in between 7.3 \% and 6.3 \%. SOSEX($W, v_c$) outperforms SOSEX($W(0), W(0)$) by far for the hydrogen-bonded complexes, and is even slightly more accurate than the double-hybrid DSD-PBE-P86-D3(BJ),\cite{Santra2019a} one of the best double hybrid functionals for weak interactions.\cite{Mehta2018} For dispersion interactions, the performance of SOSEX($W(0), W(0)$) and SOSEX($W, v_c$) is comparable. Here, the empirically dispersion corrected\cite{Grimme2010, Stefan2011} functionals, the hybrid PBE0-D3(BJ) and DSD-PBE-P86-D3(BJ), are much more accurate than all MBPT based methods. 
A few exceptions aside, fig.~\ref{fig::s66} shows that RPA understabilizes the complexes in the S66 database (indicated by positive errors). SOSEX corrections lower the interaction energies, i.e. the complexes are predicted to be more stable. SOSEX($W, v_c$) shows a tendency to overstabilize the hydrogen-bonded complexes. For these systems, the RPA+SOSEX($W(0), W(0)$) energies are almost identical to the ones from RPA. 

Also from the sizable differences of SOSEX($W, v_c$) (green points) to its static limit (with only a single screened interaction line, blue points) shown in figure~\ref{fig::s66} it is clear that the dynamical screening effects are important for the hydrogen-bonded complexes. As can be seen from the MAPD in table~\ref{tab::s66_values}, this does however not improve agreement with the CCSD(T) reference values. For the dispersion bound complexes, there are only negligible differences between both variants, demonstrating that the dynamical variations of the screening average out. For the last 22 complexes in the database the differences are slightly larger. In all cases, dressing the second electron-electron interaction line does not alter the results decisively.

\subsubsection{S66x8 Interaction Energy}

The S66x8 dataset contains the complexes in the S66 database at 8 different geometries.\cite{Rezac2011} The separations of the monomers in the complexes are given relative to their equilibrium distances, i.e. a relative separation of 2.0 means that the monomers separation in the complex is twice as large as the equilibrium separation. For our assessment of the SOSEX($W(0),W(0)$) correction, we divide the separations of the potential energy curve in three regions, which we denote as short (equilibrium distance scaled by a factor 0.9-0.95), middle (1.0-1.25) and long (1.5-2.0). All RPA (+SOSEX) calculation discussed here have been performed using a PBE0 Green's function.

\begin{figure}[hbt!]
    \centering
    \includegraphics[width=0.7\textwidth]{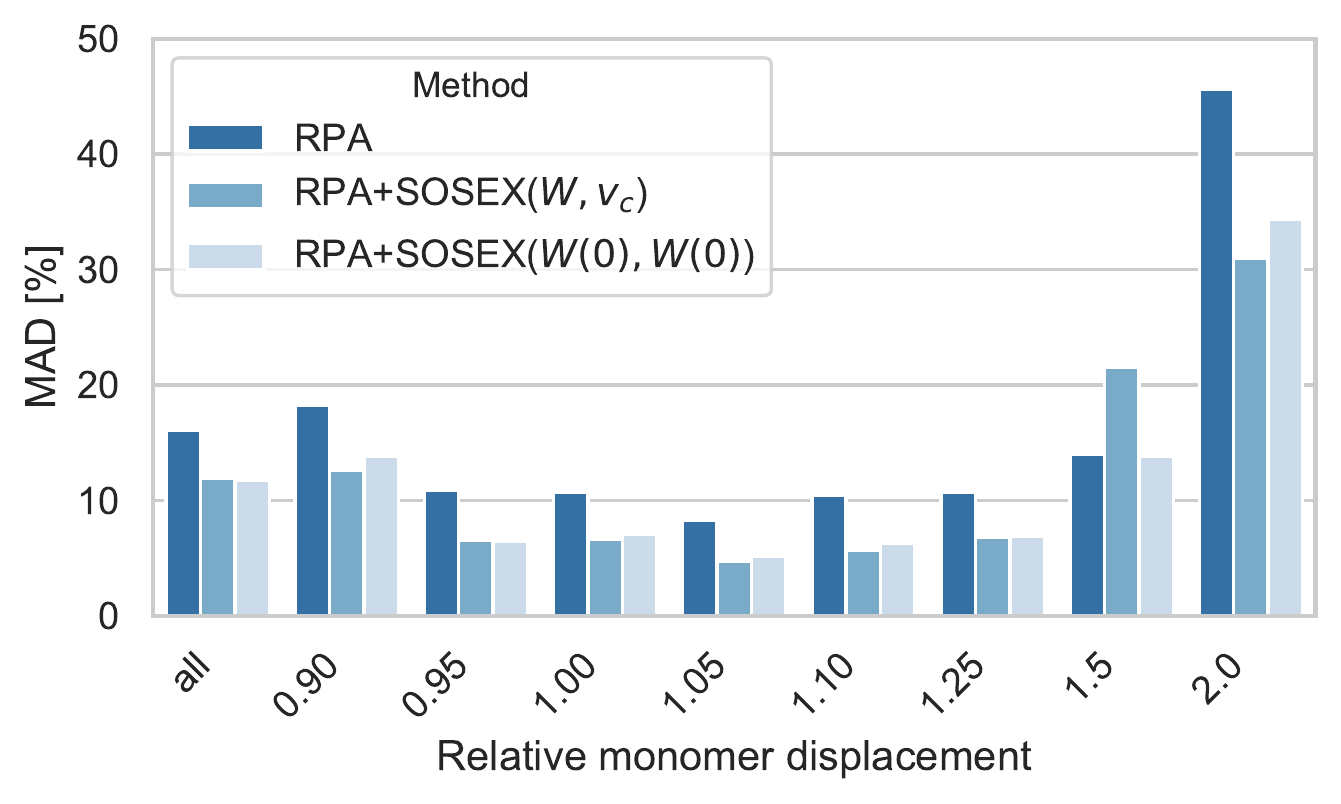}
    \caption{MADs (in percent) for the S66x8 database with respect to the CCSD(T) reference values for RPA, RPA+SOSEX$(W,v_c)$ and RPA+SOSEX$(W(0),W(0))$. MADs are shown separately for the whole database (columns on the left) and for different monomer-monomer separations.}
    \label{fig::s66x8}
\end{figure}

\begin{table}[hbt!]
    \centering
    \begin{tabular}{lccc}
    \toprule 
    & short [\%] & middle [\%] & long [\%] \\
    \midrule
        SOSEX$(W,v_c)$ & 35.2 & 42.8 & 13.5 \\
        SOSEX$(W(0),W(0))$ & 31.0 & 37.9 & 19.1 \\
    \bottomrule
    \end{tabular}
    \caption{Relative improvements obtained with different SOSEX variants over RPA for different groups of monomer-monomer separations.}
    \label{tab::mads_S66x8_groups}
\end{table}

\begin{figure}[hbt!]
    \centering
    \includegraphics[width=\textwidth]{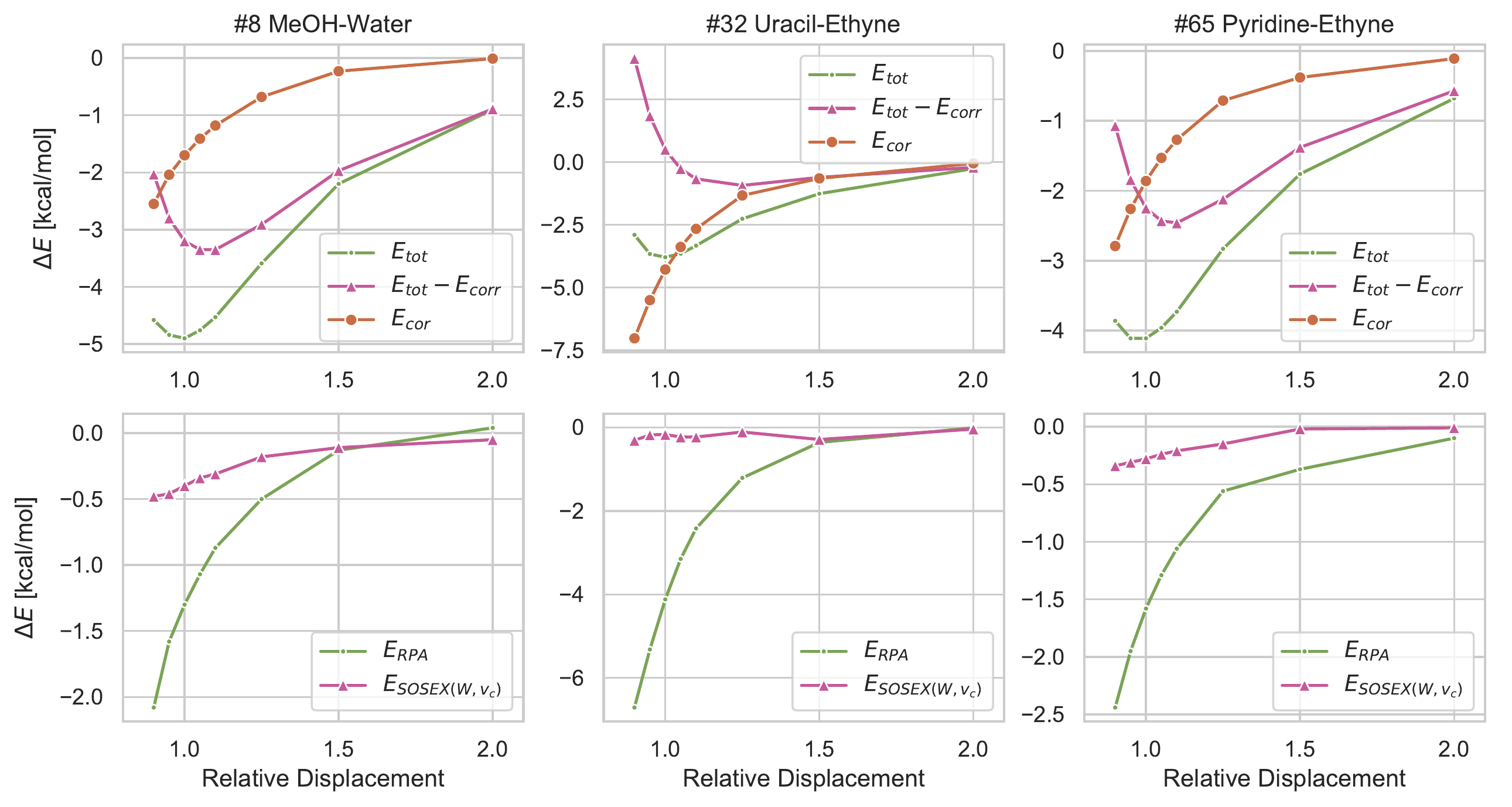}
    \caption{Upper plots: three RPA+SOSEX($W$,$v_c$)@PBE0 potential energy curves in the S66x8 database (green), separated in correlation contributions (yellow) and HF energy (evaluated with PBE0 orbitals). Lower plots: decomposition of the correlation energies into RPA and SOSEX($W$,$v_c$) contributions. All values are in kcal/mol.}
    \label{fig::distance-plots}
\end{figure}

The results of our comparison are shown in figure~\ref{fig::s66x8}, where the MAPDs with respect to CCSD(T) for the whole database as well as for the scaled monomer-monomer separations are shown. For the whole database, the average relative deviations with respect to the reference values are larger than for S66. With in between 31 and 43 \%, both SOSEX correction lead to sizable improvements over the RPA in the short and medium regime. For large monomer-monomer separations, the improvements become much smaller, with 14 \% for SOSEX$(W,v_c)$ and 19 \% for SOSEX$(W(0),W(0))$. This can be rationalized by observing that for large electron-electron distances the correlation contributions to the interaction energies quickly decay to zero. This is shown in figure~\ref{fig::distance-plots} where we have plotted three of the RPA+SOSEX($W$,$v_c$) potential energy curves (Green curves in the upper plots) in the S66x8 database and separated the correlation contributions (The Green curves are the sums of the red and yellow curves). The lower plots separately show the RPA and SOSEX($W$,$v_c$) contributions to the correlation energy differences. 

In all three plots, the potential energy curves are dominated by the difference of the correlation energy of the dimer and the sum of correlation energies of the monomers. Therefore, the approximation used for the calculation of the correlation energy plays a large role. However, this difference quickly goes to zero for larger separations. At two times of the equilibrium distance, the correlation contributions to the potential energy curves are almost zero in all three considered examples. Therefore, the expression used for the correlation energy becomes less and less important with increasing monomer separation. This argument also holds if one expresses the contributions in \% of the total interaction energy. 

One would expect the SOSEX contribution to decay faster than the RPA one, since the former is of exchange nature and therefore fundamentally short-ranged.\cite{Jiang2007} However, the plots in the lower part of figure~\ref{fig::distance-plots} shows that this is only the case for the potential energy curve on the right, but not for the two curves on the left, where SOSEX and RPA contributions seem to decay equally fast.

\section{\label{sec::conclusions}Conclusions} 
The accuracy of the RPA can in principle be improved by including vertex correction in the self-energy. This can be done either directly, or indirectly through the solution of the BSE. Including the first-order vertex in the self-energy, different variants of SOSEX are obtained. These are the well-known AC-SOSEX, herein termed SOSEX$(W,v_c)$, first introduced by Jansen \emph{et. al},\cite{Jansen2010} in which only one of the Coulomb interaction lines is dynamically screened, as well as an energy expression which is obtained from the statically screened $G3W2$ correction to the $GW$ self-energy.\cite{Gruneis2014, Forster2022} This energy expression has already been introduced in our earlier work\cite{Forster2022}, albeit without a rigorous derivation. Especially, we have implicitly assumed that the integral over the coupling strength is evaluated using a trapezoidal rule. Here, we have derived this expression (referred to as SOSEX$(W(0),W(0))$ in this work) taking into account its $\lambda$-dependence and highlighted the differences to SOSEX$(W,v_c)$. We have then assessed the accuracy of the SOSEX$(W(0),W(0))$ correction to RPA correlation energies for a wide range of chemical problems including bond dissociation, thermochemistry, kinetics, and non-covalent interactions. 

The main conclusion we can draw from our work is that in situation where the addition of SOSEX$(W,v_c)$ leads to major improvements over the RPA, the addition of SOSEX$(W(0),W(0))$ does as well. This is the case for the calculation of ionization potentials and electron affinities where RPA+SOSEX approaches challenge the accuracy of modern double-hybrid functionals.\cite{Goerigk2017} Also for non-covalent interactions both SOSEX variants lead to the same substantial improvements over RPA. SOSEX$(W,v_c)$ is most accurate for the hydrogen-bonded complexes while SOSEX$(W(0),W(0))$ is slightly more accurate for dispersion interactions. We also showed that the frequency-dependence of the screened interactions does seem to be an important factor for hydrogen-bonding but not for dispersion interactions.

Differences between both SOSEX variants have been observed in the dissociation of diatomic molecules. As RPA and unlike RPA+SOSEX($W$,$v_c$),\cite{Paier2010,Henderson2010} RPA+SOSEX$(W(0),W(0))$ dissociates the Hydrogen molecule correctly. RPA does so because the self-correlation error effectively describes static correlation.\cite{Henderson2010} The situation seems to be similar for RPA+SOSEX\newline$(W(0),W(0))$ since in contrast to RPA+SOSEX($W$,$v_c$) it is not completely self-correlation free for 1-electron systems. We have also shown that this qualitative difference is due to the screening of the second electron-electron interaction line. 

The incomplete cancellation of self-correlation error does however negatively affect the dissociation of charged dimers for which RPA+SOSEX($W$,$v_c$) fixes most of the deficiencies of RPA.\cite{Paier2010,Chen2018} Here, RPA+SOSEX$(W(0),W(0))$ performs even worse than RPA. Furthermore, the good dissociation of diatomic molecules does not automatically carry over to accurate barrier heights\cite{Zhao2005a, Zhao2005b} where both SOSEX variants considerably worsen the RPA results.

In summary, our results suggest that the statically screened SOSEX is a suitable alternative to dynamically screened SOSEX. While both formally scale as $N^5$ with system size, the computation of the SOSEX$(W,v_c)$ correction requires a numerical imaginary frequency integration. The calculation of the SOSEX($W(0),W(0)$) correction is therefore much cheaper, comparable to MP2. MP2 is however inadequate for large molecules since it neglects screening effects entirely.\cite{Macke1950, Nguyen2020} RPA+SOSEX$(W(0),W(0))$ is in principle applicable also to large molecules. A stochastic linear scaling implementation of the SOSEX self-energy has already been developed\cite{Vlcek2019} and a recent RPA+SOSEX implementation by Ochsenfeld and co-workers\cite{Beuerle2018} allowed applications to the L7 dataset,\cite{Sedlak2013} albeit with small basis sets. Other low-scaling MP2 implementations\cite{Doser2009a,Pinski2015, Nagy2016} could potentially be generalized to SOSEX as well. 

Finally, it should be mentioned that the accuracy of the  dynamically screened SOSEX correction to the RPA can be improved upon by the addition of renormalized single excitations.\cite{Ren2012a, Ren2013} Other methods which have been shown to outperform SOSEX, in particular for barrier heights, are the AXK kernel method\cite{Bates2013,Chen2018, Mezei2019} or a SOSEX variant in which the terms of RPA and SOSEX beyond second order in $v_c$ are scaled down.\cite{Mezei2019} It remains to be investigated whether the concept of static screening can also be combined with those approaches and leads to good results. 

\begin{acknowledgement}
This research received funding (project number 731.017.417) from the Netherlands Organization for Scientific Research (NWO) in the framework of the Innovation Fund for Chemistry and from the Ministry of Economic Affairs in the framework of the \enquote{\emph{TKI/PPS-Toeslagregeling}}. We thank Mauricio Rodríguez-Mayorga and Timothy Daas for fruitful discussions.
\end{acknowledgement}


\begin{suppinfo}
Derivation of \eqref{sigma}, discussion of the crossing symmetries of the 4-point vertices, assessment of the dependence of the $\lambda$-integration on the number of Gauss-Legendre points for the S66 database, .csv files with all calculated energies at the extrapolated CBS TZ3P, and QZ6P level and explanations of those files.
\end{suppinfo}


\bibliography{all}

\providecommand{\latin}[1]{#1}
\makeatletter
\providecommand{\doi}
  {\begingroup\let\do\@makeother\dospecials
  \catcode`\{=1 \catcode`\}=2 \doi@aux}
\providecommand{\doi@aux}[1]{\endgroup\texttt{#1}}
\makeatother
\providecommand*\mcitethebibliography{\thebibliography}
\csname @ifundefined\endcsname{endmcitethebibliography}
  {\let\endmcitethebibliography\endthebibliography}{}
\begin{mcitethebibliography}{186}
\providecommand*\natexlab[1]{#1}
\providecommand*\mciteSetBstSublistMode[1]{}
\providecommand*\mciteSetBstMaxWidthForm[2]{}
\providecommand*\mciteBstWouldAddEndPuncttrue
  {\def\EndOfBibitem{\unskip.}}
\providecommand*\mciteBstWouldAddEndPunctfalse
  {\let\EndOfBibitem\relax}
\providecommand*\mciteSetBstMidEndSepPunct[3]{}
\providecommand*\mciteSetBstSublistLabelBeginEnd[3]{}
\providecommand*\EndOfBibitem{}
\mciteSetBstSublistMode{f}
\mciteSetBstMaxWidthForm{subitem}{(\alph{mcitesubitemcount})}
\mciteSetBstSublistLabelBeginEnd
  {\mcitemaxwidthsubitemform\space}
  {\relax}
  {\relax}

\bibitem[Macke(1950)]{Macke1950}
Macke,~W. {{\"{U}}ber die Wechselwirkungen im Fermi-Gas}. \emph{Zeitschrift fur
  Naturforsch.} \textbf{1950}, \emph{5}, 192--208\relax
\mciteBstWouldAddEndPuncttrue
\mciteSetBstMidEndSepPunct{\mcitedefaultmidpunct}
{\mcitedefaultendpunct}{\mcitedefaultseppunct}\relax
\EndOfBibitem
\bibitem[Bohm and Pines(1953)Bohm, and Pines]{Bohm1953}
Bohm,~D.; Pines,~D. {A collective description of electron interactions: III.
  Coulomb interactions in a degenerate electron gas}. \emph{Phys. Rev.}
  \textbf{1953}, \emph{92}, 609--625\relax
\mciteBstWouldAddEndPuncttrue
\mciteSetBstMidEndSepPunct{\mcitedefaultmidpunct}
{\mcitedefaultendpunct}{\mcitedefaultseppunct}\relax
\EndOfBibitem
\bibitem[He{\ss}elmann and G{\"{o}}rling(2011)He{\ss}elmann, and
  G{\"{o}}rling]{Hesselmann2011}
He{\ss}elmann,~A.; G{\"{o}}rling,~A. {Random-phase approximation correlation
  methods for molecules and solids}. \emph{Mol. Phys.} \textbf{2011},
  \emph{109}, 2473--2500\relax
\mciteBstWouldAddEndPuncttrue
\mciteSetBstMidEndSepPunct{\mcitedefaultmidpunct}
{\mcitedefaultendpunct}{\mcitedefaultseppunct}\relax
\EndOfBibitem
\bibitem[Eshuis and Furche(2011)Eshuis, and Furche]{Eshuis2011}
Eshuis,~H.; Furche,~F. {A parameter-free density functional that works for
  noncovalent interactions}. \emph{J. Phys. Chem. Lett.} \textbf{2011},
  \emph{2}, 983--989\relax
\mciteBstWouldAddEndPuncttrue
\mciteSetBstMidEndSepPunct{\mcitedefaultmidpunct}
{\mcitedefaultendpunct}{\mcitedefaultseppunct}\relax
\EndOfBibitem
\bibitem[Eshuis \latin{et~al.}(2012)Eshuis, Bates, and Furche]{Eshuis2012a}
Eshuis,~H.; Bates,~J.~E.; Furche,~F. {Electron correlation methods based on the
  random phase approximation}. \emph{Theor. Chem. Acc.} \textbf{2012},
  \emph{131}, 1084\relax
\mciteBstWouldAddEndPuncttrue
\mciteSetBstMidEndSepPunct{\mcitedefaultmidpunct}
{\mcitedefaultendpunct}{\mcitedefaultseppunct}\relax
\EndOfBibitem
\bibitem[Ren \latin{et~al.}(2012)Ren, Rinke, Joas, and Scheffler]{Ren2012a}
Ren,~X.; Rinke,~P.; Joas,~C.; Scheffler,~M. {Random-phase approximation and its
  applications in computational chemistry and materials science}. \emph{J.
  Mater. Sci.} \textbf{2012}, \emph{47}, 7447--7471\relax
\mciteBstWouldAddEndPuncttrue
\mciteSetBstMidEndSepPunct{\mcitedefaultmidpunct}
{\mcitedefaultendpunct}{\mcitedefaultseppunct}\relax
\EndOfBibitem
\bibitem[Chen \latin{et~al.}(2017)Chen, Voora, Agee, Balasubramani, and
  Furche]{Chen2017}
Chen,~G.~P.; Voora,~V.~K.; Agee,~M.~M.; Balasubramani,~S.~G.; Furche,~F.
  {Random-Phase Approximation Methods}. \emph{Annu. Rev. Phys. Chem.}
  \textbf{2017}, \emph{68}, 421--445\relax
\mciteBstWouldAddEndPuncttrue
\mciteSetBstMidEndSepPunct{\mcitedefaultmidpunct}
{\mcitedefaultendpunct}{\mcitedefaultseppunct}\relax
\EndOfBibitem
\bibitem[Chedid \latin{et~al.}(2018)Chedid, Ferrara, and Eshuis]{Chedid2018}
Chedid,~J.; Ferrara,~N.~M.; Eshuis,~H. {Describing transition metal homogeneous
  catalysis using the random phase approximation}. \emph{Theor. Chem. Acc.}
  \textbf{2018}, \emph{137:158}, 1--11\relax
\mciteBstWouldAddEndPuncttrue
\mciteSetBstMidEndSepPunct{\mcitedefaultmidpunct}
{\mcitedefaultendpunct}{\mcitedefaultseppunct}\relax
\EndOfBibitem
\bibitem[Kreppel \latin{et~al.}(2020)Kreppel, Graf, Laqua, and
  Ochsenfeld]{Kreppel2020}
Kreppel,~A.; Graf,~D.; Laqua,~H.; Ochsenfeld,~C. {Range-Separated
  Density-Functional Theory in Combination with the Random Phase Approximation:
  An Accuracy Benchmark}. \emph{J. Chem. Theory Comput.} \textbf{2020},
  \emph{16}, 2985--2994\relax
\mciteBstWouldAddEndPuncttrue
\mciteSetBstMidEndSepPunct{\mcitedefaultmidpunct}
{\mcitedefaultendpunct}{\mcitedefaultseppunct}\relax
\EndOfBibitem
\bibitem[Modrzejewski \latin{et~al.}(2020)Modrzejewski, Yourdkhani, and
  Klime{\v{s}}]{Modrzejewski2020}
Modrzejewski,~M.; Yourdkhani,~S.; Klime{\v{s}},~J. {Random Phase Approximation
  Applied to Many-Body Noncovalent Systems}. \emph{J. Chem. Theory Comput.}
  \textbf{2020}, \emph{16}, 427--442\relax
\mciteBstWouldAddEndPuncttrue
\mciteSetBstMidEndSepPunct{\mcitedefaultmidpunct}
{\mcitedefaultendpunct}{\mcitedefaultseppunct}\relax
\EndOfBibitem
\bibitem[Langreth and Perdew(1975)Langreth, and Perdew]{Langreth1975}
Langreth,~D.~C.; Perdew,~J.~P. {The Exhange-Correlation Energy of a metalic
  surface}. \emph{Solid State Commun.} \textbf{1975}, \emph{17},
  1425--1429\relax
\mciteBstWouldAddEndPuncttrue
\mciteSetBstMidEndSepPunct{\mcitedefaultmidpunct}
{\mcitedefaultendpunct}{\mcitedefaultseppunct}\relax
\EndOfBibitem
\bibitem[Langreth and Perdew(1977)Langreth, and Perdew]{Langreth1977}
Langreth,~D.~C.; Perdew,~J.~P. {Exchange-correlation energy of a metallic
  surface: Wave-vector analysis}. \emph{Phys. Rev. B} \textbf{1977}, \emph{15},
  2884--2901\relax
\mciteBstWouldAddEndPuncttrue
\mciteSetBstMidEndSepPunct{\mcitedefaultmidpunct}
{\mcitedefaultendpunct}{\mcitedefaultseppunct}\relax
\EndOfBibitem
\bibitem[Harris and Griffin(1975)Harris, and Griffin]{Harris1975}
Harris,~J.; Griffin,~A. {Correlation energy and van der Waals interaction of
  coupled metal films}. \emph{Phys. Rev. B} \textbf{1975}, \emph{11},
  3669--3677\relax
\mciteBstWouldAddEndPuncttrue
\mciteSetBstMidEndSepPunct{\mcitedefaultmidpunct}
{\mcitedefaultendpunct}{\mcitedefaultseppunct}\relax
\EndOfBibitem
\bibitem[Coester(1958)]{Coester1958}
Coester,~F. {Bound states of a many-particle system}. \emph{Nucl. Phys.}
  \textbf{1958}, \emph{7}, 421--424\relax
\mciteBstWouldAddEndPuncttrue
\mciteSetBstMidEndSepPunct{\mcitedefaultmidpunct}
{\mcitedefaultendpunct}{\mcitedefaultseppunct}\relax
\EndOfBibitem
\bibitem[Coester and K{\"{u}}mmel(1960)Coester, and K{\"{u}}mmel]{Coester1960}
Coester,~F.; K{\"{u}}mmel,~H. {Short-range correlations in nuclear wave
  functions}. \emph{Nucl. Phys.} \textbf{1960}, \emph{17}, 477--485\relax
\mciteBstWouldAddEndPuncttrue
\mciteSetBstMidEndSepPunct{\mcitedefaultmidpunct}
{\mcitedefaultendpunct}{\mcitedefaultseppunct}\relax
\EndOfBibitem
\bibitem[{\v{C}}{\'{i}}{\v{z}}ek(1966)]{Cizek1966}
{\v{C}}{\'{i}}{\v{z}}ek,~J. {On the Correlation Problem in Atomic and Molecular
  Systems. Calculation of Wavefunction Components in Ursell-Type Expansion
  Using Quantum-Field Theoretical Methods}. \emph{J. Chem. Phys.}
  \textbf{1966}, \emph{45}, 4256--4266\relax
\mciteBstWouldAddEndPuncttrue
\mciteSetBstMidEndSepPunct{\mcitedefaultmidpunct}
{\mcitedefaultendpunct}{\mcitedefaultseppunct}\relax
\EndOfBibitem
\bibitem[{\v{C}}{\'{i}}{\v{z}}ek(1969)]{Cizek1969}
{\v{C}}{\'{i}}{\v{z}}ek,~J. {On the Use of the Cluster Expansion and the
  Technique of Diagrams in Calculations of Correlation Effects in Atoms and
  Molecules}. \emph{Adv. Chem. Physics, Vol. XIV} \textbf{1969}, \emph{XIV},
  35--89\relax
\mciteBstWouldAddEndPuncttrue
\mciteSetBstMidEndSepPunct{\mcitedefaultmidpunct}
{\mcitedefaultendpunct}{\mcitedefaultseppunct}\relax
\EndOfBibitem
\bibitem[Paldus \latin{et~al.}(1972)Paldus, {\v{C}}{\'{i}}{\v{z}}ek, and
  Shavitt]{Paldus1972}
Paldus,~J.; {\v{C}}{\'{i}}{\v{z}}ek,~J.; Shavitt,~I. {Correlation Problems in
  Atomic and Molecular Systems. IV. Extended Coupled-Pair Many-Electron Theory
  and Its Application to the BH3 Molecule}. \emph{Phys. Rev. A} \textbf{1972},
  \emph{5}, 50--67\relax
\mciteBstWouldAddEndPuncttrue
\mciteSetBstMidEndSepPunct{\mcitedefaultmidpunct}
{\mcitedefaultendpunct}{\mcitedefaultseppunct}\relax
\EndOfBibitem
\bibitem[Scuseria \latin{et~al.}(2008)Scuseria, Henderson, and
  Sorensen]{Scuseria2008}
Scuseria,~G.~E.; Henderson,~T.~M.; Sorensen,~D.~C. {The ground state
  correlation energy of the random phase approximation from a ring coupled
  cluster doubles approach}. \emph{J. Chem. Phys.} \textbf{2008}, \emph{129},
  231101\relax
\mciteBstWouldAddEndPuncttrue
\mciteSetBstMidEndSepPunct{\mcitedefaultmidpunct}
{\mcitedefaultendpunct}{\mcitedefaultseppunct}\relax
\EndOfBibitem
\bibitem[Scuseria \latin{et~al.}(2013)Scuseria, Henderson, and
  Bulik]{Scuseria2013}
Scuseria,~G.~E.; Henderson,~T.~M.; Bulik,~I.~W. {Particle-particle and
  quasiparticle random phase approximations: Connections to coupled cluster
  theory}. \emph{J. Chem. Phys.} \textbf{2013}, \emph{139}, 104113\relax
\mciteBstWouldAddEndPuncttrue
\mciteSetBstMidEndSepPunct{\mcitedefaultmidpunct}
{\mcitedefaultendpunct}{\mcitedefaultseppunct}\relax
\EndOfBibitem
\bibitem[Abrikosov \latin{et~al.}(1975)Abrikosov, Gorkov, and
  Dzyaloshinski]{Abrikosov1975}
Abrikosov,~A.~A.; Gorkov,~P.~L.; Dzyaloshinski,~I.~E. \emph{{Methods of quantum
  field theory in statistical physics}}; Dover Publications INC. New York,
  1975\relax
\mciteBstWouldAddEndPuncttrue
\mciteSetBstMidEndSepPunct{\mcitedefaultmidpunct}
{\mcitedefaultendpunct}{\mcitedefaultseppunct}\relax
\EndOfBibitem
\bibitem[Mattuck(1992)]{Mattuck1992}
Mattuck,~R.~D. \emph{{A Guide to Feynman Diagrams in the Many-body Problem}},
  2nd ed.; Dover Publications INC. New York, 1992\relax
\mciteBstWouldAddEndPuncttrue
\mciteSetBstMidEndSepPunct{\mcitedefaultmidpunct}
{\mcitedefaultendpunct}{\mcitedefaultseppunct}\relax
\EndOfBibitem
\bibitem[Bruus and Flensberg(2004)Bruus, and Flensberg]{Bruus2004}
Bruus,~H.; Flensberg,~K. \emph{{Many-Body Quantum Theory in Condensed Matter
  Physics: An Introduction}}; OUP Oxford: Oxford, 2004\relax
\mciteBstWouldAddEndPuncttrue
\mciteSetBstMidEndSepPunct{\mcitedefaultmidpunct}
{\mcitedefaultendpunct}{\mcitedefaultseppunct}\relax
\EndOfBibitem
\bibitem[Martin \latin{et~al.}(2016)Martin, Reining, and Ceperley]{martin2016}
Martin,~R.~M.; Reining,~L.; Ceperley,~D.~M. \emph{{Interacting electrons}};
  Cambridge University Press, 2016\relax
\mciteBstWouldAddEndPuncttrue
\mciteSetBstMidEndSepPunct{\mcitedefaultmidpunct}
{\mcitedefaultendpunct}{\mcitedefaultseppunct}\relax
\EndOfBibitem
\bibitem[Klein(1961)]{Klein1961}
Klein,~A. {Perturbation theory for an infinite medium of fermions}. \emph{Phys.
  Rev.} \textbf{1961}, \emph{121}, 950--956\relax
\mciteBstWouldAddEndPuncttrue
\mciteSetBstMidEndSepPunct{\mcitedefaultmidpunct}
{\mcitedefaultendpunct}{\mcitedefaultseppunct}\relax
\EndOfBibitem
\bibitem[Luttinger and Ward(1960)Luttinger, and Ward]{Luttinger1960a}
Luttinger,~J.~M.; Ward,~J.~C. {Ground-state energy of a many-fermion system.
  II}. \emph{Phys. Rev.} \textbf{1960}, \emph{118}, 1417--1427\relax
\mciteBstWouldAddEndPuncttrue
\mciteSetBstMidEndSepPunct{\mcitedefaultmidpunct}
{\mcitedefaultendpunct}{\mcitedefaultseppunct}\relax
\EndOfBibitem
\bibitem[Kohn and {Sham. L. J.}(1965)Kohn, and {Sham. L. J.}]{Kohn1965}
Kohn,~W.; {Sham. L. J.}, {Self-Consistent Equations Including Exchange and
  Correlation Effects}. \emph{Phys. Rev.} \textbf{1965}, \emph{140},
  A1133\relax
\mciteBstWouldAddEndPuncttrue
\mciteSetBstMidEndSepPunct{\mcitedefaultmidpunct}
{\mcitedefaultendpunct}{\mcitedefaultseppunct}\relax
\EndOfBibitem
\bibitem[Hohenberg and Kohn(1964)Hohenberg, and Kohn]{Hohenberg1964}
Hohenberg,~P.; Kohn,~W. {Inhomogeneous Electron Gas}. \emph{Phys. Rev.}
  \textbf{1964}, \emph{136}, 864--871\relax
\mciteBstWouldAddEndPuncttrue
\mciteSetBstMidEndSepPunct{\mcitedefaultmidpunct}
{\mcitedefaultendpunct}{\mcitedefaultseppunct}\relax
\EndOfBibitem
\bibitem[Casida(1995)]{Casida1995}
Casida,~M.~E. {Generalization of the optimized-effective-potential model to
  include electron correlation: A variational derivation of the
  Sham-Schl{\"{u}}ter equation for the exact exchange-correlation potential}.
  \emph{Phys. Rev. A} \textbf{1995}, \emph{51}, 2005--2013\relax
\mciteBstWouldAddEndPuncttrue
\mciteSetBstMidEndSepPunct{\mcitedefaultmidpunct}
{\mcitedefaultendpunct}{\mcitedefaultseppunct}\relax
\EndOfBibitem
\bibitem[Dahlen \latin{et~al.}(2006)Dahlen, van Leeuwen, and von
  Barth]{Dahlen2006}
Dahlen,~N.~E.; van Leeuwen,~R.; von Barth,~U. {Variational energy functionals
  of the Green function and of the density tested on molecules}. \emph{Phys.
  Rev. A - At. Mol. Opt. Phys.} \textbf{2006}, \emph{73}, 012511\relax
\mciteBstWouldAddEndPuncttrue
\mciteSetBstMidEndSepPunct{\mcitedefaultmidpunct}
{\mcitedefaultendpunct}{\mcitedefaultseppunct}\relax
\EndOfBibitem
\bibitem[Hedin(1965)]{Hedin1965}
Hedin,~L. {New method for calculating the one-particle Green's function with
  application to the electron-gas problem}. \emph{Phys. Rev.} \textbf{1965},
  \emph{139}, A796\relax
\mciteBstWouldAddEndPuncttrue
\mciteSetBstMidEndSepPunct{\mcitedefaultmidpunct}
{\mcitedefaultendpunct}{\mcitedefaultseppunct}\relax
\EndOfBibitem
\bibitem[Not()]{Note-1}
The pair bubble approximation is typically also denoted as RPA. To avoid
  potential confusion with the expression for the correlation energy, we will
  use the term bubble approximation when referring to the screening.\relax
\mciteBstWouldAddEndPunctfalse
\mciteSetBstMidEndSepPunct{\mcitedefaultmidpunct}
{}{\mcitedefaultseppunct}\relax
\EndOfBibitem
\bibitem[Zhang and Gr{\"{u}}neis(2019)Zhang, and Gr{\"{u}}neis]{Zhang2019}
Zhang,~I.~Y.; Gr{\"{u}}neis,~A. {Coupled cluster theory in materials science}.
  \emph{Front. Mater.} \textbf{2019}, \emph{6}, 00123\relax
\mciteBstWouldAddEndPuncttrue
\mciteSetBstMidEndSepPunct{\mcitedefaultmidpunct}
{\mcitedefaultendpunct}{\mcitedefaultseppunct}\relax
\EndOfBibitem
\bibitem[Keller \latin{et~al.}(2022)Keller, Tsatsoulis, Reuter, and
  Margraf]{Keller2022}
Keller,~E.; Tsatsoulis,~T.; Reuter,~K.; Margraf,~J.~T. {Regularized
  second-order correlation methods for extended systems}. \emph{J. Chem. Phys.}
  \textbf{2022}, \emph{156}, 024106\relax
\mciteBstWouldAddEndPuncttrue
\mciteSetBstMidEndSepPunct{\mcitedefaultmidpunct}
{\mcitedefaultendpunct}{\mcitedefaultseppunct}\relax
\EndOfBibitem
\bibitem[Eshuis \latin{et~al.}(2010)Eshuis, Yarkony, and Furche]{Eshuis2010}
Eshuis,~H.; Yarkony,~J.; Furche,~F. {Fast computation of molecular random phase
  approximation correlation energies using resolution of the identity and
  imaginary frequency integration}. \emph{J. Chem. Phys.} \textbf{2010},
  \emph{132}, 234114\relax
\mciteBstWouldAddEndPuncttrue
\mciteSetBstMidEndSepPunct{\mcitedefaultmidpunct}
{\mcitedefaultendpunct}{\mcitedefaultseppunct}\relax
\EndOfBibitem
\bibitem[Wilhelm \latin{et~al.}(2016)Wilhelm, Seewald, {Del Ben}, and
  Hutter]{Wilhelm2016}
Wilhelm,~J.; Seewald,~P.; {Del Ben},~M.; Hutter,~J. {Large-Scale Cubic-Scaling
  Random Phase Approximation Correlation Energy Calculations Using a Gaussian
  Basis}. \emph{J. Chem. Theory Comput.} \textbf{2016}, \emph{12},
  5851--5859\relax
\mciteBstWouldAddEndPuncttrue
\mciteSetBstMidEndSepPunct{\mcitedefaultmidpunct}
{\mcitedefaultendpunct}{\mcitedefaultseppunct}\relax
\EndOfBibitem
\bibitem[Wilhelm \latin{et~al.}(2018)Wilhelm, Golze, Talirz, Hutter, and
  Pignedoli]{Wilhelm2018}
Wilhelm,~J.; Golze,~D.; Talirz,~L.; Hutter,~J.; Pignedoli,~C.~A. {Toward GW
  Calculations on Thousands of Atoms}. \emph{J. Phys. Chem. Lett.}
  \textbf{2018}, \emph{9}, 306--312\relax
\mciteBstWouldAddEndPuncttrue
\mciteSetBstMidEndSepPunct{\mcitedefaultmidpunct}
{\mcitedefaultendpunct}{\mcitedefaultseppunct}\relax
\EndOfBibitem
\bibitem[Wilhelm \latin{et~al.}(2021)Wilhelm, Seewald, and Golze]{Wilhelm2021}
Wilhelm,~J.; Seewald,~P.; Golze,~D. {Low-scaling GW with benchmark accuracy and
  application to phosphorene nanosheets}. \emph{J. Chem. Theory Comput.}
  \textbf{2021}, \emph{17}, 1662--1677\relax
\mciteBstWouldAddEndPuncttrue
\mciteSetBstMidEndSepPunct{\mcitedefaultmidpunct}
{\mcitedefaultendpunct}{\mcitedefaultseppunct}\relax
\EndOfBibitem
\bibitem[F{\"{o}}rster and Visscher(2020)F{\"{o}}rster, and
  Visscher]{Forster2020b}
F{\"{o}}rster,~A.; Visscher,~L. {Low-Order Scaling G0W0 by Pair Atomic Density
  Fitting}. \emph{J. Chem. Theory Comput.} \textbf{2020}, \emph{16},
  7381--7399\relax
\mciteBstWouldAddEndPuncttrue
\mciteSetBstMidEndSepPunct{\mcitedefaultmidpunct}
{\mcitedefaultendpunct}{\mcitedefaultseppunct}\relax
\EndOfBibitem
\bibitem[Duchemin and Blase(2019)Duchemin, and Blase]{Duchemin2019}
Duchemin,~I.; Blase,~X. {Separable resolution-of-the-identity with all-electron
  Gaussian bases: Application to cubic-scaling RPA}. \emph{J. Chem. Phys.}
  \textbf{2019}, \emph{150}, 174120\relax
\mciteBstWouldAddEndPuncttrue
\mciteSetBstMidEndSepPunct{\mcitedefaultmidpunct}
{\mcitedefaultendpunct}{\mcitedefaultseppunct}\relax
\EndOfBibitem
\bibitem[Duchemin and Blase(2021)Duchemin, and Blase]{Duchemin2021a}
Duchemin,~I.; Blase,~X. {Cubic-Scaling All-Electron GW Calculations with a
  Separable Density-Fitting Space-Time Approach}. \emph{J. Chem. Theory
  Comput.} \textbf{2021}, \emph{17}, 2383--2393\relax
\mciteBstWouldAddEndPuncttrue
\mciteSetBstMidEndSepPunct{\mcitedefaultmidpunct}
{\mcitedefaultendpunct}{\mcitedefaultseppunct}\relax
\EndOfBibitem
\bibitem[Schurkus and Ochsenfeld(2016)Schurkus, and Ochsenfeld]{Schurkus2016}
Schurkus,~H.~F.; Ochsenfeld,~C. {Communication: An effective linear-scaling
  atomic-orbital reformulation of the random-phase approximation using a
  contracted double-Laplace transformation}. \emph{J. Chem. Phys.}
  \textbf{2016}, \emph{144}, 031101\relax
\mciteBstWouldAddEndPuncttrue
\mciteSetBstMidEndSepPunct{\mcitedefaultmidpunct}
{\mcitedefaultendpunct}{\mcitedefaultseppunct}\relax
\EndOfBibitem
\bibitem[Luenser \latin{et~al.}(2017)Luenser, Schurkus, and
  Ochsenfeld]{Luenser2017}
Luenser,~A.; Schurkus,~H.~F.; Ochsenfeld,~C. {Vanishing-Overhead Linear-Scaling
  Random Phase Approximation by Cholesky Decomposition and an Attenuated
  Coulomb-Metric}. \emph{J. Chem. Theory Comput.} \textbf{2017}, \emph{13},
  1647--1655\relax
\mciteBstWouldAddEndPuncttrue
\mciteSetBstMidEndSepPunct{\mcitedefaultmidpunct}
{\mcitedefaultendpunct}{\mcitedefaultseppunct}\relax
\EndOfBibitem
\bibitem[Vl{\v{c}}ek \latin{et~al.}(2017)Vl{\v{c}}ek, Rabani, Neuhauser, and
  Baer]{Vlcek2017}
Vl{\v{c}}ek,~V.; Rabani,~E.; Neuhauser,~D.; Baer,~R. {Stochastic GW
  Calculations for Molecules}. \emph{J. Chem. Theory Comput.} \textbf{2017},
  \emph{13}, 4997--5003\relax
\mciteBstWouldAddEndPuncttrue
\mciteSetBstMidEndSepPunct{\mcitedefaultmidpunct}
{\mcitedefaultendpunct}{\mcitedefaultseppunct}\relax
\EndOfBibitem
\bibitem[Graf \latin{et~al.}(2018)Graf, Beuerle, Schurkus, Luenser, Savasci,
  and Ochsenfeld]{Graf2018}
Graf,~D.; Beuerle,~M.; Schurkus,~H.~F.; Luenser,~A.; Savasci,~G.;
  Ochsenfeld,~C. {Accurate and Efficient Parallel Implementation of an
  Effective Linear-Scaling Direct Random Phase Approximation Method}. \emph{J.
  Chem. Theory Comput.} \textbf{2018}, \emph{14}, 2505--2515\relax
\mciteBstWouldAddEndPuncttrue
\mciteSetBstMidEndSepPunct{\mcitedefaultmidpunct}
{\mcitedefaultendpunct}{\mcitedefaultseppunct}\relax
\EndOfBibitem
\bibitem[Lu \latin{et~al.}(2010)Lu, Nguyen, and Galli]{Lu2010}
Lu,~D.; Nguyen,~H.~V.; Galli,~G. {Power series expansion of the random phase
  approximation correlation energy: The role of the third- and higher-order
  contributions}. \emph{J. Chem. Phys.} \textbf{2010}, \emph{133}\relax
\mciteBstWouldAddEndPuncttrue
\mciteSetBstMidEndSepPunct{\mcitedefaultmidpunct}
{\mcitedefaultendpunct}{\mcitedefaultseppunct}\relax
\EndOfBibitem
\bibitem[Leb{\`{e}}gue \latin{et~al.}(2010)Leb{\`{e}}gue, Harl, Gould,
  {\'{A}}ngy{\'{a}}n, Kresse, and Dobson]{Lebegue2010}
Leb{\`{e}}gue,~S.; Harl,~J.; Gould,~T.; {\'{A}}ngy{\'{a}}n,~J.~G.; Kresse,~G.;
  Dobson,~J.~F. {Cohesive properties and asymptotics of the dispersion
  interaction in graphite by the random phase approximation}. \emph{Phys. Rev.
  Lett.} \textbf{2010}, \emph{105}, 1--4\relax
\mciteBstWouldAddEndPuncttrue
\mciteSetBstMidEndSepPunct{\mcitedefaultmidpunct}
{\mcitedefaultendpunct}{\mcitedefaultseppunct}\relax
\EndOfBibitem
\bibitem[Nguyen \latin{et~al.}(2020)Nguyen, Chen, Agee, Burow, Tang, and
  Furche]{Nguyen2020}
Nguyen,~B.~D.; Chen,~G.~P.; Agee,~M.~M.; Burow,~A.~M.; Tang,~M.~P.; Furche,~F.
  {Divergence of Many-Body Perturbation Theory for Noncovalent Interactions of
  Large Molecules}. \emph{J. Chem. Theory Comput.} \textbf{2020}, \emph{16},
  2258--2273\relax
\mciteBstWouldAddEndPuncttrue
\mciteSetBstMidEndSepPunct{\mcitedefaultmidpunct}
{\mcitedefaultendpunct}{\mcitedefaultseppunct}\relax
\EndOfBibitem
\bibitem[Irmler \latin{et~al.}(2019)Irmler, Gallo, Hummel, and
  Gr{\"{u}}neis]{Irmler2019a}
Irmler,~A.; Gallo,~A.; Hummel,~F.; Gr{\"{u}}neis,~A. {Duality of Ring and
  Ladder Diagrams and Its Importance for Many-Electron Perturbation Theories}.
  \emph{Phys. Rev. Lett.} \textbf{2019}, \emph{123}, 156401\relax
\mciteBstWouldAddEndPuncttrue
\mciteSetBstMidEndSepPunct{\mcitedefaultmidpunct}
{\mcitedefaultendpunct}{\mcitedefaultseppunct}\relax
\EndOfBibitem
\bibitem[Hummel \latin{et~al.}(2019)Hummel, Gr{\"{u}}neis, Kresse, and
  Ziesche]{Hummel2019}
Hummel,~F.; Gr{\"{u}}neis,~A.; Kresse,~G.; Ziesche,~P. {Screened Exchange
  Corrections to the Random Phase Approximation from Many-Body Perturbation
  Theory}. \emph{J. Chem. Theory Comput.} \textbf{2019}, \emph{15},
  3223--3236\relax
\mciteBstWouldAddEndPuncttrue
\mciteSetBstMidEndSepPunct{\mcitedefaultmidpunct}
{\mcitedefaultendpunct}{\mcitedefaultseppunct}\relax
\EndOfBibitem
\bibitem[Singwi \latin{et~al.}(1968)Singwi, Tosi, Land, and
  Sj{\"{o}}lander]{Singwi1968}
Singwi,~K.~S.; Tosi,~M.~P.; Land,~R.~H.; Sj{\"{o}}lander,~A. {Electron
  correlations at metallic densities}. \emph{Phys. Rev.} \textbf{1968},
  \emph{176}, 589--599\relax
\mciteBstWouldAddEndPuncttrue
\mciteSetBstMidEndSepPunct{\mcitedefaultmidpunct}
{\mcitedefaultendpunct}{\mcitedefaultseppunct}\relax
\EndOfBibitem
\bibitem[Jiang and Engel(2007)Jiang, and Engel]{Jiang2007}
Jiang,~H.; Engel,~E. {Random-phase-approximation-based correlation energy
  functionals: Benchmark results for atoms}. \emph{J. Chem. Phys.}
  \textbf{2007}, \emph{127}, 184108\relax
\mciteBstWouldAddEndPuncttrue
\mciteSetBstMidEndSepPunct{\mcitedefaultmidpunct}
{\mcitedefaultendpunct}{\mcitedefaultseppunct}\relax
\EndOfBibitem
\bibitem[Lochan and Head-Gordon(2007)Lochan, and Head-Gordon]{Lochan2007}
Lochan,~R.~C.; Head-Gordon,~M. {Orbital-optimized opposite-spin scaled
  second-order correlation: An economical method to improve the description of
  open-shell molecules}. \emph{J. Chem. Phys.} \textbf{2007}, \emph{126},
  164101\relax
\mciteBstWouldAddEndPuncttrue
\mciteSetBstMidEndSepPunct{\mcitedefaultmidpunct}
{\mcitedefaultendpunct}{\mcitedefaultseppunct}\relax
\EndOfBibitem
\bibitem[Neese \latin{et~al.}(2009)Neese, Schwabe, Kossmann, Schirmer, and
  Grimme]{Neese2009d}
Neese,~F.; Schwabe,~T.; Kossmann,~S.; Schirmer,~B.; Grimme,~S. {Assessment of
  orbital-optimized, spin-component scaled second-order many-body perturbation
  theory for thermochemistry and kinetics}. \emph{J. Chem. Theory Comput.}
  \textbf{2009}, \emph{5}, 3060--3073\relax
\mciteBstWouldAddEndPuncttrue
\mciteSetBstMidEndSepPunct{\mcitedefaultmidpunct}
{\mcitedefaultendpunct}{\mcitedefaultseppunct}\relax
\EndOfBibitem
\bibitem[Kossmann and Neese(2010)Kossmann, and Neese]{Kossmann2010}
Kossmann,~S.; Neese,~F. {Correlated ab initio spin densities for larger
  molecules: Orbital-optimized spin-component-scaled MP2 method}. \emph{J.
  Phys. Chem. A} \textbf{2010}, \emph{114}, 11768--11781\relax
\mciteBstWouldAddEndPuncttrue
\mciteSetBstMidEndSepPunct{\mcitedefaultmidpunct}
{\mcitedefaultendpunct}{\mcitedefaultseppunct}\relax
\EndOfBibitem
\bibitem[Gr{\"{u}}neis \latin{et~al.}(2010)Gr{\"{u}}neis, Marsman, and
  Kresse]{Gruneis2010a}
Gr{\"{u}}neis,~A.; Marsman,~M.; Kresse,~G. {Second-order M{\o}ller-Plesset
  perturbation theory applied to extended systems. II. Structural and energetic
  properties}. \emph{J. Chem. Phys.} \textbf{2010}, \emph{133}, 074107\relax
\mciteBstWouldAddEndPuncttrue
\mciteSetBstMidEndSepPunct{\mcitedefaultmidpunct}
{\mcitedefaultendpunct}{\mcitedefaultseppunct}\relax
\EndOfBibitem
\bibitem[van Schilfgaarde \latin{et~al.}(2006)van Schilfgaarde, Kotani, and
  Faleev]{VanSchilfgaarde2006}
van Schilfgaarde,~M.; Kotani,~T.; Faleev,~S. {Quasiparticle self-consistent GW
  theory}. \emph{Phys. Rev. Lett.} \textbf{2006}, \emph{96}, 226402\relax
\mciteBstWouldAddEndPuncttrue
\mciteSetBstMidEndSepPunct{\mcitedefaultmidpunct}
{\mcitedefaultendpunct}{\mcitedefaultseppunct}\relax
\EndOfBibitem
\bibitem[Tal \latin{et~al.}(2021)Tal, Chen, and Pasquarello]{Tal2021}
Tal,~A.; Chen,~W.; Pasquarello,~A. {Vertex function compliant with the Ward
  identity for quasiparticle self-consistent calculations beyond GW}.
  \emph{Phys. Rev. B} \textbf{2021}, \emph{103}, 161104\relax
\mciteBstWouldAddEndPuncttrue
\mciteSetBstMidEndSepPunct{\mcitedefaultmidpunct}
{\mcitedefaultendpunct}{\mcitedefaultseppunct}\relax
\EndOfBibitem
\bibitem[Jung \latin{et~al.}(2004)Jung, Lochan, Dutoi, and
  Head-Gordon]{Jung2004}
Jung,~Y.; Lochan,~R.~C.; Dutoi,~A.~D.; Head-Gordon,~M. {Scaled opposite-spin
  second order moller-plesset correlation energy: An economical electronic
  structure method}. \emph{J. Chem. Phys.} \textbf{2004}, \emph{121},
  9793--9802\relax
\mciteBstWouldAddEndPuncttrue
\mciteSetBstMidEndSepPunct{\mcitedefaultmidpunct}
{\mcitedefaultendpunct}{\mcitedefaultseppunct}\relax
\EndOfBibitem
\bibitem[Lochan \latin{et~al.}(2005)Lochan, Jung, and Head-Gordon]{Lochan2005}
Lochan,~R.~C.; Jung,~Y.; Head-Gordon,~M. {Scaled opposite spin second order
  M{\o}ller-Plesset theory with improved physical description of long-range
  dispersion interactions}. \emph{J. Phys. Chem. A} \textbf{2005}, \emph{109},
  7598--7605\relax
\mciteBstWouldAddEndPuncttrue
\mciteSetBstMidEndSepPunct{\mcitedefaultmidpunct}
{\mcitedefaultendpunct}{\mcitedefaultseppunct}\relax
\EndOfBibitem
\bibitem[Grimme(2003)]{Grimme2003}
Grimme,~S. {Improved second-order M{\o}ller-Plesset perturbation theory by
  separate scaling of parallel- and antiparallel-spin pair correlation
  energies}. \emph{J. Chem. Phys.} \textbf{2003}, \emph{118}, 9095--9102\relax
\mciteBstWouldAddEndPuncttrue
\mciteSetBstMidEndSepPunct{\mcitedefaultmidpunct}
{\mcitedefaultendpunct}{\mcitedefaultseppunct}\relax
\EndOfBibitem
\bibitem[Szabados(2006)]{Szabados2006}
Szabados,~{\'{A}}. {Theoretical interpretation of Grimme's
  spin-component-scaled second order M{\o}ller-Plesset theory}. \emph{J. Chem.
  Phys.} \textbf{2006}, \emph{125}, 214105\relax
\mciteBstWouldAddEndPuncttrue
\mciteSetBstMidEndSepPunct{\mcitedefaultmidpunct}
{\mcitedefaultendpunct}{\mcitedefaultseppunct}\relax
\EndOfBibitem
\bibitem[Pitoň{\'{a}}k \latin{et~al.}(2009)Pitoň{\'{a}}k, Neogr{\'{a}}dy,
  {\v{C}}ern{\'{y}}, Grimme, and Hobza]{Pitonak2009}
Pitoň{\'{a}}k,~M.; Neogr{\'{a}}dy,~P.; {\v{C}}ern{\'{y}},~J.; Grimme,~S.;
  Hobza,~P. {Scaled MP3 non-covalent interaction energies agree closely with
  accurate CCSD(T) benchmark data}. \emph{ChemPhysChem} \textbf{2009},
  \emph{10}, 282--289\relax
\mciteBstWouldAddEndPuncttrue
\mciteSetBstMidEndSepPunct{\mcitedefaultmidpunct}
{\mcitedefaultendpunct}{\mcitedefaultseppunct}\relax
\EndOfBibitem
\bibitem[Sedlak \latin{et~al.}(2013)Sedlak, Riley, Řez{\'{a}}{\v{c}},
  Pitoň{\'{a}}k, and Hobza]{Sedlak2013a}
Sedlak,~R.; Riley,~K.~E.; Řez{\'{a}}{\v{c}},~J.; Pitoň{\'{a}}k,~M.; Hobza,~P.
  {MP2.5 and MP2.X: Approaching CCSD(T) quality description of noncovalent
  interaction at the cost of a single CCSD iteration}. \emph{ChemPhysChem}
  \textbf{2013}, \emph{14}, 698--707\relax
\mciteBstWouldAddEndPuncttrue
\mciteSetBstMidEndSepPunct{\mcitedefaultmidpunct}
{\mcitedefaultendpunct}{\mcitedefaultseppunct}\relax
\EndOfBibitem
\bibitem[Goldey and Head-Gordon(2012)Goldey, and Head-Gordon]{Goldey2012}
Goldey,~M.; Head-Gordon,~M. {Attenuating away the errors in inter- and
  intramolecular interactions from second-order M{\o}ller-plesset calculations
  in the small aug-cc-pVDZ basis set}. \emph{J. Phys. Chem. Lett.}
  \textbf{2012}, \emph{3}, 3592--3598\relax
\mciteBstWouldAddEndPuncttrue
\mciteSetBstMidEndSepPunct{\mcitedefaultmidpunct}
{\mcitedefaultendpunct}{\mcitedefaultseppunct}\relax
\EndOfBibitem
\bibitem[Goldey \latin{et~al.}(2013)Goldey, Dutoi, and Head-Gordon]{Goldey2013}
Goldey,~M.; Dutoi,~A.; Head-Gordon,~M. {Attenuated second-order
  M{\o}ller-Plesset perturbation theory: Performance within the aug-cc-pVTZ
  basis}. \emph{Phys. Chem. Chem. Phys.} \textbf{2013}, \emph{15},
  15869--15875\relax
\mciteBstWouldAddEndPuncttrue
\mciteSetBstMidEndSepPunct{\mcitedefaultmidpunct}
{\mcitedefaultendpunct}{\mcitedefaultseppunct}\relax
\EndOfBibitem
\bibitem[Goldey \latin{et~al.}(2015)Goldey, Belzunces, and
  Head-Gordon]{Goldey2015}
Goldey,~M.~B.; Belzunces,~B.; Head-Gordon,~M. {Attenuated MP2 with a Long-Range
  Dispersion Correction for Treating Nonbonded Interactions}. \emph{J. Chem.
  Theory Comput.} \textbf{2015}, \emph{11}, 4159--4168\relax
\mciteBstWouldAddEndPuncttrue
\mciteSetBstMidEndSepPunct{\mcitedefaultmidpunct}
{\mcitedefaultendpunct}{\mcitedefaultseppunct}\relax
\EndOfBibitem
\bibitem[Lee and Head-Gordon(2018)Lee, and Head-Gordon]{Lee2018}
Lee,~J.; Head-Gordon,~M. {Regularized Orbital-Optimized Second-Order
  M{\o}ller-Plesset Perturbation Theory: A Reliable Fifth-Order-Scaling
  Electron Correlation Model with Orbital Energy Dependent Regularizers}.
  \emph{J. Chem. Theory Comput.} \textbf{2018}, \emph{14}, 5203--5219\relax
\mciteBstWouldAddEndPuncttrue
\mciteSetBstMidEndSepPunct{\mcitedefaultmidpunct}
{\mcitedefaultendpunct}{\mcitedefaultseppunct}\relax
\EndOfBibitem
\bibitem[Monino and Loos(2022)Monino, and Loos]{Monino2022}
Monino,~E.; Loos,~P.-F. {Unphysical Discontinuities, Intruder States and
  Regularization in GW Methods}. \emph{J. Chem. Phys.} \textbf{2022},
  \emph{156}, 231101\relax
\mciteBstWouldAddEndPuncttrue
\mciteSetBstMidEndSepPunct{\mcitedefaultmidpunct}
{\mcitedefaultendpunct}{\mcitedefaultseppunct}\relax
\EndOfBibitem
\bibitem[Pittner(2003)]{Pittner2003}
Pittner,~J. {Continuous transition between Brillouin-Wigner and
  Rayleigh-Schr{\"{o}}dinger perturbation theory, generalized Bloch equation,
  and Hilbert space multireference coupled cluster}. \emph{J. Chem. Phys.}
  \textbf{2003}, \emph{118}, 10876\relax
\mciteBstWouldAddEndPuncttrue
\mciteSetBstMidEndSepPunct{\mcitedefaultmidpunct}
{\mcitedefaultendpunct}{\mcitedefaultseppunct}\relax
\EndOfBibitem
\bibitem[Engel and Jiang(2006)Engel, and Jiang]{Engel2006}
Engel,~E.; Jiang,~H. {Orbital-Dependent Representation of the Correlation
  Energy Functional: Properties of Second-Order Kohn–Sham Perturbation
  Expansion}. \emph{Int. J. Quantum Chem.} \textbf{2006}, \emph{106},
  3242--3259\relax
\mciteBstWouldAddEndPuncttrue
\mciteSetBstMidEndSepPunct{\mcitedefaultmidpunct}
{\mcitedefaultendpunct}{\mcitedefaultseppunct}\relax
\EndOfBibitem
\bibitem[Jiang and Engel(2006)Jiang, and Engel]{Jiang2006}
Jiang,~H.; Engel,~E. {Kohn-Sham perturbation theory: Simple solution to
  variational instability of second order correlation energy functional}.
  \emph{J. Chem. Phys.} \textbf{2006}, \emph{125}, 184108\relax
\mciteBstWouldAddEndPuncttrue
\mciteSetBstMidEndSepPunct{\mcitedefaultmidpunct}
{\mcitedefaultendpunct}{\mcitedefaultseppunct}\relax
\EndOfBibitem
\bibitem[Daas \latin{et~al.}(2021)Daas, Fabiano, {Della Sala}, Gori-Giorgi, and
  Vuckovic]{Daas2021}
Daas,~T.~J.; Fabiano,~E.; {Della Sala},~F.; Gori-Giorgi,~P.; Vuckovic,~S.
  {Noncovalent Interactions from Models for the M{\o}ller-Plesset Adiabatic
  Connection}. \emph{J. Phys. Chem. Lett.} \textbf{2021}, \emph{12},
  4867--4875\relax
\mciteBstWouldAddEndPuncttrue
\mciteSetBstMidEndSepPunct{\mcitedefaultmidpunct}
{\mcitedefaultendpunct}{\mcitedefaultseppunct}\relax
\EndOfBibitem
\bibitem[Daas \latin{et~al.}(2022)Daas, Kooi, Grooteman, Seidl, and
  Gori-Giorgi]{Daas2022}
Daas,~T.~J.; Kooi,~D.~P.; Grooteman,~A. J. A.~F.; Seidl,~M.; Gori-Giorgi,~P.
  {Gradient Expansions for the Large-Coupling Strength Limit of the
  M{\o}ller–Plesset Adiabatic Connection}. \emph{J. Chem. Theory Comput.}
  \textbf{2022}, \emph{18}, 1584--1594\relax
\mciteBstWouldAddEndPuncttrue
\mciteSetBstMidEndSepPunct{\mcitedefaultmidpunct}
{\mcitedefaultendpunct}{\mcitedefaultseppunct}\relax
\EndOfBibitem
\bibitem[Kurth \latin{et~al.}(1999)Kurth, Perdew, and Blaha]{Kurth1999}
Kurth,~S.; Perdew,~J.~P.; Blaha,~P. {Molecular and solid-state tests of density
  functional approximations: LSD, GGAs, and Meta-GGAs}. \emph{Int. J. Quantum
  Chem.} \textbf{1999}, \emph{75}, 889--909\relax
\mciteBstWouldAddEndPuncttrue
\mciteSetBstMidEndSepPunct{\mcitedefaultmidpunct}
{\mcitedefaultendpunct}{\mcitedefaultseppunct}\relax
\EndOfBibitem
\bibitem[Yan \latin{et~al.}(2000)Yan, Perdew, and Kurth]{Yan2000}
Yan,~Z.; Perdew,~J.~P.; Kurth,~S. {Density functional for short-range
  correlation: Accuracy of the random-phase approximation for isoelectronic
  energy changes}. \emph{Phys. Rev. B - Condens. Matter Mater. Phys.}
  \textbf{2000}, \emph{61}, 16430--16439\relax
\mciteBstWouldAddEndPuncttrue
\mciteSetBstMidEndSepPunct{\mcitedefaultmidpunct}
{\mcitedefaultendpunct}{\mcitedefaultseppunct}\relax
\EndOfBibitem
\bibitem[Angyan \latin{et~al.}(2005)Angyan, Gerber, Savin, and
  Toulouse]{Angyan2005}
Angyan,~J.~G.; Gerber,~I.~C.; Savin,~A.; Toulouse,~J. {Van der Waals forces in
  density functional theory: Perturbational long-range electron-interaction
  corrections}. \emph{Phys. Rev. A} \textbf{2005}, \emph{72}, 012510\relax
\mciteBstWouldAddEndPuncttrue
\mciteSetBstMidEndSepPunct{\mcitedefaultmidpunct}
{\mcitedefaultendpunct}{\mcitedefaultseppunct}\relax
\EndOfBibitem
\bibitem[Janesko \latin{et~al.}(2009)Janesko, Henderson, and
  Scuseria]{Janesko2009a}
Janesko,~B.~G.; Henderson,~T.~M.; Scuseria,~G.~E. {Long-range-corrected hybrids
  including random phase approximation correlation}. \emph{J. Chem. Phys.}
  \textbf{2009}, \emph{130}, 081105\relax
\mciteBstWouldAddEndPuncttrue
\mciteSetBstMidEndSepPunct{\mcitedefaultmidpunct}
{\mcitedefaultendpunct}{\mcitedefaultseppunct}\relax
\EndOfBibitem
\bibitem[Janesko \latin{et~al.}(2009)Janesko, Henderson, and
  Scuseria]{Janesko2009b}
Janesko,~B.~G.; Henderson,~T.~M.; Scuseria,~G.~E. {Long-range-corrected hybrid
  density functionals including random phase approximation correlation:
  Application to noncovalent interactions}. \emph{J. Chem. Phys.}
  \textbf{2009}, \emph{131}, 034110\relax
\mciteBstWouldAddEndPuncttrue
\mciteSetBstMidEndSepPunct{\mcitedefaultmidpunct}
{\mcitedefaultendpunct}{\mcitedefaultseppunct}\relax
\EndOfBibitem
\bibitem[Toulouse \latin{et~al.}(2009)Toulouse, Gerber, Jansen, Savin, and
  {\'{A}}ngy{\'{a}}n]{Toulouse2009}
Toulouse,~J.; Gerber,~I.~C.; Jansen,~G.; Savin,~A.; {\'{A}}ngy{\'{a}}n,~J.~G.
  {Adiabatic-connection fluctuation-dissipation density-functional theory based
  on range separation}. \emph{Phys. Rev. Lett.} \textbf{2009}, \emph{102},
  096404\relax
\mciteBstWouldAddEndPuncttrue
\mciteSetBstMidEndSepPunct{\mcitedefaultmidpunct}
{\mcitedefaultendpunct}{\mcitedefaultseppunct}\relax
\EndOfBibitem
\bibitem[Zhu \latin{et~al.}(2010)Zhu, Toulouse, Savin, and
  {\'{A}}ngy{\'{a}}n]{Zhu2010}
Zhu,~W.; Toulouse,~J.; Savin,~A.; {\'{A}}ngy{\'{a}}n,~J.~G. {Range-separated
  density-functional theory with random phase approximation applied to
  noncovalent intermolecular interactions}. \emph{J. Chem. Phys.}
  \textbf{2010}, \emph{132}, 244108\relax
\mciteBstWouldAddEndPuncttrue
\mciteSetBstMidEndSepPunct{\mcitedefaultmidpunct}
{\mcitedefaultendpunct}{\mcitedefaultseppunct}\relax
\EndOfBibitem
\bibitem[Toulouse \latin{et~al.}(2010)Toulouse, Zhu, {\'{A}}ngy{\'{a}}n, and
  Savin]{Toulouse2010}
Toulouse,~J.; Zhu,~W.; {\'{A}}ngy{\'{a}}n,~J.~G.; Savin,~A. {Range-separated
  density-functional theory with the random-phase approximation: Detailed
  formalism and illustrative applications}. \emph{Phys. Rev. A - At. Mol. Opt.
  Phys.} \textbf{2010}, \emph{82}, 032502\relax
\mciteBstWouldAddEndPuncttrue
\mciteSetBstMidEndSepPunct{\mcitedefaultmidpunct}
{\mcitedefaultendpunct}{\mcitedefaultseppunct}\relax
\EndOfBibitem
\bibitem[Toulouse \latin{et~al.}(2011)Toulouse, Zhu, Savin, Jansen, and
  {\'{A}}ngy{\'{a}}n]{Toulouse2011b}
Toulouse,~J.; Zhu,~W.; Savin,~A.; Jansen,~G.; {\'{A}}ngy{\'{a}}n,~J.~G.
  {Closed-shell ring coupled cluster doubles theory with range separation
  applied on weak intermolecular interactions}. \emph{J. Chem. Phys.}
  \textbf{2011}, \emph{135}, 084119\relax
\mciteBstWouldAddEndPuncttrue
\mciteSetBstMidEndSepPunct{\mcitedefaultmidpunct}
{\mcitedefaultendpunct}{\mcitedefaultseppunct}\relax
\EndOfBibitem
\bibitem[Beuerle and Ochsenfeld(2017)Beuerle, and Ochsenfeld]{Beuerle2017}
Beuerle,~M.; Ochsenfeld,~C. {Short-range second order screened exchange
  correction to RPA correlation energies}. \emph{J. Chem. Phys.} \textbf{2017},
  \emph{147}, 204107\relax
\mciteBstWouldAddEndPuncttrue
\mciteSetBstMidEndSepPunct{\mcitedefaultmidpunct}
{\mcitedefaultendpunct}{\mcitedefaultseppunct}\relax
\EndOfBibitem
\bibitem[Ren \latin{et~al.}(2011)Ren, Tkatchenko, Rinke, and
  Scheffler]{Ren2011}
Ren,~X.; Tkatchenko,~A.; Rinke,~P.; Scheffler,~M. {Beyond the random-phase
  approximation for the electron correlation energy: The importance of single
  excitations}. \emph{Phys. Rev. Lett.} \textbf{2011}, \emph{106}, 153003\relax
\mciteBstWouldAddEndPuncttrue
\mciteSetBstMidEndSepPunct{\mcitedefaultmidpunct}
{\mcitedefaultendpunct}{\mcitedefaultseppunct}\relax
\EndOfBibitem
\bibitem[Paier \latin{et~al.}(2012)Paier, Ren, Rinke, Scuseria, Gr{\"{u}}neis,
  Kresse, and Scheffler]{Paier2012}
Paier,~J.; Ren,~X.; Rinke,~P.; Scuseria,~G.~E.; Gr{\"{u}}neis,~A.; Kresse,~G.;
  Scheffler,~M. {Assessment of correlation energies based on the random-phase
  approximation}. \emph{New J. Phys.} \textbf{2012}, \emph{14}, 043002\relax
\mciteBstWouldAddEndPuncttrue
\mciteSetBstMidEndSepPunct{\mcitedefaultmidpunct}
{\mcitedefaultendpunct}{\mcitedefaultseppunct}\relax
\EndOfBibitem
\bibitem[Ren \latin{et~al.}(2013)Ren, Rinke, Scuseria, and Scheffler]{Ren2013}
Ren,~X.; Rinke,~P.; Scuseria,~G.~E.; Scheffler,~M. {Renormalized second-order
  perturbation theory for the electron correlation energy: Concept,
  implementation, and benchmarks}. \emph{Phys. Rev. B - Condens. Matter Mater.
  Phys.} \textbf{2013}, \emph{88}, 035120\relax
\mciteBstWouldAddEndPuncttrue
\mciteSetBstMidEndSepPunct{\mcitedefaultmidpunct}
{\mcitedefaultendpunct}{\mcitedefaultseppunct}\relax
\EndOfBibitem
\bibitem[Hellgren and {Von Barth}(2007)Hellgren, and {Von Barth}]{Hellgren2007}
Hellgren,~M.; {Von Barth},~U. {Correlation potential in density functional
  theory at the GWA level: Spherical atoms}. \emph{Phys. Rev. B} \textbf{2007},
  \emph{76}, 075107\relax
\mciteBstWouldAddEndPuncttrue
\mciteSetBstMidEndSepPunct{\mcitedefaultmidpunct}
{\mcitedefaultendpunct}{\mcitedefaultseppunct}\relax
\EndOfBibitem
\bibitem[Hellgren and {Von Barth}(2008)Hellgren, and {Von Barth}]{Hellgren2008}
Hellgren,~M.; {Von Barth},~U. {Linear density response function within the
  time-dependent exact-exchange approximation}. \emph{Phys. Rev. B - Condens.
  Matter Mater. Phys.} \textbf{2008}, \emph{78}, 115107\relax
\mciteBstWouldAddEndPuncttrue
\mciteSetBstMidEndSepPunct{\mcitedefaultmidpunct}
{\mcitedefaultendpunct}{\mcitedefaultseppunct}\relax
\EndOfBibitem
\bibitem[He{\ss}elmann and G{\"{o}}rling(2010)He{\ss}elmann, and
  G{\"{o}}rling]{Hesselmann2010}
He{\ss}elmann,~A.; G{\"{o}}rling,~A. {Random phase approximation correlation
  energies with exact Kohn-Sham exchange}. \emph{Mol. Phys.} \textbf{2010},
  \emph{108}, 359--372\relax
\mciteBstWouldAddEndPuncttrue
\mciteSetBstMidEndSepPunct{\mcitedefaultmidpunct}
{\mcitedefaultendpunct}{\mcitedefaultseppunct}\relax
\EndOfBibitem
\bibitem[He{\ss}elmann and G{\"{o}}rling(2011)He{\ss}elmann, and
  G{\"{o}}rling]{Hesselmann2011a}
He{\ss}elmann,~A.; G{\"{o}}rling,~A. {Correct description of the bond
  dissociation limit without breaking spin symmetry by a
  random-phase-approximation correlation functional}. \emph{Phys. Rev. Lett.}
  \textbf{2011}, \emph{106}, 093001\relax
\mciteBstWouldAddEndPuncttrue
\mciteSetBstMidEndSepPunct{\mcitedefaultmidpunct}
{\mcitedefaultendpunct}{\mcitedefaultseppunct}\relax
\EndOfBibitem
\bibitem[Bleiziffer \latin{et~al.}(2015)Bleiziffer, Krug, and
  G{\"{o}}rling]{Bleiziffer2015}
Bleiziffer,~P.; Krug,~M.; G{\"{o}}rling,~A. {Self-consistent Kohn-Sham method
  based on the adiabatic-connection fluctuation-dissipation theorem and the
  exact-exchange kernel}. \emph{J. Chem. Phys.} \textbf{2015}, \emph{142},
  244108\relax
\mciteBstWouldAddEndPuncttrue
\mciteSetBstMidEndSepPunct{\mcitedefaultmidpunct}
{\mcitedefaultendpunct}{\mcitedefaultseppunct}\relax
\EndOfBibitem
\bibitem[Mussard \latin{et~al.}(2016)Mussard, Rocca, Jansen, and
  {\'{A}}ngy{\'{a}}n]{Mussard2016}
Mussard,~B.; Rocca,~D.; Jansen,~G.; {\'{A}}ngy{\'{a}}n,~J.~G. {Dielectric
  Matrix Formulation of Correlation Energies in the Random Phase Approximation:
  Inclusion of Exchange Effects}. \emph{J. Chem. Theory Comput.} \textbf{2016},
  \emph{12}, 2191--2202\relax
\mciteBstWouldAddEndPuncttrue
\mciteSetBstMidEndSepPunct{\mcitedefaultmidpunct}
{\mcitedefaultendpunct}{\mcitedefaultseppunct}\relax
\EndOfBibitem
\bibitem[Bates and Furche(2013)Bates, and Furche]{Bates2013}
Bates,~J.~E.; Furche,~F. {Communication: Random phase approximation
  renormalized many-body perturbation theory}. \emph{J. Chem. Phys.}
  \textbf{2013}, \emph{139}, 171103\relax
\mciteBstWouldAddEndPuncttrue
\mciteSetBstMidEndSepPunct{\mcitedefaultmidpunct}
{\mcitedefaultendpunct}{\mcitedefaultseppunct}\relax
\EndOfBibitem
\bibitem[Erhard \latin{et~al.}(2016)Erhard, Bleiziffer, and
  G{\"{o}}rling]{Erhard2016}
Erhard,~J.; Bleiziffer,~P.; G{\"{o}}rling,~A. {Power Series Approximation for
  the Correlation Kernel Leading to Kohn-Sham Methods Combining Accuracy,
  Computational Efficiency, and General Applicability}. \emph{Phys. Rev. Lett.}
  \textbf{2016}, \emph{117}, 143002\relax
\mciteBstWouldAddEndPuncttrue
\mciteSetBstMidEndSepPunct{\mcitedefaultmidpunct}
{\mcitedefaultendpunct}{\mcitedefaultseppunct}\relax
\EndOfBibitem
\bibitem[Olsen \latin{et~al.}(2019)Olsen, Patrick, Bates, Ruzsinszky, and
  Thygesen]{Olsen2019}
Olsen,~T.; Patrick,~C.~E.; Bates,~J.~E.; Ruzsinszky,~A.; Thygesen,~K.~S.
  {Beyond the RPA and GW methods with adiabatic xc-kernels for accurate ground
  state and quasiparticle energies}. \emph{Nat. Comput. Mater.} \textbf{2019},
  \emph{5}, 106\relax
\mciteBstWouldAddEndPuncttrue
\mciteSetBstMidEndSepPunct{\mcitedefaultmidpunct}
{\mcitedefaultendpunct}{\mcitedefaultseppunct}\relax
\EndOfBibitem
\bibitem[G{\"{o}}rling(2019)]{Gorling2019}
G{\"{o}}rling,~A. {Hierarchies of methods towards the exact Kohn-Sham
  correlation energy based on the adiabatic-connection fluctuation-dissipation
  theorem}. \emph{Phys. Rev. B} \textbf{2019}, \emph{99}, 235120\relax
\mciteBstWouldAddEndPuncttrue
\mciteSetBstMidEndSepPunct{\mcitedefaultmidpunct}
{\mcitedefaultendpunct}{\mcitedefaultseppunct}\relax
\EndOfBibitem
\bibitem[Maggio and Kresse(2016)Maggio, and Kresse]{Maggio2016}
Maggio,~E.; Kresse,~G. {Correlation energy for the homogeneous electron gas:
  Exact Bethe-Salpeter solution and an approximate evaluation}. \emph{Phys.
  Rev. B} \textbf{2016}, \emph{93}, 235113\relax
\mciteBstWouldAddEndPuncttrue
\mciteSetBstMidEndSepPunct{\mcitedefaultmidpunct}
{\mcitedefaultendpunct}{\mcitedefaultseppunct}\relax
\EndOfBibitem
\bibitem[Holzer \latin{et~al.}(2018)Holzer, Gui, Harding, Kresse, Helgaker, and
  Klopper]{Holzer2018a}
Holzer,~C.; Gui,~X.; Harding,~M.~E.; Kresse,~G.; Helgaker,~T.; Klopper,~W.
  {Bethe-Salpeter correlation energies of atoms and molecules}. \emph{J. Chem.
  Phys.} \textbf{2018}, \emph{149}, 144106 (2018);\relax
\mciteBstWouldAddEndPuncttrue
\mciteSetBstMidEndSepPunct{\mcitedefaultmidpunct}
{\mcitedefaultendpunct}{\mcitedefaultseppunct}\relax
\EndOfBibitem
\bibitem[Loos \latin{et~al.}(2020)Loos, Scemama, Duchemin, Jacquemin, and
  Blase]{Loos2020}
Loos,~P.-F.; Scemama,~A.; Duchemin,~I.; Jacquemin,~D.; Blase,~X. {Pros and Cons
  of the Bethe-Salpeter Formalism for Ground-State Energies}. \emph{J. Phys.
  Chem. Lett.} \textbf{2020}, \emph{11}, 3536--3545\relax
\mciteBstWouldAddEndPuncttrue
\mciteSetBstMidEndSepPunct{\mcitedefaultmidpunct}
{\mcitedefaultendpunct}{\mcitedefaultseppunct}\relax
\EndOfBibitem
\bibitem[Trushin \latin{et~al.}(2021)Trushin, Thierbach, and
  G{\"{o}}rling]{Trushin2021}
Trushin,~E.; Thierbach,~A.; G{\"{o}}rling,~A. {Toward chemical accuracy at low
  computational cost: Density-functional theory with $\sigma$-functionals for
  the correlation energy}. \emph{J. Chem. Phys.} \textbf{2021}, \emph{154},
  014104\relax
\mciteBstWouldAddEndPuncttrue
\mciteSetBstMidEndSepPunct{\mcitedefaultmidpunct}
{\mcitedefaultendpunct}{\mcitedefaultseppunct}\relax
\EndOfBibitem
\bibitem[Fauser \latin{et~al.}(2021)Fauser, Trushin, Neiss, and
  G{\"{o}}rling]{Fauser2021}
Fauser,~S.; Trushin,~E.; Neiss,~C.; G{\"{o}}rling,~A. {Chemical accuracy with
  $\sigma$-functionals for the Kohn-Sham correlation energy optimized for
  different input orbitals and eigenvalues}. \emph{J. Chem. Phys.}
  \textbf{2021}, \emph{155}, 134111\relax
\mciteBstWouldAddEndPuncttrue
\mciteSetBstMidEndSepPunct{\mcitedefaultmidpunct}
{\mcitedefaultendpunct}{\mcitedefaultseppunct}\relax
\EndOfBibitem
\bibitem[Colonna \latin{et~al.}(2014)Colonna, Hellgren, and {De
  Gironcoli}]{Colonna2014}
Colonna,~N.; Hellgren,~M.; {De Gironcoli},~S. {Correlation energy within
  exact-exchange adiabatic connection fluctuation-dissipation theory:
  Systematic development and simple approximations}. \emph{Phys. Rev. B -
  Condens. Matter Mater. Phys.} \textbf{2014}, \emph{90}, 1--10\relax
\mciteBstWouldAddEndPuncttrue
\mciteSetBstMidEndSepPunct{\mcitedefaultmidpunct}
{\mcitedefaultendpunct}{\mcitedefaultseppunct}\relax
\EndOfBibitem
\bibitem[Colonna \latin{et~al.}(2016)Colonna, Hellgren, and {De
  Gironcoli}]{Colonna2016}
Colonna,~N.; Hellgren,~M.; {De Gironcoli},~S. {Molecular bonding with the RPAx:
  From weak dispersion forces to strong correlation}. \emph{Phys. Rev. B}
  \textbf{2016}, \emph{93}, 1--11\relax
\mciteBstWouldAddEndPuncttrue
\mciteSetBstMidEndSepPunct{\mcitedefaultmidpunct}
{\mcitedefaultendpunct}{\mcitedefaultseppunct}\relax
\EndOfBibitem
\bibitem[Hellgren \latin{et~al.}(2018)Hellgren, Colonna, and {De
  Gironcoli}]{Hellgren2018}
Hellgren,~M.; Colonna,~N.; {De Gironcoli},~S. {Beyond the random phase
  approximation with a local exchange vertex}. \emph{Phys. Rev. B}
  \textbf{2018}, \emph{98}, 1--12\relax
\mciteBstWouldAddEndPuncttrue
\mciteSetBstMidEndSepPunct{\mcitedefaultmidpunct}
{\mcitedefaultendpunct}{\mcitedefaultseppunct}\relax
\EndOfBibitem
\bibitem[Hellgren and Baguet(2021)Hellgren, and Baguet]{Hellgren2021a}
Hellgren,~M.; Baguet,~L. {Random phase approximation with exchange for an
  accurate description of crystalline polymorphism}. \emph{Phys. Rev. Res.}
  \textbf{2021}, \emph{3}\relax
\mciteBstWouldAddEndPuncttrue
\mciteSetBstMidEndSepPunct{\mcitedefaultmidpunct}
{\mcitedefaultendpunct}{\mcitedefaultseppunct}\relax
\EndOfBibitem
\bibitem[Sharp and Horton(1953)Sharp, and Horton]{Sharp1953}
Sharp,~T.; Horton,~G. {A Variational Approach to the Unipotential Many-Electron
  Problem}. \emph{Phys. Rev.} \textbf{1953}, \emph{90}, 317\relax
\mciteBstWouldAddEndPuncttrue
\mciteSetBstMidEndSepPunct{\mcitedefaultmidpunct}
{\mcitedefaultendpunct}{\mcitedefaultseppunct}\relax
\EndOfBibitem
\bibitem[Talman and Shadwick(1976)Talman, and Shadwick]{Talman1976}
Talman,~J.~D.; Shadwick,~W.~F. {Optimized effective atomic central potential}.
  \emph{Phys. Rev. A} \textbf{1976}, \emph{14}, 36--40\relax
\mciteBstWouldAddEndPuncttrue
\mciteSetBstMidEndSepPunct{\mcitedefaultmidpunct}
{\mcitedefaultendpunct}{\mcitedefaultseppunct}\relax
\EndOfBibitem
\bibitem[{Engel, Eberhard and Dreizler}(2013)]{EngelEberhardandDreizler2013}
{Engel, Eberhard and Dreizler},~R.~M. \emph{{Density functional theory An
  Advanced Course}}; Springer, 2013\relax
\mciteBstWouldAddEndPuncttrue
\mciteSetBstMidEndSepPunct{\mcitedefaultmidpunct}
{\mcitedefaultendpunct}{\mcitedefaultseppunct}\relax
\EndOfBibitem
\bibitem[Freeman(1977)]{DavidL.Freeman1977}
Freeman,~D. {Coupled-cluster expansion applied to the electron gas: Inclusion
  of ring and exchange effects}. \emph{Phys. Rev. B} \textbf{1977}, \emph{15},
  5512--5521\relax
\mciteBstWouldAddEndPuncttrue
\mciteSetBstMidEndSepPunct{\mcitedefaultmidpunct}
{\mcitedefaultendpunct}{\mcitedefaultseppunct}\relax
\EndOfBibitem
\bibitem[Jansen \latin{et~al.}(2010)Jansen, Liu, and
  {\'{A}}ngy{\'{a}}n]{Jansen2010}
Jansen,~G.; Liu,~R.~F.; {\'{A}}ngy{\'{a}}n,~J.~G. {On the equivalence of
  ring-coupled cluster and adiabatic connection fluctuation-dissipation theorem
  random phase approximation correlation energy expressions}. \emph{J. Chem.
  Phys.} \textbf{2010}, \emph{133}, 154106\relax
\mciteBstWouldAddEndPuncttrue
\mciteSetBstMidEndSepPunct{\mcitedefaultmidpunct}
{\mcitedefaultendpunct}{\mcitedefaultseppunct}\relax
\EndOfBibitem
\bibitem[Paier \latin{et~al.}(2010)Paier, Janesko, Henderson, Scuseria,
  Gr{\"{u}}neis, and Kresse]{Paier2010}
Paier,~J.; Janesko,~B.~G.; Henderson,~T.~M.; Scuseria,~G.~E.;
  Gr{\"{u}}neis,~A.; Kresse,~G. {Hybrid functionals including random phase
  approximation correlation and second-order screened exchange}. \emph{J. Chem.
  Phys.} \textbf{2010}, \emph{132}, 094103\relax
\mciteBstWouldAddEndPuncttrue
\mciteSetBstMidEndSepPunct{\mcitedefaultmidpunct}
{\mcitedefaultendpunct}{\mcitedefaultseppunct}\relax
\EndOfBibitem
\bibitem[Gr{\"{u}}neis \latin{et~al.}(2009)Gr{\"{u}}neis, Marsman, Harl,
  Schimka, and Kresse]{Gruneis2009}
Gr{\"{u}}neis,~A.; Marsman,~M.; Harl,~J.; Schimka,~L.; Kresse,~G. {Making the
  random phase approximation to electronic correlation accurate}. \emph{J.
  Chem. Phys.} \textbf{2009}, \emph{131}, 154115\relax
\mciteBstWouldAddEndPuncttrue
\mciteSetBstMidEndSepPunct{\mcitedefaultmidpunct}
{\mcitedefaultendpunct}{\mcitedefaultseppunct}\relax
\EndOfBibitem
\bibitem[F{\"{o}}rster and Visscher(2022)F{\"{o}}rster, and
  Visscher]{Forster2022}
F{\"{o}}rster,~A.; Visscher,~L. {Exploring the statically screened G3W2
  correction to the GW self-energy : Charged excitations and total energies of
  finite systems}. \emph{Phys. Rev. B} \textbf{2022}, \emph{105}, 125121\relax
\mciteBstWouldAddEndPuncttrue
\mciteSetBstMidEndSepPunct{\mcitedefaultmidpunct}
{\mcitedefaultendpunct}{\mcitedefaultseppunct}\relax
\EndOfBibitem
\bibitem[Gr{\"{u}}neis \latin{et~al.}(2014)Gr{\"{u}}neis, Kresse, Hinuma, and
  Oba]{Gruneis2014}
Gr{\"{u}}neis,~A.; Kresse,~G.; Hinuma,~Y.; Oba,~F. {Ionization potentials of
  solids: The importance of vertex corrections}. \emph{Phys. Rev. Lett.}
  \textbf{2014}, \emph{112}, 096401\relax
\mciteBstWouldAddEndPuncttrue
\mciteSetBstMidEndSepPunct{\mcitedefaultmidpunct}
{\mcitedefaultendpunct}{\mcitedefaultseppunct}\relax
\EndOfBibitem
\bibitem[Dyson(1949)]{Dyson1949}
Dyson,~F.~J. {The S matrix in quantum electrodynamics}. \emph{Phys. Rev.}
  \textbf{1949}, \emph{75}, 1736--1755\relax
\mciteBstWouldAddEndPuncttrue
\mciteSetBstMidEndSepPunct{\mcitedefaultmidpunct}
{\mcitedefaultendpunct}{\mcitedefaultseppunct}\relax
\EndOfBibitem
\bibitem[Rieger \latin{et~al.}(1999)Rieger, Steinbeck, White, Rojas, and
  Godby]{Rieger1999}
Rieger,~M.~M.; Steinbeck,~L.; White,~I.~D.; Rojas,~H.~N.; Godby,~R.~W. {GW
  space-time method for the self-energy of large systems}. \emph{Comput. Phys.
  Commun.} \textbf{1999}, \emph{117}, 211--228\relax
\mciteBstWouldAddEndPuncttrue
\mciteSetBstMidEndSepPunct{\mcitedefaultmidpunct}
{\mcitedefaultendpunct}{\mcitedefaultseppunct}\relax
\EndOfBibitem
\bibitem[Maggio and Kresse(2017)Maggio, and Kresse]{Maggio2017}
Maggio,~E.; Kresse,~G. {GW Vertex Corrected Calculations for Molecular
  Systems}. \emph{J. Chem. Theory Comput.} \textbf{2017}, \emph{13},
  4765--4778\relax
\mciteBstWouldAddEndPuncttrue
\mciteSetBstMidEndSepPunct{\mcitedefaultmidpunct}
{\mcitedefaultendpunct}{\mcitedefaultseppunct}\relax
\EndOfBibitem
\bibitem[Baym and Kadanoff(1961)Baym, and Kadanoff]{Baym1961}
Baym,~G.; Kadanoff,~L.~P. {Conservation laws and correlation functions}.
  \emph{Phys. Rev.} \textbf{1961}, \emph{124}, 287--299\relax
\mciteBstWouldAddEndPuncttrue
\mciteSetBstMidEndSepPunct{\mcitedefaultmidpunct}
{\mcitedefaultendpunct}{\mcitedefaultseppunct}\relax
\EndOfBibitem
\bibitem[Salpeter and Bethe(1951)Salpeter, and Bethe]{Salpeter1951}
Salpeter,~E.~E.; Bethe,~H.~A. {A relativistic equation for bound-state
  problems}. \emph{Phys. Rev.} \textbf{1951}, \emph{84}, 1232--1242\relax
\mciteBstWouldAddEndPuncttrue
\mciteSetBstMidEndSepPunct{\mcitedefaultmidpunct}
{\mcitedefaultendpunct}{\mcitedefaultseppunct}\relax
\EndOfBibitem
\bibitem[Starke and Kresse(2012)Starke, and Kresse]{Starke2012}
Starke,~R.; Kresse,~G. {Self-consistent Green function equations and the
  hierarchy of approximations for the four-point propagator}. \emph{Phys. Rev.
  B} \textbf{2012}, \emph{85}, 075119\relax
\mciteBstWouldAddEndPuncttrue
\mciteSetBstMidEndSepPunct{\mcitedefaultmidpunct}
{\mcitedefaultendpunct}{\mcitedefaultseppunct}\relax
\EndOfBibitem
\bibitem[Held \latin{et~al.}(2011)Held, Taranto, Rohringer, and
  Toschi]{Held2011}
Held,~K.; Taranto,~C.; Rohringer,~G.; Toschi,~A. In \emph{LDA+DMFT approach to
  strongly Correl. Mater.}; Pavarini,~E., Koch,~E., Vollhardt,~D.,
  Lichtenstein,~A., Eds.; 2011; Chapter 13, pp 13.1--13.22\relax
\mciteBstWouldAddEndPuncttrue
\mciteSetBstMidEndSepPunct{\mcitedefaultmidpunct}
{\mcitedefaultendpunct}{\mcitedefaultseppunct}\relax
\EndOfBibitem
\bibitem[Caruso \latin{et~al.}(2013)Caruso, Rohr, Hellgren, Ren, Rinke, Rubio,
  and Scheffler]{Caruso2013b}
Caruso,~F.; Rohr,~D.~R.; Hellgren,~M.; Ren,~X.; Rinke,~P.; Rubio,~A.;
  Scheffler,~M. {Bond breaking and bond formation: How electron correlation is
  captured in many-body perturbation theory and density-functional theory}.
  \emph{Phys. Rev. Lett.} \textbf{2013}, \emph{110}, 146403\relax
\mciteBstWouldAddEndPuncttrue
\mciteSetBstMidEndSepPunct{\mcitedefaultmidpunct}
{\mcitedefaultendpunct}{\mcitedefaultseppunct}\relax
\EndOfBibitem
\bibitem[Rohringer \latin{et~al.}(2012)Rohringer, Valli, and
  Toschi]{Rohringer2012}
Rohringer,~G.; Valli,~A.; Toschi,~A. {Local electronic correlation at the
  two-particle level}. \emph{Phys. Rev. B - Condens. Matter Mater. Phys.}
  \textbf{2012}, \emph{86}, 125114\relax
\mciteBstWouldAddEndPuncttrue
\mciteSetBstMidEndSepPunct{\mcitedefaultmidpunct}
{\mcitedefaultendpunct}{\mcitedefaultseppunct}\relax
\EndOfBibitem
\bibitem[Rohringer \latin{et~al.}(2018)Rohringer, Hafermann, Toschi, Katanin,
  Antipov, Katsnelson, Lichtenstein, Rubtsov, and Held]{Rohringer2018}
Rohringer,~G.; Hafermann,~H.; Toschi,~A.; Katanin,~A.~A.; Antipov,~A.~E.;
  Katsnelson,~M.~I.; Lichtenstein,~A.~I.; Rubtsov,~A.~N.; Held,~K.
  {Diagrammatic routes to nonlocal correlations beyond dynamical mean field
  theory}. \emph{Rev. Mod. Phys.} \textbf{2018}, \emph{90}, 25003\relax
\mciteBstWouldAddEndPuncttrue
\mciteSetBstMidEndSepPunct{\mcitedefaultmidpunct}
{\mcitedefaultendpunct}{\mcitedefaultseppunct}\relax
\EndOfBibitem
\bibitem[Krien \latin{et~al.}(2021)Krien, Kauch, and Held]{Krien2021}
Krien,~F.; Kauch,~A.; Held,~K. {Tiling with triangles: Parquet and GW$\gamma$
  methods unified}. \emph{Phys. Rev. Res.} \textbf{2021}, \emph{3}, 13149\relax
\mciteBstWouldAddEndPuncttrue
\mciteSetBstMidEndSepPunct{\mcitedefaultmidpunct}
{\mcitedefaultendpunct}{\mcitedefaultseppunct}\relax
\EndOfBibitem
\bibitem[Ren \latin{et~al.}(2015)Ren, Marom, Caruso, Scheffler, and
  Rinke]{Ren2015}
Ren,~X.; Marom,~N.; Caruso,~F.; Scheffler,~M.; Rinke,~P. {Beyond the GW
  approximation: A second-order screened exchange correction}. \emph{Phys. Rev.
  B - Condens. Matter Mater. Phys.} \textbf{2015}, \emph{92}, 081104(R)\relax
\mciteBstWouldAddEndPuncttrue
\mciteSetBstMidEndSepPunct{\mcitedefaultmidpunct}
{\mcitedefaultendpunct}{\mcitedefaultseppunct}\relax
\EndOfBibitem
\bibitem[van Leeuwen \latin{et~al.}(2015)van Leeuwen, Dahlen, Stefanucci,
  Almbladh, and {Von Barth}]{VanLeeuwen2015}
van Leeuwen,~R.; Dahlen,~N.~E.; Stefanucci,~G.; Almbladh,~C.~O.; {Von
  Barth},~U. In \emph{Time-Dependent Density Funct. Theory}; Marques,~M.~A.,
  Ullrich,~C.~A., Nogueira,~F., Rubio,~A., Burke,~K., Gross,~E.~K., Eds.;
  Springer Heidelberg, 2015; pp 185--217\relax
\mciteBstWouldAddEndPuncttrue
\mciteSetBstMidEndSepPunct{\mcitedefaultmidpunct}
{\mcitedefaultendpunct}{\mcitedefaultseppunct}\relax
\EndOfBibitem
\bibitem[Stefanucci \latin{et~al.}(2014)Stefanucci, Pavlyukh, Uimonen, and van
  Leeuwen]{Stefanucci2014}
Stefanucci,~G.; Pavlyukh,~Y.; Uimonen,~A.~M.; van Leeuwen,~R. {Diagrammatic
  expansion for positive spectral functions beyond GW: Application to vertex
  corrections in the electron gas}. \emph{Phys. Rev. B} \textbf{2014},
  \emph{90}, 115134\relax
\mciteBstWouldAddEndPuncttrue
\mciteSetBstMidEndSepPunct{\mcitedefaultmidpunct}
{\mcitedefaultendpunct}{\mcitedefaultseppunct}\relax
\EndOfBibitem
\bibitem[Rodr{\'{i}}guez-Mayorga \latin{et~al.}(2021)Rodr{\'{i}}guez-Mayorga,
  Mitxelena, Bruneval, and Piris]{Rodriguez-Mayorga2021}
Rodr{\'{i}}guez-Mayorga,~M.; Mitxelena,~I.; Bruneval,~F.; Piris,~M. {Coupling
  Natural Orbital Functional Theory and Many-Body Perturbation Theory by Using
  Nondynamically Correlated Canonical Orbitals}. \emph{J. Chem. Theory Comput.}
  \textbf{2021}, \emph{17}, 7562--7574\relax
\mciteBstWouldAddEndPuncttrue
\mciteSetBstMidEndSepPunct{\mcitedefaultmidpunct}
{\mcitedefaultendpunct}{\mcitedefaultseppunct}\relax
\EndOfBibitem
\bibitem[Becke(2014)]{Becke2014}
Becke,~A.~D. {Perspective: Fifty years of density-functional theory in chemical
  physics}. \emph{J. Chem. Phys.} \textbf{2014}, \emph{140}, 18A301\relax
\mciteBstWouldAddEndPuncttrue
\mciteSetBstMidEndSepPunct{\mcitedefaultmidpunct}
{\mcitedefaultendpunct}{\mcitedefaultseppunct}\relax
\EndOfBibitem
\bibitem[Vuckovic \latin{et~al.}(2020)Vuckovic, Fabiano, Gori-Giorgi, and
  Burke]{Vuckovic2020}
Vuckovic,~S.; Fabiano,~E.; Gori-Giorgi,~P.; Burke,~K. {MAP: An MP2 Accuracy
  Predictor for Weak Interactions from Adiabatic Connection Theory}. \emph{J.
  Chem. Theory Comput.} \textbf{2020}, \emph{16}, 4141--4149\relax
\mciteBstWouldAddEndPuncttrue
\mciteSetBstMidEndSepPunct{\mcitedefaultmidpunct}
{\mcitedefaultendpunct}{\mcitedefaultseppunct}\relax
\EndOfBibitem
\bibitem[Baerends \latin{et~al.}(2022)Baerends, Ziegler, Atkins, Autschbach,
  Baseggio, Bashford, B{\'{e}}rces, Bickelhaupt, Bo, Boerrigter, Cavallo, Daul,
  Chong, Chulhai, Deng, Dickson, Dieterich, Ellis, van Faassen, Fan, Fischer,
  F{\"{o}}rster, Guerra, Franchini, Ghysels, Giammona, van Gisbergen, Goez,
  G{\"{o}}tz, Groeneveld, Gritsenko, Gr{\"{u}}ning, Gusarov, Harris, van~den
  Hoek, Hu, Jacob, Jacobsen, Jensen, Joubert, Kaminski, van Kessel,
  K{\"{o}}nig, Kootstra, Kovalenko, Krykunov, van Lenthe, McCormack, Michalak,
  Mitoraj, Morton, Neugebauer, Nicu, Noodleman, Osinga, Patchkovskii,
  Pavanello, Peeples, Philipsen, Post, Pye, Ramanantoanina, Ramos, Ravenek,
  Reimann, Rodr{\'{i}}guez, Ros, R{\"{u}}ger, Schipper, Schl{\"{u}}ns, van
  Schoot, Schreckenbach, Seldenthuis, Seth, Snijders, Sol{\`{a}}, Stener,
  Swart, Swerhone, Tognetti, te~Velde, Vernooijs, Versluis, Visscher, Visser,
  Wang, Wesolowski, van Wezenbeek, Wiesenekker, Wolff, Woo, and
  Yakovlev]{adf2022}
Baerends,~E.; Ziegler,~T.; Atkins,~A.; Autschbach,~J.; Baseggio,~O.;
  Bashford,~D.; B{\'{e}}rces,~A.; Bickelhaupt,~F.; Bo,~C.; Boerrigter,~P.;
  Cavallo,~L.; Daul,~C.; Chong,~D.; Chulhai,~D.; Deng,~L.; Dickson,~R.;
  Dieterich,~J.; Ellis,~D.; van Faassen,~M.; Fan,~L.; Fischer,~T.;
  F{\"{o}}rster,~A.; Guerra,~C.~F.; Franchini,~M.; Ghysels,~A.; Giammona,~A.;
  van Gisbergen,~S.; Goez,~A.; G{\"{o}}tz,~A.; Groeneveld,~J.; Gritsenko,~O.;
  Gr{\"{u}}ning,~M.; Gusarov,~S.; Harris,~F.; van~den Hoek,~P.; Hu,~Z.;
  Jacob,~C.; Jacobsen,~H.; Jensen,~L.; Joubert,~L.; Kaminski,~J.; van
  Kessel,~G.; K{\"{o}}nig,~C.; Kootstra,~F.; Kovalenko,~A.; Krykunov,~M.; van
  Lenthe,~E.; McCormack,~D.; Michalak,~A.; Mitoraj,~M.; Morton,~S.;
  Neugebauer,~J.; Nicu,~V.; Noodleman,~L.; Osinga,~V.; Patchkovskii,~S.;
  Pavanello,~M.; Peeples,~C.; Philipsen,~P.; Post,~D.; Pye,~C.;
  Ramanantoanina,~H.; Ramos,~P.; Ravenek,~W.; Reimann,~M.; Rodr{\'{i}}guez,~J.;
  Ros,~P.; R{\"{u}}ger,~R.; Schipper,~P.; Schl{\"{u}}ns,~D.; van Schoot,~H.;
  Schreckenbach,~G.; Seldenthuis,~J.; Seth,~M.; Snijders,~J.; Sol{\`{a}},~M.;
  Stener,~M.; Swart,~M.; Swerhone,~D.; Tognetti,~V.; te~Velde,~G.;
  Vernooijs,~P.; Versluis,~L.; Visscher,~L.; Visser,~O.; Wang,~F.;
  Wesolowski,~T.; van Wezenbeek,~E.; Wiesenekker,~G.; Wolff,~S.; Woo,~T.;
  Yakovlev,~A. {ADF2022.1, locally modified development version}. 2022\relax
\mciteBstWouldAddEndPuncttrue
\mciteSetBstMidEndSepPunct{\mcitedefaultmidpunct}
{\mcitedefaultendpunct}{\mcitedefaultseppunct}\relax
\EndOfBibitem
\bibitem[F{\"{o}}rster and Visscher(2021)F{\"{o}}rster, and
  Visscher]{Forster2021}
F{\"{o}}rster,~A.; Visscher,~L. {GW100: A Slater-Type Orbital Perspective}.
  \emph{J. Chem. Theory Comput.} \textbf{2021}, \emph{17}, 5080--5097\relax
\mciteBstWouldAddEndPuncttrue
\mciteSetBstMidEndSepPunct{\mcitedefaultmidpunct}
{\mcitedefaultendpunct}{\mcitedefaultseppunct}\relax
\EndOfBibitem
\bibitem[Krykunov \latin{et~al.}(2009)Krykunov, Ziegler, and {Van
  Lenthe}]{Krykunov2009}
Krykunov,~M.; Ziegler,~T.; {Van Lenthe},~E. {Hybrid density functional
  calculations of nuclear magnetic shieldings using slater-type orbitals and
  the zeroth- order regular approximation}. \emph{Int. J. Quantum Chem.}
  \textbf{2009}, \emph{109}, 1676--1683\relax
\mciteBstWouldAddEndPuncttrue
\mciteSetBstMidEndSepPunct{\mcitedefaultmidpunct}
{\mcitedefaultendpunct}{\mcitedefaultseppunct}\relax
\EndOfBibitem
\bibitem[Wirz \latin{et~al.}(2017)Wirz, Reine, and Pedersen]{Wirz2017}
Wirz,~L.~N.; Reine,~S.~S.; Pedersen,~T.~B. {On Resolution-of-the-Identity
  Electron Repulsion Integral Approximations and Variational Stability}.
  \emph{J. Chem. Theory Comput.} \textbf{2017}, \emph{13}, 4897--4906\relax
\mciteBstWouldAddEndPuncttrue
\mciteSetBstMidEndSepPunct{\mcitedefaultmidpunct}
{\mcitedefaultendpunct}{\mcitedefaultseppunct}\relax
\EndOfBibitem
\bibitem[F{\"{o}}rster \latin{et~al.}(2020)F{\"{o}}rster, Franchini, van
  Lenthe, and Visscher]{Forster2020}
F{\"{o}}rster,~A.; Franchini,~M.; van Lenthe,~E.; Visscher,~L. {A Quadratic
  Pair Atomic Resolution of the Identity Based SOS-AO-MP2 Algorithm Using
  Slater Type Orbitals}. \emph{J. Chem. Theory Comput.} \textbf{2020},
  \emph{16}, 875 -- 891\relax
\mciteBstWouldAddEndPuncttrue
\mciteSetBstMidEndSepPunct{\mcitedefaultmidpunct}
{\mcitedefaultendpunct}{\mcitedefaultseppunct}\relax
\EndOfBibitem
\bibitem[Kaltak \latin{et~al.}(2014)Kaltak, Klime{\v{s}}, and
  Kresse]{Kaltak2014}
Kaltak,~M.; Klime{\v{s}},~J.; Kresse,~G. {Low scaling algorithms for the random
  phase approximation: Imaginary time and laplace transformations}. \emph{J.
  Chem. Theory Comput.} \textbf{2014}, \emph{10}, 2498--2507\relax
\mciteBstWouldAddEndPuncttrue
\mciteSetBstMidEndSepPunct{\mcitedefaultmidpunct}
{\mcitedefaultendpunct}{\mcitedefaultseppunct}\relax
\EndOfBibitem
\bibitem[Kaltak \latin{et~al.}(2014)Kaltak, Klime{\v{s}}, and
  Kresse]{Kaltak2014a}
Kaltak,~M.; Klime{\v{s}},~J.; Kresse,~G. {Cubic scaling algorithm for the
  random phase approximation: Self-interstitials and vacancies in Si}.
  \emph{Phys. Rev. B} \textbf{2014}, \emph{90}, 054115\relax
\mciteBstWouldAddEndPuncttrue
\mciteSetBstMidEndSepPunct{\mcitedefaultmidpunct}
{\mcitedefaultendpunct}{\mcitedefaultseppunct}\relax
\EndOfBibitem
\bibitem[Liu \latin{et~al.}(2016)Liu, Kaltak, Klime{\v{s}}, and
  Kresse]{Liu2016}
Liu,~P.; Kaltak,~M.; Klime{\v{s}},~J.; Kresse,~G. {Cubic scaling GW: Towards
  fast quasiparticle calculations}. \emph{Phys. Rev. B} \textbf{2016},
  \emph{94}, 165109\relax
\mciteBstWouldAddEndPuncttrue
\mciteSetBstMidEndSepPunct{\mcitedefaultmidpunct}
{\mcitedefaultendpunct}{\mcitedefaultseppunct}\relax
\EndOfBibitem
\bibitem[Helgaker \latin{et~al.}(1997)Helgaker, Klopper, Koch, and
  Noga]{Helgaker1997}
Helgaker,~T.; Klopper,~W.; Koch,~H.; Noga,~J. {Basis-set convergence of
  correlated calculations on water}. \emph{J. Chem. Phys.} \textbf{1997},
  \emph{106}, 9639--9646\relax
\mciteBstWouldAddEndPuncttrue
\mciteSetBstMidEndSepPunct{\mcitedefaultmidpunct}
{\mcitedefaultendpunct}{\mcitedefaultseppunct}\relax
\EndOfBibitem
\bibitem[Jensen(2013)]{Jensen2013}
Jensen,~F. {Atomic orbital basis sets}. \emph{Wiley Interdiscip. Rev. Comput.
  Mol. Sci.} \textbf{2013}, \emph{3}, 273--295\relax
\mciteBstWouldAddEndPuncttrue
\mciteSetBstMidEndSepPunct{\mcitedefaultmidpunct}
{\mcitedefaultendpunct}{\mcitedefaultseppunct}\relax
\EndOfBibitem
\bibitem[Knowles and Handy(1984)Knowles, and Handy]{Knowles1984}
Knowles,~P.~J.; Handy,~N.~C. {A new determinant-based full configuration
  interaction method}. \emph{Chem. Phys. Lett.} \textbf{1984}, \emph{111},
  315--321\relax
\mciteBstWouldAddEndPuncttrue
\mciteSetBstMidEndSepPunct{\mcitedefaultmidpunct}
{\mcitedefaultendpunct}{\mcitedefaultseppunct}\relax
\EndOfBibitem
\bibitem[Knowles and Handy(1989)Knowles, and Handy]{Knowles1989}
Knowles,~P.~J.; Handy,~N.~C. {A determinant based full configuration
  interaction program}. \emph{Comput. Phys. Commun.} \textbf{1989}, \emph{54},
  75--83\relax
\mciteBstWouldAddEndPuncttrue
\mciteSetBstMidEndSepPunct{\mcitedefaultmidpunct}
{\mcitedefaultendpunct}{\mcitedefaultseppunct}\relax
\EndOfBibitem
\bibitem[Henderson and Scuseria(2010)Henderson, and Scuseria]{Henderson2010}
Henderson,~T.~M.; Scuseria,~G.~E. {The connection between self-interaction and
  static correlation: A random phase approximation perspective}. \emph{Mol.
  Phys.} \textbf{2010}, \emph{108}, 2511--2517\relax
\mciteBstWouldAddEndPuncttrue
\mciteSetBstMidEndSepPunct{\mcitedefaultmidpunct}
{\mcitedefaultendpunct}{\mcitedefaultseppunct}\relax
\EndOfBibitem
\bibitem[Peterson \latin{et~al.}(1993)Peterson, Kendall, and
  Dunning]{Peterson1993}
Peterson,~K.~A.; Kendall,~R.~A.; Dunning,~T.~H. {Benchmark calculations with
  correlated molecular wave functions. III. Configuration interaction
  calculations on first row homonuclear diatomics}. \emph{J. Chem. Phys.}
  \textbf{1993}, \emph{99}, 9790--9805\relax
\mciteBstWouldAddEndPuncttrue
\mciteSetBstMidEndSepPunct{\mcitedefaultmidpunct}
{\mcitedefaultendpunct}{\mcitedefaultseppunct}\relax
\EndOfBibitem
\bibitem[H{\"{o}}nisch \latin{et~al.}(2009)H{\"{o}}nisch, Hemming, Archer,
  Sidall, and McManus]{Honisch2009}
H{\"{o}}nisch,~B.; Hemming,~G.~N.; Archer,~D.; Sidall,~M.; McManus,~J.~F.
  {Beryllium Dimer — Caught in the Act of Bonding}. \emph{Science (80-. ).}
  \textbf{2009}, \emph{324}, 1548--1552\relax
\mciteBstWouldAddEndPuncttrue
\mciteSetBstMidEndSepPunct{\mcitedefaultmidpunct}
{\mcitedefaultendpunct}{\mcitedefaultseppunct}\relax
\EndOfBibitem
\bibitem[R{\o}eggen and Veseth(2005)R{\o}eggen, and Veseth]{Roeggen2005}
R{\o}eggen,~I.; Veseth,~L. {Interatomic potential for the X1g+ state of Be2,
  revisited}. 2005\relax
\mciteBstWouldAddEndPuncttrue
\mciteSetBstMidEndSepPunct{\mcitedefaultmidpunct}
{\mcitedefaultendpunct}{\mcitedefaultseppunct}\relax
\EndOfBibitem
\bibitem[Goerigk \latin{et~al.}(2017)Goerigk, Hansen, Bauer, Ehrlich, Najibi,
  and Grimme]{Goerigk2017}
Goerigk,~L.; Hansen,~A.; Bauer,~C.; Ehrlich,~S.; Najibi,~A.; Grimme,~S. {A look
  at the density functional theory zoo with the advanced GMTKN55 database for
  general main group thermochemistry, kinetics and noncovalent interactions}.
  \emph{Phys. Chem. Chem. Phys.} \textbf{2017}, \emph{19}, 32184--32215\relax
\mciteBstWouldAddEndPuncttrue
\mciteSetBstMidEndSepPunct{\mcitedefaultmidpunct}
{\mcitedefaultendpunct}{\mcitedefaultseppunct}\relax
\EndOfBibitem
\bibitem[Chen \latin{et~al.}(2018)Chen, Agee, and Furche]{Chen2018}
Chen,~G.~P.; Agee,~M.~M.; Furche,~F. {Performance and Scope of Perturbative
  Corrections to Random-Phase Approximation Energies}. \emph{J. Chem. Theory
  Comput.} \textbf{2018}, \emph{14}, 5701--5714\relax
\mciteBstWouldAddEndPuncttrue
\mciteSetBstMidEndSepPunct{\mcitedefaultmidpunct}
{\mcitedefaultendpunct}{\mcitedefaultseppunct}\relax
\EndOfBibitem
\bibitem[Karton \latin{et~al.}(2011)Karton, Daon, and Martin]{Karton2011}
Karton,~A.; Daon,~S.; Martin,~J.~M. {W4-11: A high-confidence benchmark dataset
  for computational thermochemistry derived from first-principles W4 data}.
  \emph{Chem. Phys. Lett.} \textbf{2011}, \emph{510}, 165--178\relax
\mciteBstWouldAddEndPuncttrue
\mciteSetBstMidEndSepPunct{\mcitedefaultmidpunct}
{\mcitedefaultendpunct}{\mcitedefaultseppunct}\relax
\EndOfBibitem
\bibitem[Karton \latin{et~al.}(2006)Karton, Rabinovich, and Martin]{Karton2006}
Karton,~A.; Rabinovich,~E.; Martin,~J.~M. {W4 theory for computational
  thermochemistry: In pursuit of confident sub-kJ/mol predictions}. \emph{J.
  Chem. Phys.} \textbf{2006}, \emph{125}, 144108\relax
\mciteBstWouldAddEndPuncttrue
\mciteSetBstMidEndSepPunct{\mcitedefaultmidpunct}
{\mcitedefaultendpunct}{\mcitedefaultseppunct}\relax
\EndOfBibitem
\bibitem[Zhao \latin{et~al.}(2005)Zhao, Lynch, and Truhlar]{Zhao2005a}
Zhao,~Y.; Lynch,~B.~J.; Truhlar,~D.~G. {Multi-coefficient extrapolated density
  functional theory for thermochemistry and thermochemical kinetics}.
  \emph{Phys. Chem. Chem. Phys.} \textbf{2005}, \emph{7}, 43--52\relax
\mciteBstWouldAddEndPuncttrue
\mciteSetBstMidEndSepPunct{\mcitedefaultmidpunct}
{\mcitedefaultendpunct}{\mcitedefaultseppunct}\relax
\EndOfBibitem
\bibitem[Zhao \latin{et~al.}(2005)Zhao, Gonz{\'{a}}lez-Garda, and
  Truhlar]{Zhao2005b}
Zhao,~Y.; Gonz{\'{a}}lez-Garda,~N.; Truhlar,~D.~G. {Benchmark database of
  barrier heights for heavy atom transfer, nucleophilic substitution,
  association, and unimolecular reactions and its use to test theoretical
  methods}. \emph{J. Phys. Chem. A} \textbf{2005}, \emph{109}, 2012--2018\relax
\mciteBstWouldAddEndPuncttrue
\mciteSetBstMidEndSepPunct{\mcitedefaultmidpunct}
{\mcitedefaultendpunct}{\mcitedefaultseppunct}\relax
\EndOfBibitem
\bibitem[Karton and Martin(2012)Karton, and Martin]{Karton2012b}
Karton,~A.; Martin,~J.~M. {Explicitly correlated Wn theory: W1-F12 and W2-F12}.
  \emph{J. Chem. Phys.} \textbf{2012}, \emph{136}, 124114\relax
\mciteBstWouldAddEndPuncttrue
\mciteSetBstMidEndSepPunct{\mcitedefaultmidpunct}
{\mcitedefaultendpunct}{\mcitedefaultseppunct}\relax
\EndOfBibitem
\bibitem[Curtiss \latin{et~al.}(1991)Curtiss, Raghavachari, Trucks, and
  Pople]{Curtiss1991}
Curtiss,~L.~A.; Raghavachari,~K.; Trucks,~G.~W.; Pople,~J.~A. {Gaussian-2
  theory for molecular energies of first- and second-row compounds}. \emph{J.
  Chem. Phys.} \textbf{1991}, \emph{94}, 7221--7230\relax
\mciteBstWouldAddEndPuncttrue
\mciteSetBstMidEndSepPunct{\mcitedefaultmidpunct}
{\mcitedefaultendpunct}{\mcitedefaultseppunct}\relax
\EndOfBibitem
\bibitem[Hellgren \latin{et~al.}(2012)Hellgren, Rohr, and Gross]{Hellgren2012}
Hellgren,~M.; Rohr,~D.~R.; Gross,~E.~K. {Correlation potentials for molecular
  bond dissociation within the self-consistent random phase approximation}.
  \emph{J. Chem. Phys.} \textbf{2012}, \emph{136}, 034106\relax
\mciteBstWouldAddEndPuncttrue
\mciteSetBstMidEndSepPunct{\mcitedefaultmidpunct}
{\mcitedefaultendpunct}{\mcitedefaultseppunct}\relax
\EndOfBibitem
\bibitem[Verma and Bartlett(2012)Verma, and Bartlett]{Verma2012}
Verma,~P.; Bartlett,~R.~J. {Increasing the applicability of density functional
  theory. II. Correlation potentials from the random phase approximation and
  beyond}. \emph{J. Chem. Phys.} \textbf{2012}, \emph{136}, 044105\relax
\mciteBstWouldAddEndPuncttrue
\mciteSetBstMidEndSepPunct{\mcitedefaultmidpunct}
{\mcitedefaultendpunct}{\mcitedefaultseppunct}\relax
\EndOfBibitem
\bibitem[Bleiziffer \latin{et~al.}(2013)Bleiziffer, He{\ss}elmann, and
  G{\"{o}}rling]{Bleiziffer2013}
Bleiziffer,~P.; He{\ss}elmann,~A.; G{\"{o}}rling,~A. {Efficient self-consistent
  treatment of electron correlation within the random phase approximation}.
  \emph{J. Chem. Phys.} \textbf{2013}, \emph{139}, 084113\relax
\mciteBstWouldAddEndPuncttrue
\mciteSetBstMidEndSepPunct{\mcitedefaultmidpunct}
{\mcitedefaultendpunct}{\mcitedefaultseppunct}\relax
\EndOfBibitem
\bibitem[Klime{\v{s}} and Kresse(2014)Klime{\v{s}}, and Kresse]{Klimes2014a}
Klime{\v{s}},~J.; Kresse,~G. {Kohn-Sham band gaps and potentials of solids from
  the optimised effective potential method within the random phase
  approximation}. \emph{J. Chem. Phys.} \textbf{2014}, \emph{140}, 054516\relax
\mciteBstWouldAddEndPuncttrue
\mciteSetBstMidEndSepPunct{\mcitedefaultmidpunct}
{\mcitedefaultendpunct}{\mcitedefaultseppunct}\relax
\EndOfBibitem
\bibitem[Hellgren \latin{et~al.}(2015)Hellgren, Caruso, Rohr, Ren, Rubio,
  Scheffler, and Rinke]{Hellgren2015}
Hellgren,~M.; Caruso,~F.; Rohr,~D.~R.; Ren,~X.; Rubio,~A.; Scheffler,~M.;
  Rinke,~P. {Static correlation and electron localization in molecular dimers
  from the self-consistent RPA and GW approximation}. \emph{Phys. Rev. B -
  Condens. Matter Mater. Phys.} \textbf{2015}, \emph{91}, 165110\relax
\mciteBstWouldAddEndPuncttrue
\mciteSetBstMidEndSepPunct{\mcitedefaultmidpunct}
{\mcitedefaultendpunct}{\mcitedefaultseppunct}\relax
\EndOfBibitem
\bibitem[Voora \latin{et~al.}(2019)Voora, Balasubramani, and Furche]{Voora2019}
Voora,~V.~K.; Balasubramani,~S.~G.; Furche,~F. {Variational generalized
  Kohn-Sham approach combining the random-phase-approximation and
  Green's-function methods}. \emph{Phys. Rev. A} \textbf{2019}, \emph{99},
  012518\relax
\mciteBstWouldAddEndPuncttrue
\mciteSetBstMidEndSepPunct{\mcitedefaultmidpunct}
{\mcitedefaultendpunct}{\mcitedefaultseppunct}\relax
\EndOfBibitem
\bibitem[Graf and Ochsenfeld(2020)Graf, and Ochsenfeld]{Graf2020}
Graf,~D.; Ochsenfeld,~C. {A range-separated generalized Kohn-Sham method
  including a long-range nonlocal random phase approximation correlation
  potential}. \emph{J. Chem. Phys.} \textbf{2020}, \emph{153}, 244118\relax
\mciteBstWouldAddEndPuncttrue
\mciteSetBstMidEndSepPunct{\mcitedefaultmidpunct}
{\mcitedefaultendpunct}{\mcitedefaultseppunct}\relax
\EndOfBibitem
\bibitem[Riemelmoser \latin{et~al.}(2021)Riemelmoser, Kaltak, and
  Kresse]{Riemelmoser2021}
Riemelmoser,~S.; Kaltak,~M.; Kresse,~G. {Optimized effective potentials from
  the random-phase approximation: Accuracy of the quasiparticle approximation}.
  \emph{J. Chem. Phys.} \textbf{2021}, \emph{154}, 154103\relax
\mciteBstWouldAddEndPuncttrue
\mciteSetBstMidEndSepPunct{\mcitedefaultmidpunct}
{\mcitedefaultendpunct}{\mcitedefaultseppunct}\relax
\EndOfBibitem
\bibitem[Yu \latin{et~al.}(2021)Yu, Tsai, Hernandez, Furche, Tsai, and
  Hernandez]{Yu2021}
Yu,~J.~M.; Tsai,~J.; Hernandez,~D.~J.; Furche,~F.; Tsai,~J.; Hernandez,~D.~J.
  {Selfconsistent random phase approximation methods}. \emph{J. Chem. Phys.}
  \textbf{2021}, \emph{155}, 040902\relax
\mciteBstWouldAddEndPuncttrue
\mciteSetBstMidEndSepPunct{\mcitedefaultmidpunct}
{\mcitedefaultendpunct}{\mcitedefaultseppunct}\relax
\EndOfBibitem
\bibitem[Caruso \latin{et~al.}(2016)Caruso, Dauth, {Van Setten}, and
  Rinke]{Caruso2016}
Caruso,~F.; Dauth,~M.; {Van Setten},~M.~J.; Rinke,~P. {Benchmark of GW
  Approaches for the GW100 Test Set}. \emph{J. Chem. Theory Comput.}
  \textbf{2016}, \emph{12}, 5076--5087\relax
\mciteBstWouldAddEndPuncttrue
\mciteSetBstMidEndSepPunct{\mcitedefaultmidpunct}
{\mcitedefaultendpunct}{\mcitedefaultseppunct}\relax
\EndOfBibitem
\bibitem[Zhang \latin{et~al.}(2022)Zhang, Shu, Xing, Chen, Sun, Huang, and
  Truhlar]{Zhang2022}
Zhang,~L.; Shu,~Y.; Xing,~C.; Chen,~X.; Sun,~S.; Huang,~Y.; Truhlar,~D.~G.
  {Recommendation of Orbitals for G 0 W 0 Calculations on Molecules and
  Crystals}. \emph{J. Chem. Theory Comput.} \textbf{2022}, \emph{18},
  3523--3537\relax
\mciteBstWouldAddEndPuncttrue
\mciteSetBstMidEndSepPunct{\mcitedefaultmidpunct}
{\mcitedefaultendpunct}{\mcitedefaultseppunct}\relax
\EndOfBibitem
\bibitem[Wang \latin{et~al.}(2021)Wang, Rinke, and Ren]{Wang2021}
Wang,~Y.; Rinke,~P.; Ren,~X. {Assessing the G 0 W 0 $\Gamma$ 0 (1) Approach:
  Beyond G 0 W 0 with Hedin's Full Second-Order Self-Energy Contribution}.
  \emph{J. Chem. Theory Comput.} \textbf{2021}, \emph{17}, 5140--5154\relax
\mciteBstWouldAddEndPuncttrue
\mciteSetBstMidEndSepPunct{\mcitedefaultmidpunct}
{\mcitedefaultendpunct}{\mcitedefaultseppunct}\relax
\EndOfBibitem
\bibitem[Hait and Head-Gordon(2018)Hait, and Head-Gordon]{Hait2018a}
Hait,~D.; Head-Gordon,~M. {Delocalization Errors in Density Functional Theory
  Are Essentially Quadratic in Fractional Occupation Number}. \emph{J. Phys.
  Chem. Lett.} \textbf{2018}, \emph{9}, 6280--6288\relax
\mciteBstWouldAddEndPuncttrue
\mciteSetBstMidEndSepPunct{\mcitedefaultmidpunct}
{\mcitedefaultendpunct}{\mcitedefaultseppunct}\relax
\EndOfBibitem
\bibitem[Furche(2001)]{Furche2001}
Furche,~F. {Molecular tests of the random phase approximation to the
  exchange-correlation energy functional}. \emph{Phys. Rev. B} \textbf{2001},
  \emph{64}, 195120\relax
\mciteBstWouldAddEndPuncttrue
\mciteSetBstMidEndSepPunct{\mcitedefaultmidpunct}
{\mcitedefaultendpunct}{\mcitedefaultseppunct}\relax
\EndOfBibitem
\bibitem[Not()]{Note-2}
Notice, that our non-counterpoise corrected calculations based on (T,Q)
  extrapolation will still include a sizable basis set incompleteness error for
  atomization energies. However, our qualitative conclusions will be
  valid.\relax
\mciteBstWouldAddEndPunctfalse
\mciteSetBstMidEndSepPunct{\mcitedefaultmidpunct}
{}{\mcitedefaultseppunct}\relax
\EndOfBibitem
\bibitem[Paier \latin{et~al.}(2010)Paier, Janesko, Henderson, Scuseria,
  Gr{\"{u}}neis, and Kresse]{Paier2010a}
Paier,~J.; Janesko,~B.~G.; Henderson,~T.~M.; Scuseria,~G.~E.;
  Gr{\"{u}}neis,~A.; Kresse,~G. {Erratum: Hybrid functionals including random
  phase approximation correlation and second-order screened exchange (Journal
  of Chemical Physics (2010) 132 (094103))}. \emph{J. Chem. Phys.}
  \textbf{2010}, \emph{133}, 2009--2011\relax
\mciteBstWouldAddEndPuncttrue
\mciteSetBstMidEndSepPunct{\mcitedefaultmidpunct}
{\mcitedefaultendpunct}{\mcitedefaultseppunct}\relax
\EndOfBibitem
\bibitem[Řez{\'{a}}{\v{c}} \latin{et~al.}(2011)Řez{\'{a}}{\v{c}}, Riley, and
  Hobza]{Rezac2011}
Řez{\'{a}}{\v{c}},~J.; Riley,~K.~E.; Hobza,~P. {S66: A well-balanced database
  of benchmark interaction energies relevant to biomolecular structures}.
  \emph{J. Chem. Theory Comput.} \textbf{2011}, \emph{7}, 2427--2438\relax
\mciteBstWouldAddEndPuncttrue
\mciteSetBstMidEndSepPunct{\mcitedefaultmidpunct}
{\mcitedefaultendpunct}{\mcitedefaultseppunct}\relax
\EndOfBibitem
\bibitem[Not()]{Note-3}
With 0.52 kcal/mol, the MAD for RPA@PBE is in excellent agreement with the 0.61
  kcal/mol MAD obtained by Nguyen \emph{et. al.} in ref.~\citen{Nguyen2020},
  which has been obtained with GTO-type basis sets and 50 \% counterpoise
  correction instead of 100 \%. This shows, that our interaction energies are
  well converged with respect to the basis set size.\relax
\mciteBstWouldAddEndPunctfalse
\mciteSetBstMidEndSepPunct{\mcitedefaultmidpunct}
{}{\mcitedefaultseppunct}\relax
\EndOfBibitem
\bibitem[Santra \latin{et~al.}(2019)Santra, Sylvetsky, and Martin]{Santra2019a}
Santra,~G.; Sylvetsky,~N.; Martin,~J.~M. {Minimally Empirical Double-Hybrid
  Functionals Trained against the GMTKN55 Database: RevDSD-PBEP86-D4,
  revDOD-PBE-D4, and DOD-SCAN-D4}. \emph{J. Phys. Chem. A} \textbf{2019},
  \emph{123}, 5129--5143\relax
\mciteBstWouldAddEndPuncttrue
\mciteSetBstMidEndSepPunct{\mcitedefaultmidpunct}
{\mcitedefaultendpunct}{\mcitedefaultseppunct}\relax
\EndOfBibitem
\bibitem[Mehta \latin{et~al.}(2018)Mehta, Casanova-P{\'{a}}ez, and
  Goerigk]{Mehta2018}
Mehta,~N.; Casanova-P{\'{a}}ez,~M.; Goerigk,~L. {Semi-empirical or
  non-empirical double-hybrid density functionals: Which are more robust?}
  \emph{Phys. Chem. Chem. Phys.} \textbf{2018}, \emph{20}, 23175--23194\relax
\mciteBstWouldAddEndPuncttrue
\mciteSetBstMidEndSepPunct{\mcitedefaultmidpunct}
{\mcitedefaultendpunct}{\mcitedefaultseppunct}\relax
\EndOfBibitem
\bibitem[Grimme \latin{et~al.}(2010)Grimme, Antony, Ehrlich, and
  Krieg]{Grimme2010}
Grimme,~S.; Antony,~J.; Ehrlich,~S.; Krieg,~H. {A consistent and accurate ab
  initio parametrization of density functional dispersion correction (DFT-D)
  for the 94 elements H-Pu}. \emph{J. Chem. Phys.} \textbf{2010}, \emph{132},
  154104\relax
\mciteBstWouldAddEndPuncttrue
\mciteSetBstMidEndSepPunct{\mcitedefaultmidpunct}
{\mcitedefaultendpunct}{\mcitedefaultseppunct}\relax
\EndOfBibitem
\bibitem[Grimme \latin{et~al.}(2011)Grimme, Ehrlich, and Goerigk]{Stefan2011}
Grimme,~S.; Ehrlich,~S.; Goerigk,~L. {Effect of the damping function in
  dispersion corrected density functional theory}. \emph{J. Comput. Chem.}
  \textbf{2011}, \emph{32}, 1456--1465\relax
\mciteBstWouldAddEndPuncttrue
\mciteSetBstMidEndSepPunct{\mcitedefaultmidpunct}
{\mcitedefaultendpunct}{\mcitedefaultseppunct}\relax
\EndOfBibitem
\bibitem[Vl{\v{c}}ek(2019)]{Vlcek2019}
Vl{\v{c}}ek,~V. {Stochastic Vertex Corrections: Linear Scaling Methods for
  Accurate Quasiparticle Energies}. \emph{J. Chem. Theory Comput.}
  \textbf{2019}, \emph{15}, 6254--6266\relax
\mciteBstWouldAddEndPuncttrue
\mciteSetBstMidEndSepPunct{\mcitedefaultmidpunct}
{\mcitedefaultendpunct}{\mcitedefaultseppunct}\relax
\EndOfBibitem
\bibitem[Beuerle \latin{et~al.}(2018)Beuerle, Graf, Schurkus, and
  Ochsenfeld]{Beuerle2018}
Beuerle,~M.; Graf,~D.; Schurkus,~H.~F.; Ochsenfeld,~C. {Efficient calculation
  of beyond RPA correlation energies in the dielectric matrix formalism}.
  \emph{J. Chem. Phys.} \textbf{2018}, \emph{148}, 204104\relax
\mciteBstWouldAddEndPuncttrue
\mciteSetBstMidEndSepPunct{\mcitedefaultmidpunct}
{\mcitedefaultendpunct}{\mcitedefaultseppunct}\relax
\EndOfBibitem
\bibitem[Sedlak \latin{et~al.}(2013)Sedlak, Janowski, Pitoň{\'{a}}k,
  Řez{\'{a}}{\v{c}}, Pulay, and Hobza]{Sedlak2013}
Sedlak,~R.; Janowski,~T.; Pitoň{\'{a}}k,~M.; Řez{\'{a}}{\v{c}},~J.;
  Pulay,~P.; Hobza,~P. {Accuracy of quantum chemical methods for large
  noncovalent complexes}. \emph{J. Chem. Theory Comput.} \textbf{2013},
  \emph{9}, 3364--3374\relax
\mciteBstWouldAddEndPuncttrue
\mciteSetBstMidEndSepPunct{\mcitedefaultmidpunct}
{\mcitedefaultendpunct}{\mcitedefaultseppunct}\relax
\EndOfBibitem
\bibitem[Doser \latin{et~al.}(2009)Doser, Lambrecht, Kussmann, and
  Ochsenfeld]{Doser2009a}
Doser,~B.; Lambrecht,~D.~S.; Kussmann,~J.; Ochsenfeld,~C. {Linear-scaling
  atomic orbital-based second-order M{\o}ller-Plesset perturbation theory by
  rigorous integral screening criteria}. \emph{J. Chem. Phys.} \textbf{2009},
  \emph{130}, 064107\relax
\mciteBstWouldAddEndPuncttrue
\mciteSetBstMidEndSepPunct{\mcitedefaultmidpunct}
{\mcitedefaultendpunct}{\mcitedefaultseppunct}\relax
\EndOfBibitem
\bibitem[Pinski \latin{et~al.}(2015)Pinski, Riplinger, Valeev, and
  Neese]{Pinski2015}
Pinski,~P.; Riplinger,~C.; Valeev,~E.~F.; Neese,~F. {Sparse maps - A systematic
  infrastructure for reduced-scaling electronic structure methods. I. An
  efficient and simple linear scaling local MP2 method that uses an
  intermediate basis of pair natural orbitals}. \emph{J. Chem. Phys.}
  \textbf{2015}, \emph{143}, 034108\relax
\mciteBstWouldAddEndPuncttrue
\mciteSetBstMidEndSepPunct{\mcitedefaultmidpunct}
{\mcitedefaultendpunct}{\mcitedefaultseppunct}\relax
\EndOfBibitem
\bibitem[Nagy \latin{et~al.}(2016)Nagy, Samu, and K{\'{a}}llay]{Nagy2016}
Nagy,~P.~R.; Samu,~G.; K{\'{a}}llay,~M. {An Integral-Direct Linear-Scaling
  Second-Order M{\o}ller-Plesset Approach}. \emph{J. Chem. Theory Comput.}
  \textbf{2016}, \emph{12}, 4897--4914\relax
\mciteBstWouldAddEndPuncttrue
\mciteSetBstMidEndSepPunct{\mcitedefaultmidpunct}
{\mcitedefaultendpunct}{\mcitedefaultseppunct}\relax
\EndOfBibitem
\bibitem[Mezei \latin{et~al.}(2019)Mezei, Ruzsinszky, and
  K{\'{a}}llay]{Mezei2019}
Mezei,~P.~D.; Ruzsinszky,~A.; K{\'{a}}llay,~M. {Reducing the Many-Electron
  Self-Interaction Error in the Second-Order Screened Exchange Method}.
  \emph{J. Chem. Theory Comput.} \textbf{2019}, \emph{15}, 6607--6616\relax
\mciteBstWouldAddEndPuncttrue
\mciteSetBstMidEndSepPunct{\mcitedefaultmidpunct}
{\mcitedefaultendpunct}{\mcitedefaultseppunct}\relax
\EndOfBibitem
\end{mcitethebibliography}


\begin{tocentry}
\includegraphics[width=\textwidth]{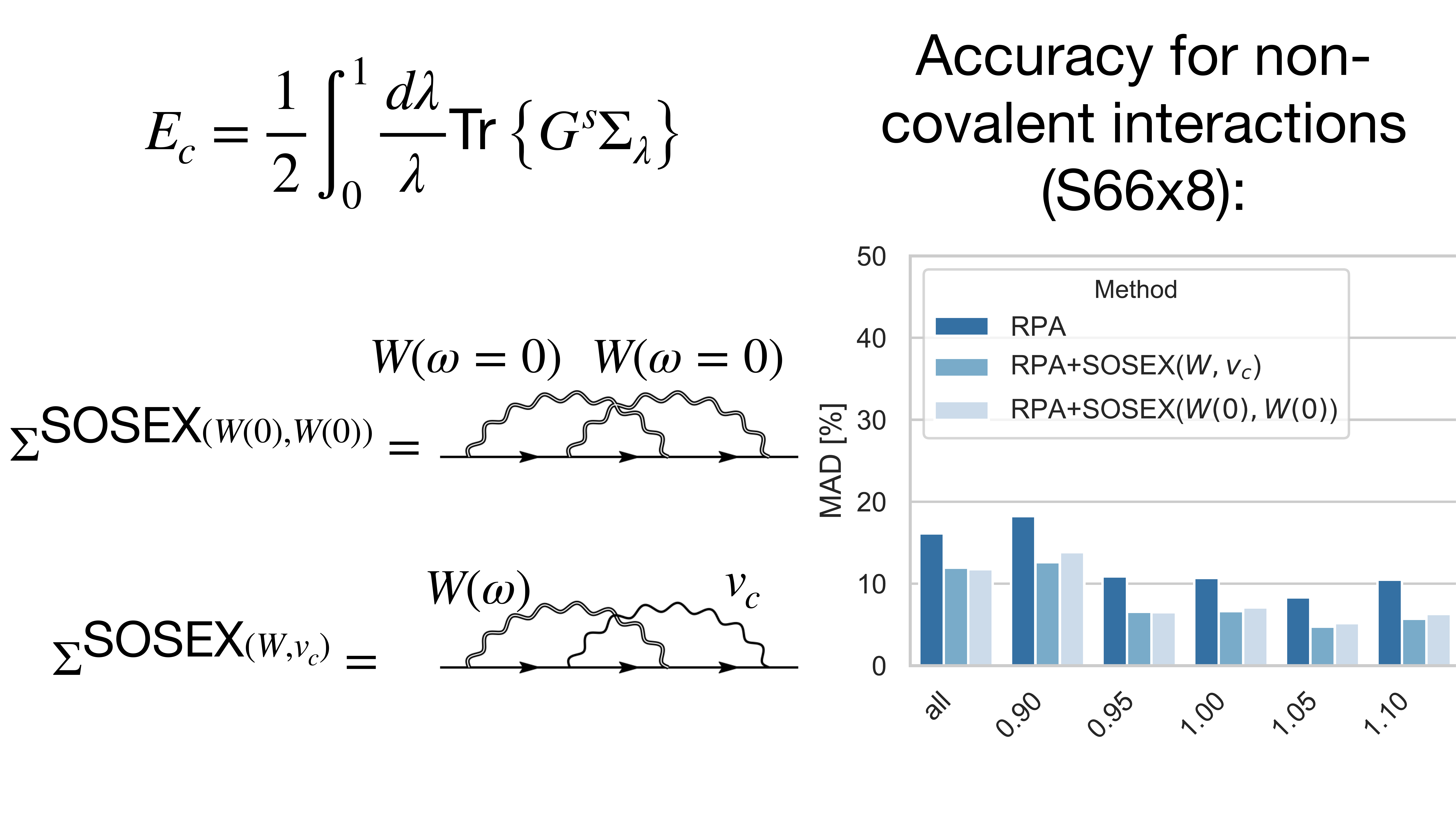}
\end{tocentry}


\end{document}